\documentclass{aa}
\usepackage{graphicx}
\usepackage[varg]{txfonts}
\usepackage{subfigure}
\usepackage{natbib}
\usepackage{multirow}
\bibpunct{(}{)}{;}{a}{}{,}
\usepackage[colorlinks=true,linkcolor=blue, citecolor=blue, filecolor=blue, urlcolor=blue]{hyperref}

\begin{document}

\title{The PSF Smoothing Effect on Concentration-Related Parameters of High Redshift Galaxies in HST and JWST}

\author{Jia-Hui Wang\inst{1,2}\and Zhao-Yu Li\inst{1,2}\thanks{Correspondence should be addressed to: lizy.astro@sjtu.edu.cn}
\and Ming-Yang Zhuang\inst{3}\and Luis C. Ho\inst{4,5} \and Li-Min Lai\inst{1,2}}

\institute{Department of Astronomy, School of Physics and Astronomy, Shanghai Jiao Tong University, 800 Dongchuan Road, Shanghai 200240, China %\email{lizy.astro@sjtu.edu.cn}
\and Key Laboratory for Particle Astrophysics and Cosmology (MOE) / Shanghai Key Laboratory for Particle Physics and Cosmology, Shanghai 200240, China
\and Department of Astronomy, University of Illinois Urbana-Champaign, Urbana, IL 61801, USA
\and Kavli Institute for Astronomy and Astrophysics, Peking University, Beijing 100871, China
\and Department of Astronomy, School of Physics, Peking University, Beijing 100871, China
}

\abstract{
We perform a comprehensive investigation of the PSF smoothing effect on the measurement of concentration-related parameters ($C$, Gini, $M_{20}$) of high redshift galaxies in the HST and JWST surveys. Our sample contains massive galaxies ($10^{9.5}{\rm M_{\odot}}\leq M_{\ast} \leq 10^{11.5}{\rm M_{\odot}}$) from the CANDELS/EGS survey (at redshift $0 < z < 2$), and the CEERS survey (at redshift $1 < z < 3$). The non-parametric concentration-related parameters ($R_{20}$, $R_{80}$, $C$, Gini, $M_{20}$) and the model-dependent parameters ($n$, $R_e$) of these galaxies are derived from {\tt Statmorph} and G{\small ALFIT}, respectively. The best-fit S\'{e}rsic index ($n$) derived from image modelling is generally robust against the PSF smoothing effect and can be used to describe the intrinsic light distribution of galaxies. On the other hand, the concentration-related parameters are significantly affected by the PSF smoothing effect since they are directly calculated from the pixels of galaxy images. We try to evaluate the PSF smoothing effect by comparing the concentration-related parameters to the S\'{e}rsic index in both observations and mock images. We find that the concentration index is generally underestimated, especially for smaller galaxies with higher S\'{e}rsic index (eventually converging to the concentration index of the PSF). However, galaxies with lower S\'{e}rsic index ($n\leq1$) or larger relative size ($R_e/{\rm FWHM} > 3$) are less affected by the PSF smoothing effect. Tests with idealised mock images reveal that overestimating the measured $R_{20}/R_e$ ratio leads to underestimating the concentration index $C$. Another commonly used concentration index $C_{59}$, derived from $R_{50}$ and $R_{90}$ values, is less affected by the PSF. The Gini coefficient and the absolute $M_{20}$ statistic also show similar behaviour as the concentration index. Caution should be taken for the possible correction of the concentration-related parameters, where both the relative size and the S\'{e}rsic index of the galaxy are important. We also generate high redshift artificial images from the low redshift HST observations and confirm that the traditional correction method which simply adds a single term to the non-parametric indicators of galaxies at higher redshifts, is unable to reliably recover the true distribution of the structural parameters. Compared to the HST images, the PSF smoothing is much less severe for images in the CEERS survey (for the short-wavelength filters) due to the much higher spatial resolution. In fact, it is better to use the S\'{e}rsic index rather than the non-parametric morphology indicators to trace the light concentration for galaxies at high redshifts.  From the single S\'{e}rsic modelling of the HST and JWST images, we also confirm that galaxies at higher redshift are more compact with smaller $R_e$. The lower mass galaxies are more disc-like ($n\sim1$) compared to the higher mass galaxies that are more spheroid dominated ($n\sim3$).} 


\titlerunning{PSF Smoothing Effect on the Concentration-related Parameters of Galaxies}
\authorrunning{Jia-Hui Wang et al. }
\maketitle

\section{Introduction}
The morphology of a galaxy encodes important information about the physical processes that contributed to its formation and evolution \citep{Conselice_2008,Mortlock_2013,Conselice_2014,Huertas-Company_2016}. In $\Lambda$CDM cosmology, galaxies grow hierarchically \citep{White_1978} and are shaped by both external and internal mechanisms such as mergers, tidal interactions, ram pressure stripping, starvation and harassment \citep{Gunn_1972,Larson_1980,Moore_1996}.
On large time scales, galaxy evolution is dominated by secular processes i.e. a slow rearrangement of energy and mass involving the bar, spiral arms and dark matter halo \citep{Kormendy_2004}.

Nearby galaxies are traditionally classified by their morphology according to the Hubble sequence \citep{Hubble_1926}, but it is less straightforward to classify high redshift galaxies because of observational limitations or biases \citep{Sandage_2005}.
In addition to the traditional visual classification, such as the Galaxy Zoo project \citep{Lintott_2011},
more quantitative methods are also widely used, such as surface brightness profile analysis, single S\'{e}rsic component fitting, and multiple component fitting with bulge-disc decomposition \citep[e.g.][]{sersic_1968,Freeman_1970,Kormendy_1977a,Kormendy_1977b,Simien_1986,Caon_1993,Peng_2002,Simard_2002,deSouza_2004,Laurikainen_2005,Gadotti_2008,Gadotti_2009,Simard_2011,Kormendy_2012,Bruce_2014,Meert_2015,Erwin_2015,Mendez-Abreu_2017,Gao_2018,Bottrell_2019,Gao_2020}. These methods can be used to quantify the morphological parameters to trace the structural evolution of galaxies at both low and high redshifts \citep{Trujillo_2007,Buitrago_2008,van-der-Wel_2014, Allen_2017,Whitney_2019,Whitney_2020}.

Another widely used galaxy morphology classification is the non-parametric method in which morphological indicators are directly calculated from image pixels, without modelling the light distribution. Therefore, the advantage of using non-parametric methods for galaxy classification is that no a-priori assumptions about a galaxy model are needed. Some of the morphological indicators that can be derived in this way are the concentration index ($C$), the asymmetry index ($A$), the clumpiness index ($S$),  the Gini coefficient ($G$) and the second-order moment of the brightest 20\% of the galaxy light ($M_{20}$) \citep{Kent_1985,Conselice_2000,Conselice_2003,Lotz_2004,Lotz_2008}. These values enable an unbiased quantification of galaxy morphology, facilitating comparison with various physical properties (mass, size, colour, starformation rate, etc.) in order to gain a better understanding of the structural growth and evolution of galaxies. \citep{statmorph_2019,Yesuf_2021,Whitney_2021,Nersesian_2023,Yao_2023}.
 The non-parametric morphology indicators can also be used to separate galaxies into spheroids, discs and merging galaxies in the $C$-$A$ and $G$-$M_{20}$ spaces \citep{Conselice_2003,Conselice_2008,Lotz_2008,Lopez-Sanjuan_2009}. 
However, for high redshift galaxies, these measurements could be biased due to the smoothing effect of the point spread function (PSF) quantifying the light spread of a point source on the detector. The `spread' is actually a blurring effect caused by various factors such as atmosphere turbulence (for ground observations), lens imperfections, mirror diffraction, detector limitations, observing filters etc.
It is important to understand the bias introduced by such effect on non-parametric indicators before any scientific analysis can be performed.

Recently, \cite{Whitney_2021} estimated the non-parametric morphology indicators in high redshift galaxies ($0.5<z<3$) in the Cosmic Assembly Near-IR Deep Extragalactic Legacy Survey (CANDELS) of the Hubble Space Telescope (HST) and confirmed that galaxies are more concentrated (have larger $C$ values) at higher redshifts. In addition, \cite{Yao_2023} analysed similar redshift galaxies ($0.8 < z < 3$) in the Cosmic Evolution Early Release Science (CEERS) survey of the James Webb Space Telescope (JWST) and found that the concentration index actually decreases for galaxies at higher redshift with $M_{\ast} > 10^{10}{\rm M_{\odot}}$, in contrast with the findings of \cite{Whitney_2021}. In fact, the JWST telescope has revealed an unexpected high fraction of discs in the early Universe, at  $z > 3$ \citep{Kartaltepe_2023,Sun_2023,Ferreira_2023}, classified by either the S\'{e}rsic index or the non-parametric morphological indicators.

To properly account for the redshift effect on the morphology indicators, previous works have used $z \sim 1$ images to create mock galaxy images at high redshift \citep{Whitney_2021}. 
The structural parameters of the mock images are then compared with the true values to derive an empirical correction for the non-parametric morphology indicators obtained from observations.
Recently, \cite{Yu_2023} improved the method by considering the ratio of the galaxy size (i.e. the Petrosian radius $R_p$\footnote{$R_p$ is defined as the radius where the surface brightness is 20\% of the average surface brightness within $R_p$ \citep{Petrosian_1976}.}) to the PSF full-width-at-half-maximum (FWHM) as an additional parameter ($R_p/{\rm FWHM}$). They used the nearby galaxy images from the Dark Energy Spectroscopic Instrument (DESI) to generate artificial images of high redshift galaxies observed in the CEERS survey, and found that the PSF smoothing leads to an underestimation of the concentration index ($C$) that is more severe for the galaxies of higher intrinsic $C$ values. They also found that the dispersion of the corrected $C$ value increases with decreasing $R_p/{\rm FWHM}$ (see their Fig. 9). 
On the other hand, since the angular resolution of the JWST galaxies at $0.8<z<3$ is roughly similar, such empirical corrections were not considered in \cite{Yao_2023}. 
In the JWST era, to better understand the evolution of high redshift galaxies, it is important to evaluate the robustness of these non-parametric structural quantities and their corrections. In fact, \cite{Andrae_2011} has already discussed the limits of the non-parametric structural measurements and concluded that they cannot be estimated independently from the images without potentially introducing  serious biases.
As shown in \cite{Davari_2014}, the S\'{e}rsic index and the effective radius of high redshift galaxies can be well recovered by 2D image modelling over a wide range of signal-to-noise ratios (S/N). Therefore, the S\'{e}rsic index can be used to reflect the intrinsic light profile shape of high redshift galaxies \citep{Trujillo_2001,Graham_2005}.

In this work, we mainly focus on the concentration-related quantities ($C$, Gini, $M_{20}$) since they are sensitive to the galaxy light distribution which suffers the most from the effect of PSF smoothing. The asymmetry ($A$) and clumpiness ($S$) indices are more affected by poor visibility of the outer region of galaxies due to surface brightness dimming. We decide to not include $A$ and $S$ in this study to avoid over-interpretation of our test results.
To better evaluate the PSF smoothing effect on the concentration-related quantities, we compare the $C$, Gini, $M_{20}$ values and the S\'{e}rsic index of high redshift galaxies from the Extended Groth Strip (EGS) field in CANDELS \citep{Grogin_2011,Koekemoer_2011} and CEERS \citep{Finkelstein_2023}. Mock images with different S\'{e}rsic indices are also constructed and analysed to help understand the reason for such bias.

The paper is organised as follows. The data and methods are described in Sect. \ref{sec:data} and \ref{sec:methods}, respectively. Sect. \ref{sec:results} presents the main results while Sect. \ref{sec:discussion} compares our results with previous works, and further investigates the effect of PSF smoothing for different light profile shapes using mock images. The results are summarised in Sect. \ref{sec:conclusion}. 
%Here we adopt the standard cosmology parameters ($H_0 = 70\,{\rm km}^{-1} {\rm s}^{-1} {\rm Mpc}^{-1}$, $\Omega_m = 0.3$, and $\Omega_{\Lambda} = 0.7$). -not here

\section{Data}\label{sec:data}

Our data sample consists of two sets of images: the first from the CANDELS/EGS survey \citep{Grogin_2011,Koekemoer_2011}, taken with the Advanced Camera for Surveys (ACS) and the Wide Field Camera 3 infrared channel (WFC3/IR) on HST; the second from the CEERS survey \citep{Finkelstein_2023}, taken with the Near Infrared Camera (NIRCam) on JWST.

\subsection{HST EGS}
The EGS mosaic was produced following \cite{Koekemoer_2011}, with a pixel scale of 0.06 \arcsec/pixel. The ACS images have $\sim$6000\,s and $\sim$12,000\,s exposure times in the F606W and F814W filters respectively, reaching the $5\sigma$ depth of 28.8 and 28.2 mag \citep{Stefanon_2017}. The WFC3/IR images have $\sim$1300\,s and $\sim$2700\,s exposure times in the F125W and F160W filters, both reaching the AB limiting magnitude of 27.6 mag. 
From the EGS catalogue \citep{Stefanon_2017}, we choose a sample of galaxies at $0 < z < 2$ with stellar masses within $10^{9.5}{\rm M_{\odot}}\leq M_{\ast} \leq 10^{11.5}{\rm M_{\odot}}$, to ensure sample completeness. According to the morphological catalogue\footnote{\url{http://rainbowx.fis.ucm.es/Rainbow_navigator_public/}} in \cite{Huertas_2015}, only the spheroids and discs are selected in the following analysis. We exclude irregular galaxies to avoid large uncertainties in the measurements.
For the photometric redshifts and stellar mass measurements, the median values of the different estimations from \cite{Stefanon_2017} are adopted. The images in different filters have different redshift ranges to ensure a similar rest-frame optical wavelength. The properties of the HST sample are shown in the top four rows of Table \ref{table:1}. We use G{\small ALAPAGOS} \citep{Barden_2012} to generate cutouts of galaxies, encompassing twice the Kron ellipse \citep{Kron_1980}. The size of the cutout is sufficient for background measurement and photometric analysis. The segmentation map of nearby objects is convolved with the {\tt Tophat2DKernel(5)} from {\tt Astropy} \citep{astropy_2022} to smooth the edges of the segment; the mask image is then generated for the F160W mosaic cutout.

Galaxies observed at different wavelengths exhibit different structures causing the non-parametric morphology indicators to vary with wavelength \citep{Baes_2020,Nersesian_2023,Yao_2023}. We measure all the structural quantities in the rest-frame wavelength at approximately $\lambda_{\rm rest}\approx 4000-6000{\rm \mathring{A}}$, to avoid wavelength dependent variations. The rest-frame wavelengths of each filter are also listed in Table \ref{table:1} for each redshift. The PSF FWHM and the number of the galaxies for each filter are also listed. The PSF is empirically constructed using the bright and non-saturated field stars (see Sect. \ref{sec:psf}). To ensure robust structural measurements, we limit our EGS sample to galaxies above a minimum S/N of 3 and $R_p>5 $ pixels and obtain 2305 galaxies. More details about the influence of the S/N on the morphology indicators are shown in Sect. \ref{sec:snr}. 

\begin{table}
%\tablenum{1}
\caption{Properties of the sample\label{table:1}}
\begin{tabular}{cccccc}
\hline\hline
$z$ & Filter Name& $\lambda_{\rm rest}$ &${\rm FWHM}$&Number & Depth \\ 
 & & $(\rm \mathring{A})$ & (arcsec)& &(mag)\\
(1) & (2) & (3) &(4) &(5) & (6) \\
\hline
0.25 & F606W & $4850$ & $0.122$ & 175 & 28.8 \\
0.75 & F814W & $4650$ & $0.123$ & 614 & 28.2 \\
1.25 & F125W & $5550$ & $0.194$ & 850 & 27.6 \\
1.75 & F160W & $5820$ & $0.195$ & 666 & 27.6 \\
\hline
1.25 & F115W & $5110$ & $0.049$ & 149 & 29.2 \\
1.75 & F150W & $5450$ & $0.054$ & 182 & 29.0 \\
2.25 & F150W & $4610$ & $0.054$ & 113 & 29.0 \\
2.75 & F200W & $5330$ & $0.070$ & 80 & 29.2 \\
\hline
\end{tabular}
\tablefoot{The top four rows with redshift range $0< z <2$ are from the EGS images. The bottom four rows with redshift range $1< z<3$ are from the CEERS images. Column 1 gives the midpoint of each redshift bin with $\Delta z = 0.5$. Column 2 gives the filter name. Column 3 gives the corresponding rest-frame wavelength. Column 4 gives the PSF FWHM for each filter. Column 5 gives the number of the galaxies. Column 6 gives the $5\sigma$ depths in the AB limiting magnitude.}
%\vspace{-3em}
\end{table}

\subsection{JWST CEERS}
The NIRCam imaging of CEERS covers a total of $100\, {\rm arcmin}^2$ in the F115W, F150W, F200W, F277W, F356W, F410M, and F444W filters down to a $5\sigma$ depth ranging from 28.8 to 29.7 mag \citep{Finkelstein_2023}. The first set of CEERS observations was taken on 21 June 2022 in four pointings in the EGS field. The image mosaic with a pixel scale of 0.03 \arcsec/pixel has additional corrections such as snowballs, wisps, 1/f noise, and background subtraction. A detailed description of the reduction process can be found in \cite{Bagley_2023}.
We choose galaxies with $1<z<3$ in the F115W, F150W, and F200W filters, which correspond to similar rest-frame optical wavelength. Finally, our JWST/CEERS sample contains 524 galaxies with ${\rm S/N}>3$ and $R_p>5$ pixels \citep{Stefanon_2017}. The properties of the CEERS images are given in the bottom four rows of Table \ref{table:1}. 

\begin{figure}
%\vspace{-2em}
\centering
\includegraphics[width=0.47\textwidth]{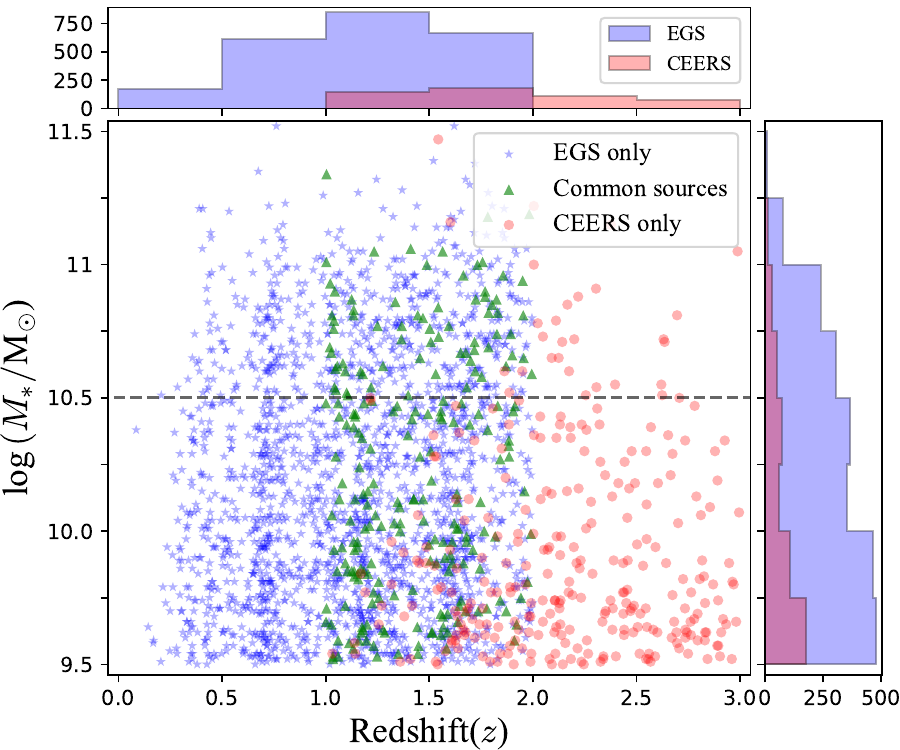}
\caption{Mass and redshift distribution of the galaxies in our sample. The blue stars indicate galaxies in the EGS images, and the red circles indicate galaxies in the CEERS images. The green triangles indicate the common sources between the EGS and CEERS images. The dashed horizontal line at ${\rm log}(M_{*}/{\rm M_{\odot}}) = 10.5$ indicates the boundary between the high-mass galaxies and low-mass galaxies in this study. \label{fig:z-m}}
%\vspace{-2em}
\end{figure}

Fig. \ref{fig:z-m} shows the mass and redshift distribution of our sample. The blue stars indicate the galaxies in the EGS images, and the red circles indicate the galaxies in the CEERS images. There are 228 galaxies  (green triangles) in common between the EGS and CEERS images. In Sect. \ref{sec:dif} , the structural parameters of the common sources are compared for consistency checks.

\section{Methods} \label{sec:methods}
This section describes the method employed to measure the structural parameters ($C$, Gini, $M_{20}$, S\'{e}rsic index $n$ and effective radius $R_e$), the PSF construction and the S/N test. % \citep{statmorph_2019}, measured with G{\small ALFIT}    

\subsection{Structural quantities measurement}
%{\tt Statmorph} is a Python package that can derive $C$, Gini, $M_{20}$, and other morphological quantities, such as ellipticity, total flux, and Petrosian radius.

The concentration index quantifies how much light is distributed in the centre of a galaxy as opposed to its outer parts \citep{Conselice_2003},

\begin{equation}
    C = 5\,\textup{log}\left(\frac{R_{80}}{R_{20}}\right),
\label{equ:c}
\end{equation}
where $R_{20}$ and $R_{80}$ are the radii of circular apertures containing 20 and 80 percent of the Petrosian flux, respectively. The Petrosian flux is defined as the flux within $1.5R_p$.

The Gini coefficient is traditionally used in economics to quantify the wealth inequality in a population; in astronomy, it is used to quantify the relative distribution of the flux within a galaxy. It is related to the light concentration but differs from the concentration index; the Gini coefficient only accounts for the clustering of bright pixels, regardless of their spatial distributions e.g. whether these bright pixels are distributed in the galaxy centre or outskirts \citep{Abraham_2003, Lotz_2004}. Assuming that a galaxy image has $n$ pixels with flux $I_i$ in each pixel, the Gini coefficient can be computed as \citep{Gerald_1962},

\begin{equation}
    G = \frac{1}{|\bar{I}|n(n-1)} \sum\limits^n\limits_{i=1}(2i - n -1 )|I_i|,
\end{equation}
where $|I_i|$ indicates the $i$-th brightest pixel, and $|\bar{I}|$ represents the mean value of the pixels. $G=1$ is obtained when all of the flux is concentrated in a single pixel, while a homogeneous brightness distribution yields $G=0$. The Gini coefficient depends on the S/N \citep{Lisker_2008}, which is tested in Sect. \ref{sec:snr}.

The second-order moment statistic is similar to the concentration index in that it indicates whether the light is concentrated within an image (not necessarily in the centre but at any location in the galaxy) \citep{Lotz_2004}. The total second-order central moment, $M_{\rm tot}$, is calculated as
\begin{equation}
    M_{\rm tot} = \sum\limits^n\limits_{i=1} M_i\equiv \sum\limits^n\limits_{i=1} I_i[(x_i-x_c)^2+(y_i-y_c)^2],
\end{equation}
where $(x_i,y_i)$, $(x_c,y_c)$ are the positions of a pixel and the galaxy centre, respectively. $I_i$ is $i$-th brightest pixel value. $M_{20}$ is defined as the normalised second-order moment of the brightest 20\% of the total flux, 
\begin{equation}
    M_{20} \equiv \textup{log}\left(\frac{\sum_i M_i}{M_{\rm tot}}\right), \,{\rm while} \sum\limits_{i} I_i <0.2I_{\rm tot}.
\end{equation}
In particular, \cite{Lotz_2008} suggested that the $G$-$M_{20}$ space can be used to roughly separate spheroids, discs, and merging galaxies at $0.2<z<0.4$. The PSF smoothing effect on the distribution of galaxies in the $G$-$M_{20}$ space will also be examined in Sect. \ref{sec:gm}.

$n$ and $R_e$ are derived by fitting the galaxy image with G{\small ALFIT} \citep{Peng_2002,Peng_2010} using a single S\'{e}rsic component \citep{sersic_1968}.
The fitting parameters include galaxy centre, total magnitude ($m$), $R_e$ measured along the major axis, $n$, axis ratio ($q$), and position angle (P.A.). Initial guesses for these parameters are taken from the {\tt Statmorph} output. We fix the sky value as the median of pixels in four background regions around the galaxy (usually $\sim0$). Our fitting results are generally consistent with \cite{van-der-Wel_2012}.

\subsection{Correlation between the concentration index and the S\'{e}rsic index}
\begin{figure}
%\vspace{-2em}
\centering
\includegraphics[width=0.47\textwidth]{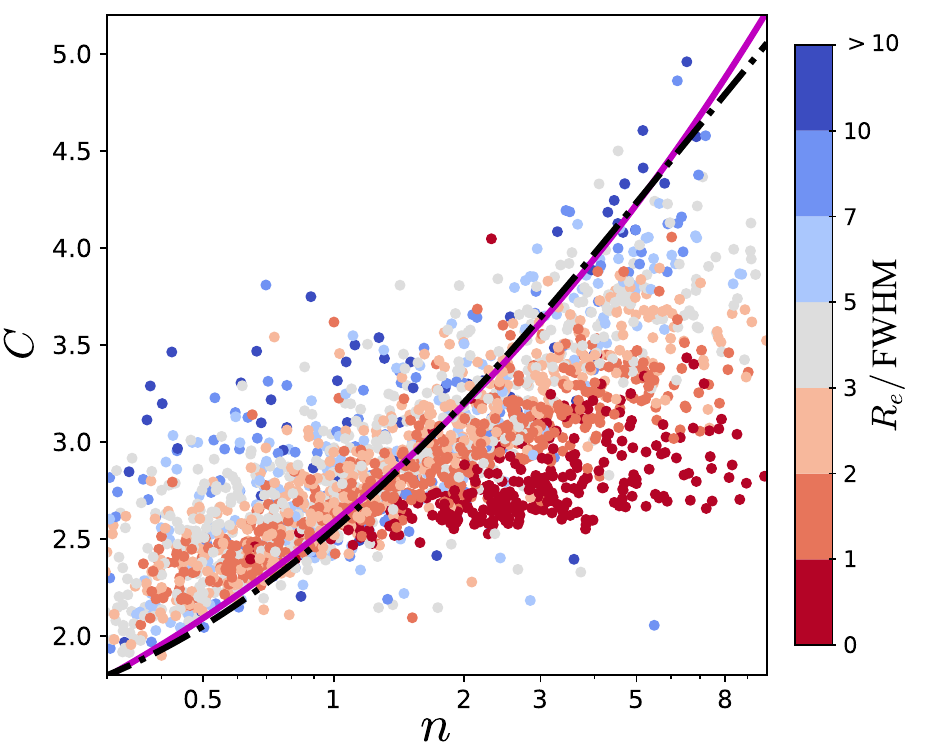}
\caption{Correlation between the concentration index and the S\'{e}rsic index for the EGS images. The black dot-dashed curve is the numerical relationship derived by integrating the best-fit S\'{e}rsic model out to 1.5$R_{p}$. The purple solid curve is the empirical relationship of Eq. (\ref{cn_equ}). The colour of the points marks the relative size of the galaxies ($R_e/{\rm FWHM}$). The data points generally agree with the numerical expectation, but showing large scatters, especially for the galaxies with the smaller relative sizes ($R_e/{\rm FWHM}\leq1$).  \label{fig:c_all}}
%\vspace{-2em}
\end{figure}

Both the S\'{e}rsic index and the concentration index reflect the shape of the light profile. Utilising the fact that $R_{20}$ and $R_{80}$ contain 20\% and 80\% of the total flux respectively, the relationship between $C$ and $n$ for a given S\'{e}rsic profile can be derived numerically. As shown in \cite{Andrae_2011}, the relationship can be approximated as
\begin{equation}
    C \approx 2.586\,n^{0.305}.
\label{cn_equ}
\end{equation}

\cite{Andrae_2011} has used galaxies in the COSMOS survey to demonstrate that $C$ and $n$ are statistically consistent, but exhibiting a large scatter. Similar to \cite{Andrae_2011}, we also plot the distribution of EGS images in the $C$-$n$ space in Fig. \ref{fig:c_all}.
We can see that the galaxies with the large relative sizes ($R_e/{\rm FWHM} >3$, blue) generally follow the expected numerical relationship. Therefore, the S\'{e}rsic index can be an estimate of the intrinsic concentration of a galaxy \citep{Trujillo_2001,Graham_2005}. However, those with the smaller sizes ($R_e/{\rm FWHM} \leq1$, dark red) have significantly underestimated concentration index.

In fact, the directly measured $C$ values (without correction) of high redshift galaxies, are usually smaller than the $C$ values of lower redshift galaxies. $R_{20}$ and $R_{80}$ are the two key parameters in the derivation of the $C$ value (see Eq.~\ref{equ:c}): since $R_{20}$ is closer to the central region of the galaxy, it is more overestimated than $R_{80}$, resulting in a lower $C$ value. Since the S\'{e}rsic index is derived from image modelling including PSF smoothing, it is a more reliable indicator of the light profile shapes of high redshift galaxies \citep{Trujillo_2007,Davari_2014}.

\subsection{PSF construction} \label{sec:psf}

\begin{figure}
\centering
\includegraphics[width=0.45\textwidth]{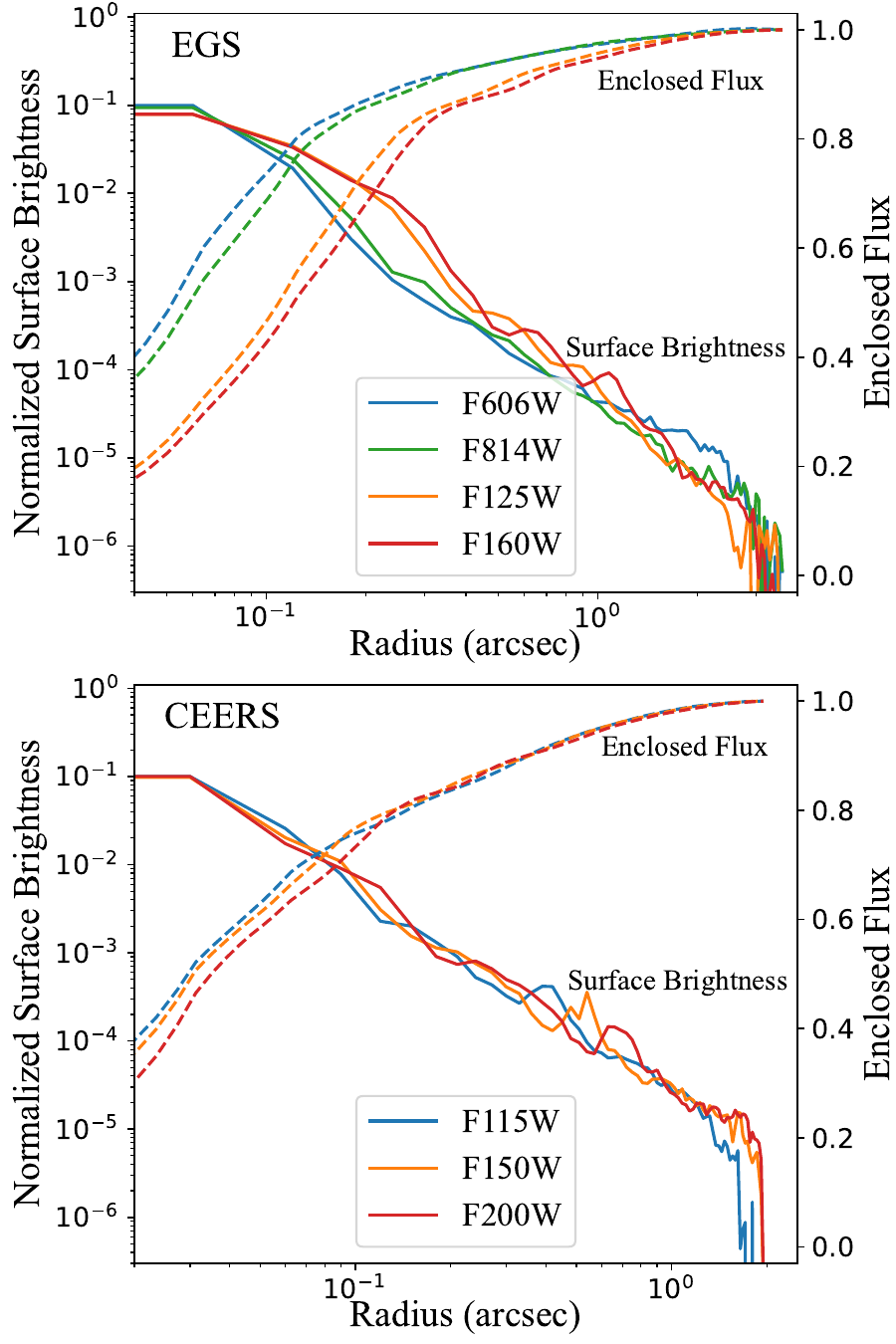}
\caption{The normalised surface brightness profiles (solid line) and the growth curves (dashed line) of the PSF models for EGS (top) and CEERS (bottom), colour-coded with different filters. The growth curves are normalised to the total flux of the PSF model within $7.2\arcsec$ (EGS) and $3.9\arcsec$ (CEERS). \label{fig:psf}}
\end{figure}

In the photometric analysis of the HST and JWST images, it is customary to build the PSF models using point sources in the image.
Following \cite{Zhuang_2023}, we use {\tt SExtractor} \citep{Bertin_1996} to perform source detection to select point sources of high signal-to-noise ratio, non-blended, non-irregular, unsaturated, and without any bad pixels. We then use {\tt PSFEx} \citep{Bertin_2011} to construct the PSF models. We generate PSF models with the cutout size of $121\times121$ pixels ($\sim7.2\arcsec\times7.2\arcsec$) for the EGS filters and $131\times131$ pixels ($\sim3.9\arcsec\times3.9\arcsec$) for the CEERS filters; the PSF images are sufficiently large, encompassing at least twice the radius that encloses 98\% of the total flux ($\sim 1.7\arcsec$ for EGS and $\sim 0.95\arcsec$ for CEERS). The PSFs of the EGS and CEERS images have the pixel scales of 0.06\arcsec/pixel and 0.03\arcsec/pixel, respectively. The FWHMs of all the PSF models are listed in Table \ref{table:1}. In Fig. \ref{fig:psf}, we show the normalised surface brightness profiles and the growth curves of the PSF models for the EGS and CEERS mosaics in the top and bottom panels, respectively. 

\subsection{S/N test} \label{sec:snr}
\begin{figure*}[h]
\includegraphics[width=1\textwidth]{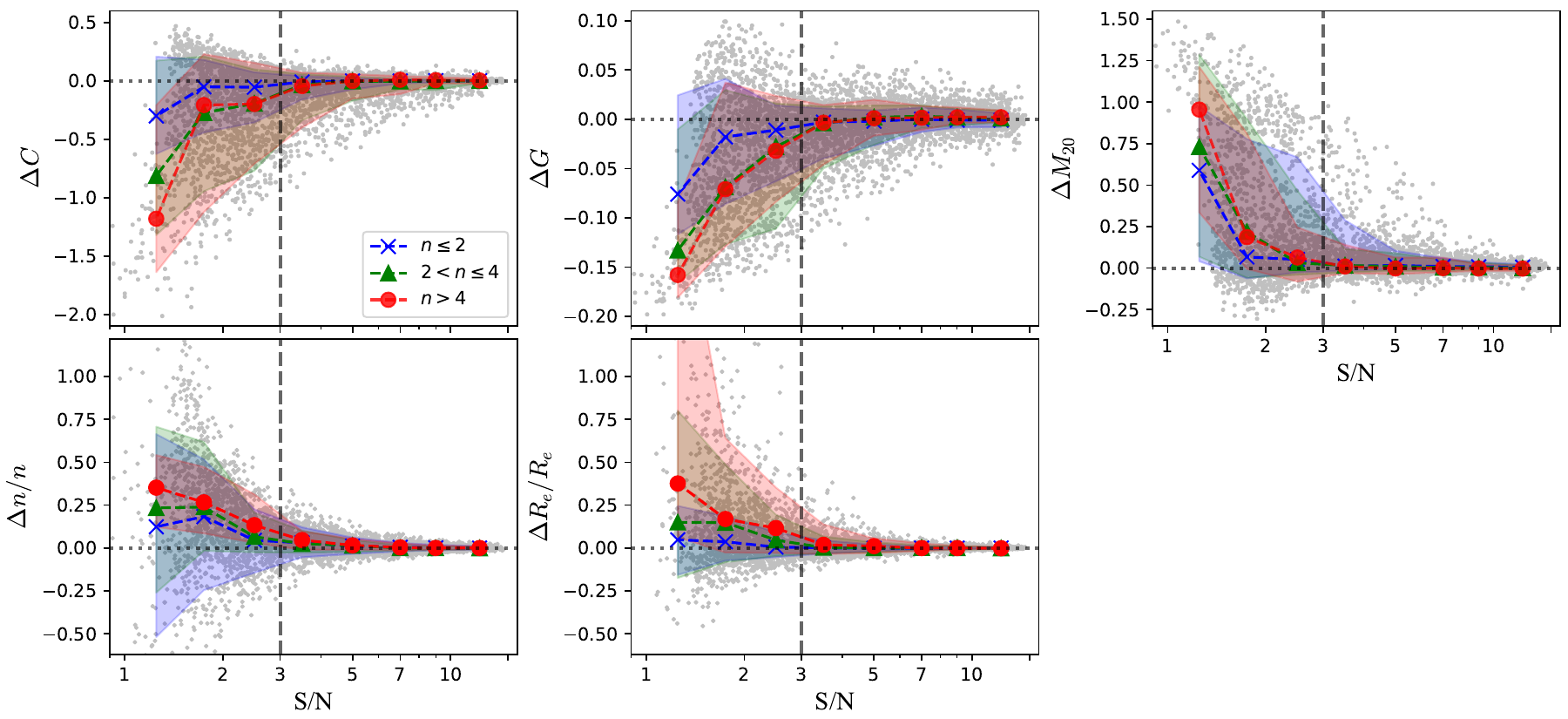}
\caption{Dependence of the morphology indicators uncertainties on S/N, estimated from mock images. The median trends of the mock galaxies with different $n$ are marked with dotted lines; the shaded region indicates the 16th to 84th percentiles. The top row shows the difference between the measured parameters and the intrinsic values, expressed as $\Delta C = C_{\rm mock}- C_{\rm int}$. The bottom row shows the difference between the measured and true $n$ and $Re$ values, relative to the true value. The horizontal dotted line at zero indicates perfect agreement between measurements and true values while the vertical dashed line marks the S/N=3 boundary beyond which the agreement considerably improves. \label{fig:snr}}
\end{figure*}

In order to evaluate the influence of S/N on the morphological indicators, we simulate 5000 mock galaxies with  $200\times200$ pixels, where the pixel scale is 0.06\arcsec/pixel. Each model is built by G{\small ALFIT} using a single S\'{e}rsic component. The structural parameters are chosen randomly in the following ranges: $1.67\leq R_e\leq35$ pixels ($\sim0.1\arcsec-2.1\arcsec$ in WFC3/F160W), $0.5\leq n\leq8$, and $0.2\leq q\leq0.9$. The magnitude range is $20\leq m_H\leq25$, corresponding to $1\leq{\rm S/N}\leq20$. We first use {\tt Statmorph} to calculate the intrinsic non-parametric morphological indicators of these idealised models. Then these images are multiplied by the exposure time of 2700s, and convolved with the PSF of the F160W band. We use {\tt apply\_poisson\_noise} from {\tt Photutils} \citep{larry_bradley_2022_6825092} to add Poisson noise with Gain = 2.5, and a real sky background from the EGS mosaics. After generating the mock galaxies, we use G{\small ALFIT} to perform the single S\'{e}rsic component fit and {\tt Statmorph} to calculate the non-parametric morphology indicators to examine the deviations.

The average S/N for each galaxy image is defined as
\begin{equation}
    {\rm S/N} = \frac{F_{\rm tot}}{\sqrt{F_{\rm tot}+q \pi R_p^2 \sigma_{\rm sky}^2}},
\end{equation}
where $F_{\rm tot}$ is the total integrated flux within the Petrosian radius, and $\sigma_{sky}$ is the standard deviation of the sky background. 
Fig. \ref{fig:snr} shows the difference between the measured structural parameters and the intrinsic values as a function of S/N. We can see that the deviations increase with decreasing S/N. At lower S/N, the S\'{e}rsic index and $R_e$ are slightly overestimated. The concentration-related parameters ($C$, $G$, $M_{20}$) on the other hand, are all underestimated. In addition to the PSF smoothing effect, the outskirts of the galaxies at low S/N are also affected by the noise which reduces the total flux of the galaxy, resulting in smaller $R_{80}$, further decreasing the measured $C$ value. For our working sample, we choose S/N>3 to reduce uncertainties in the morphology indicators and the S\'{e}rsic index.

\section{Results}\label{sec:results}
\begin{figure*}[htbp]
\centering
\large{\hspace{1em}$z\approx0.25$\hspace{8em}$z\approx0.75$\hspace{8em}$z\approx1.25$\hspace{8em}$z\approx1.75$}\\
\subfigure{
\rotatebox{90}{\scriptsize{\hspace{-28em}\large{$n>4$\hspace{8em}$2<n<4$\hspace{8.5em}$n<2$}}}
% minipage中放置4张图，通过vspace进行换行调整，但是都在subfigure下面，共享一个子标题
\begin{minipage}[t]{0.235\linewidth}
\includegraphics[width=1\linewidth]{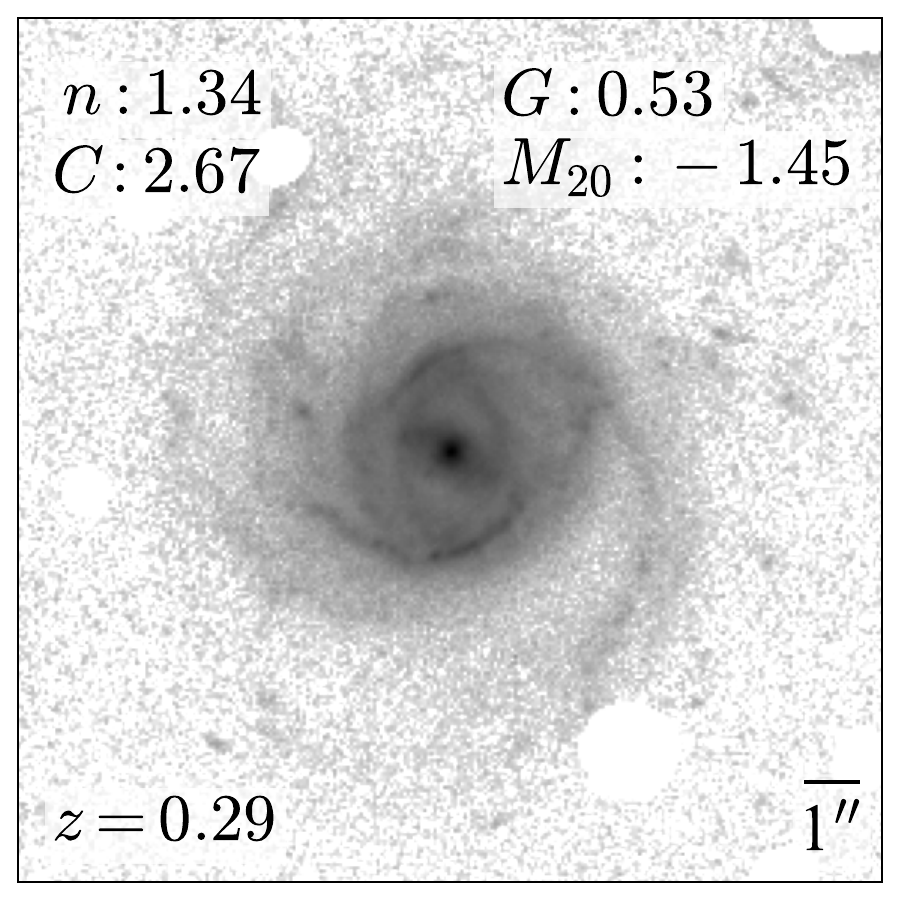}\vspace{0pt}
\includegraphics[width=1\linewidth]{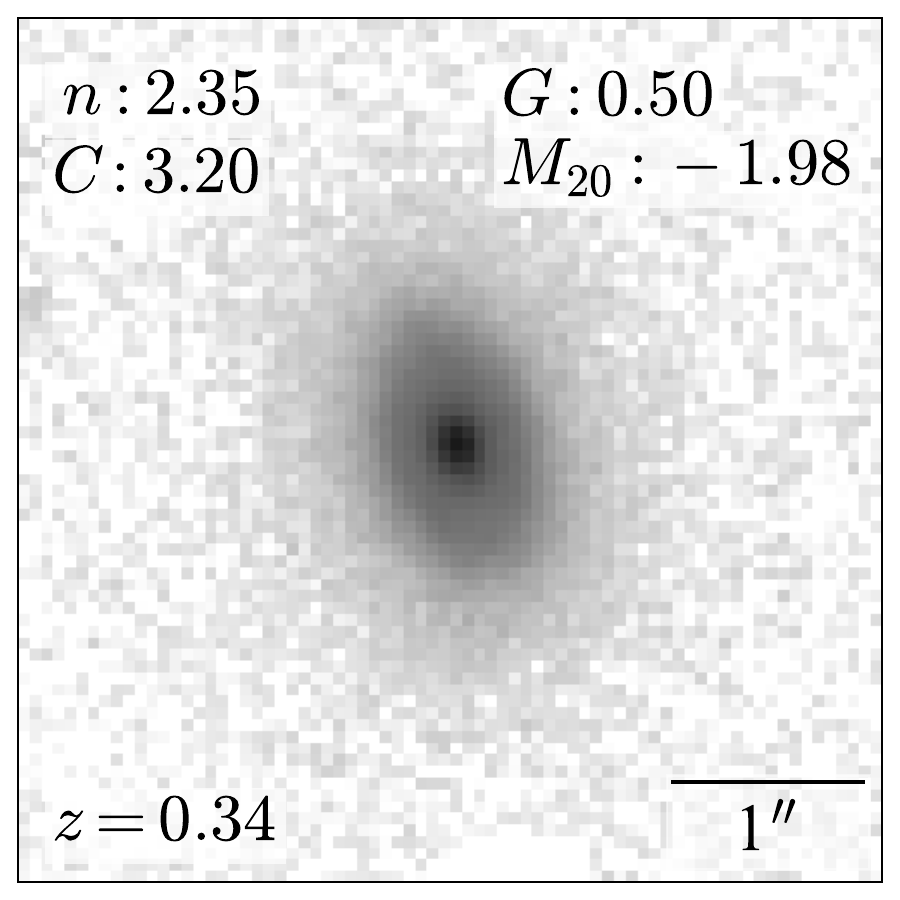}\vspace{0pt}
\includegraphics[width=1\linewidth]{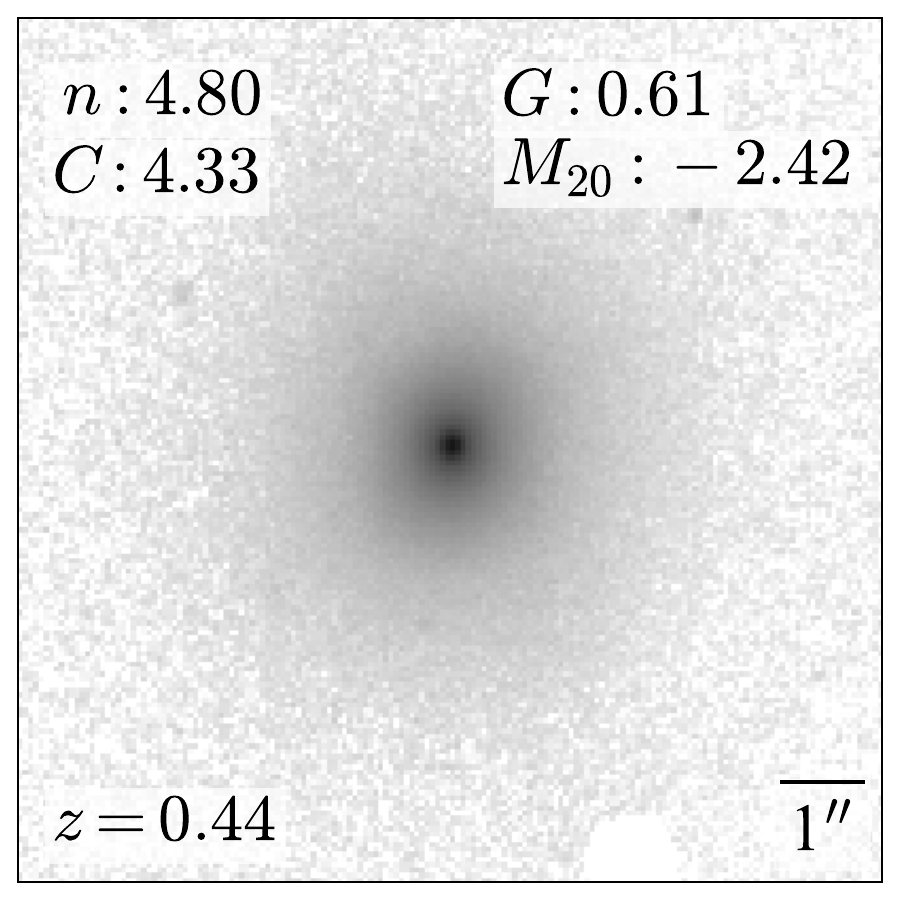}
\end{minipage}}
\hspace{-8pt}
\subfigure{
\begin{minipage}[t]{0.235\linewidth}
\includegraphics[width=1\linewidth]{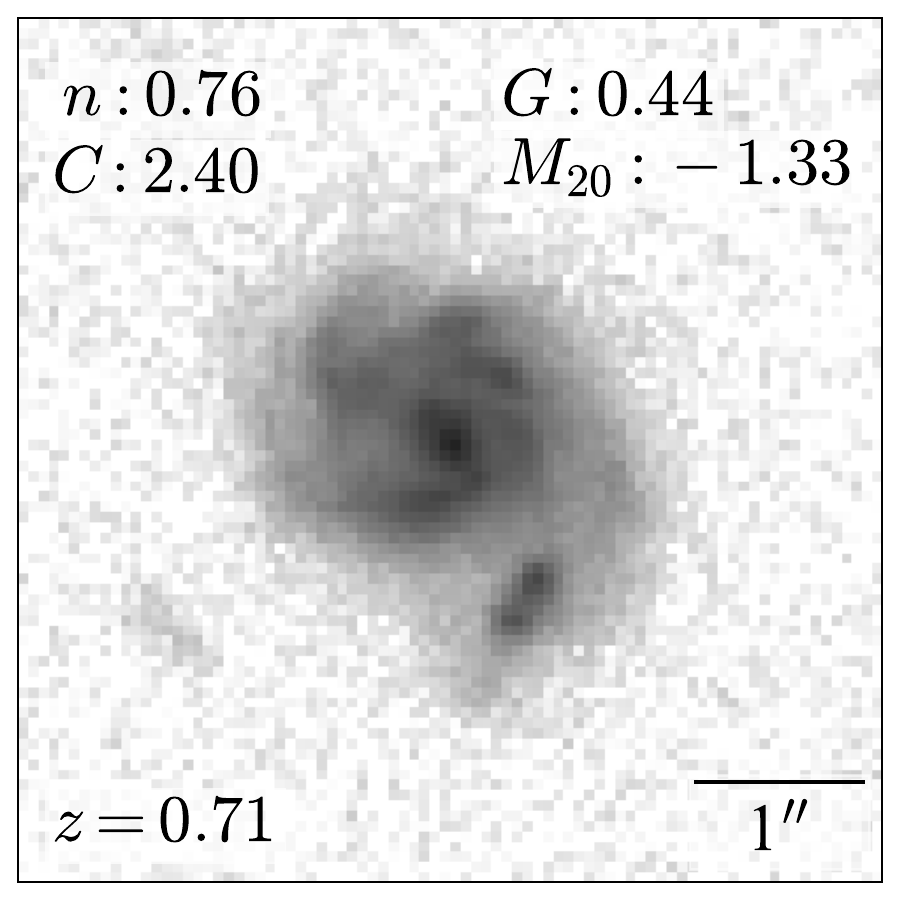}\vspace{0pt}
\includegraphics[width=1\linewidth]{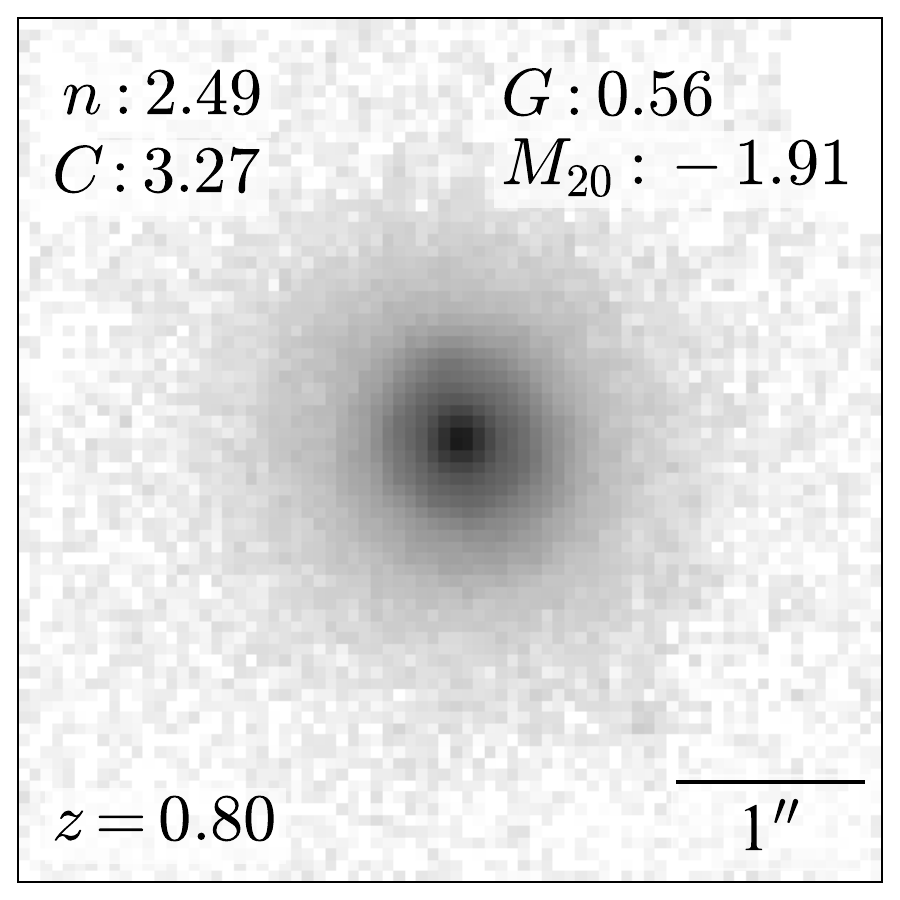}\vspace{0pt}
\includegraphics[width=1\linewidth]{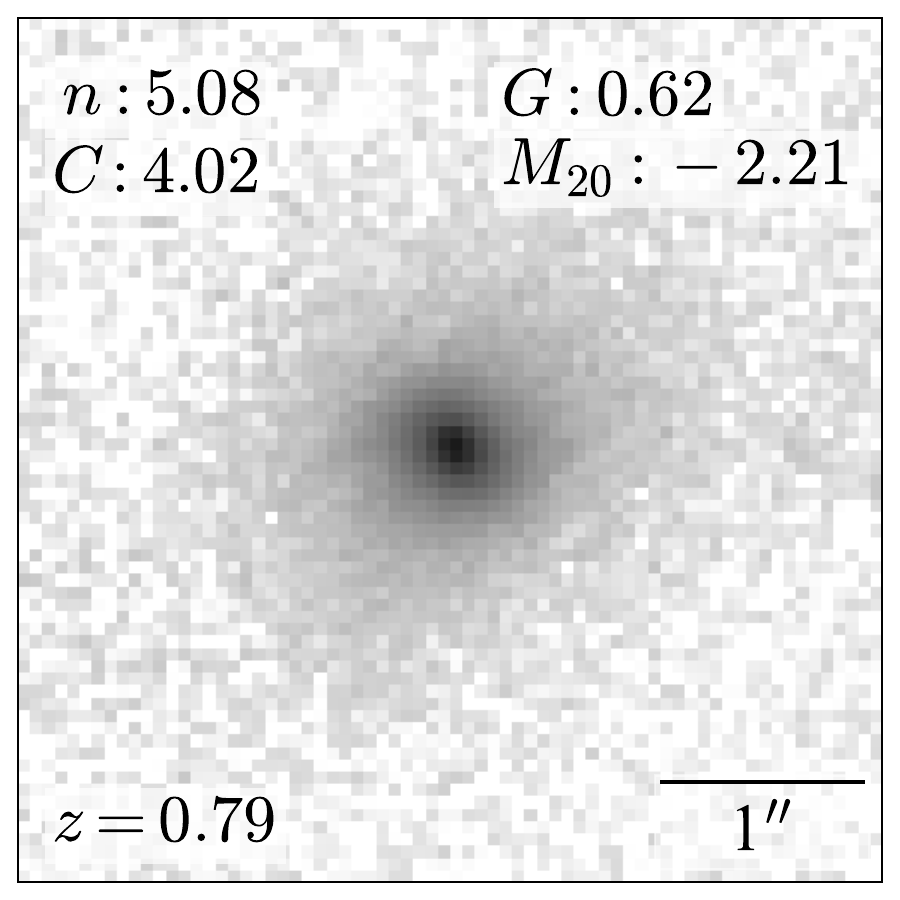}
\end{minipage}}
\hspace{-8pt}
\subfigure{
\begin{minipage}[t]{0.235\linewidth}
\includegraphics[width=1\linewidth]{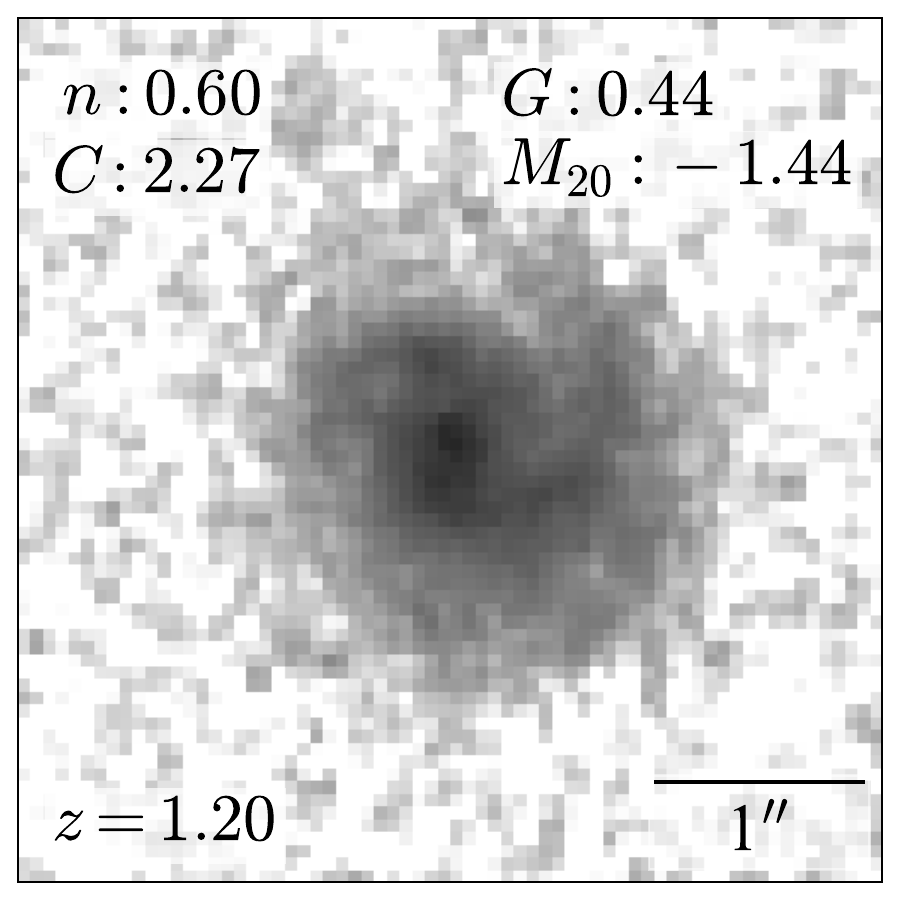}\vspace{0pt}
\includegraphics[width=1\linewidth]{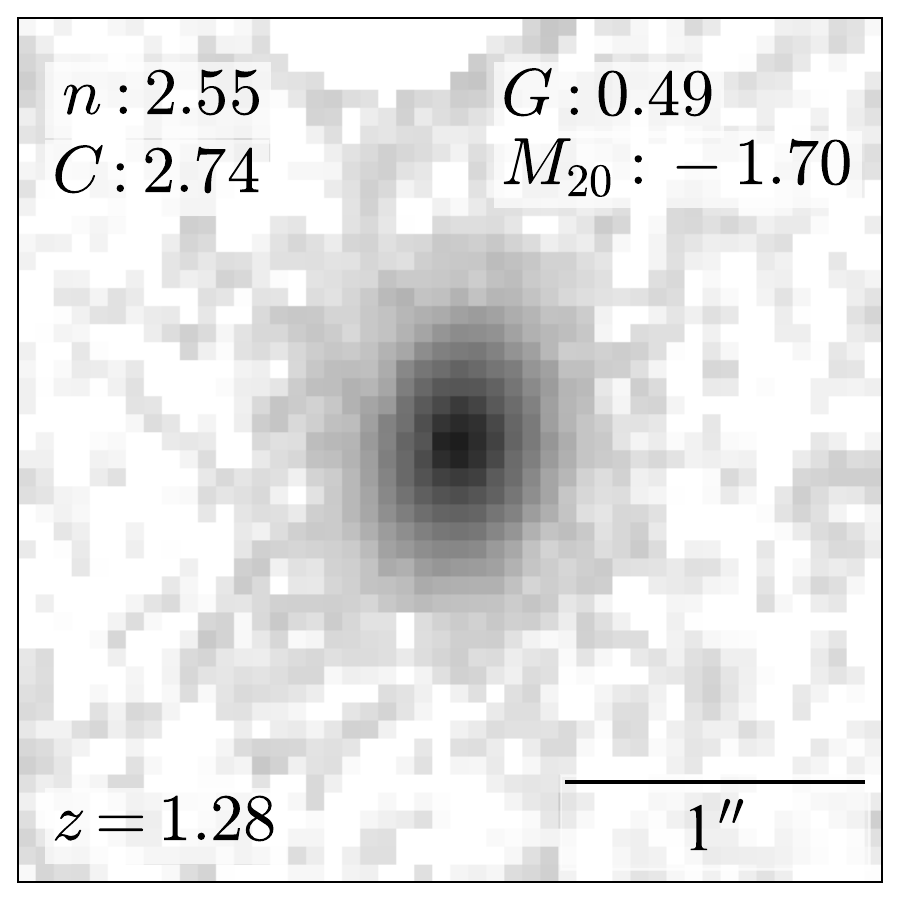}\vspace{0pt}
\includegraphics[width=1\linewidth]{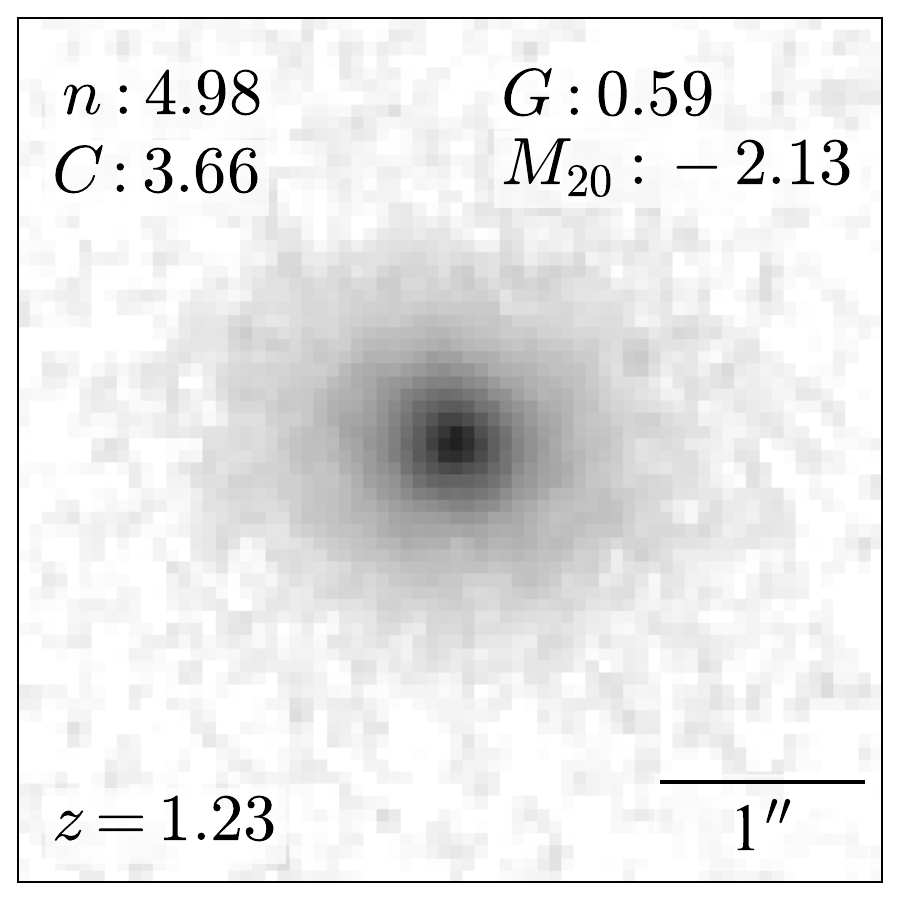}
\end{minipage}}
\hspace{-8pt}
\subfigure{
\begin{minipage}[t]{0.235\linewidth}
\includegraphics[width=1\linewidth]{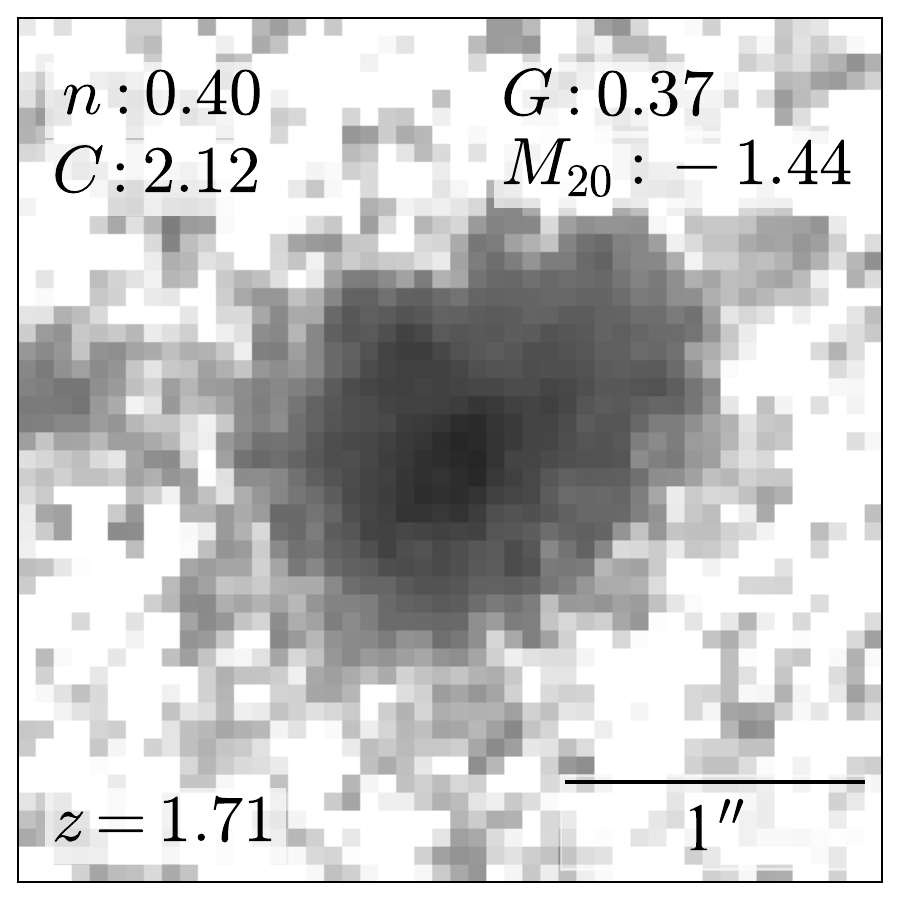}\vspace{0pt}
\includegraphics[width=1\linewidth]{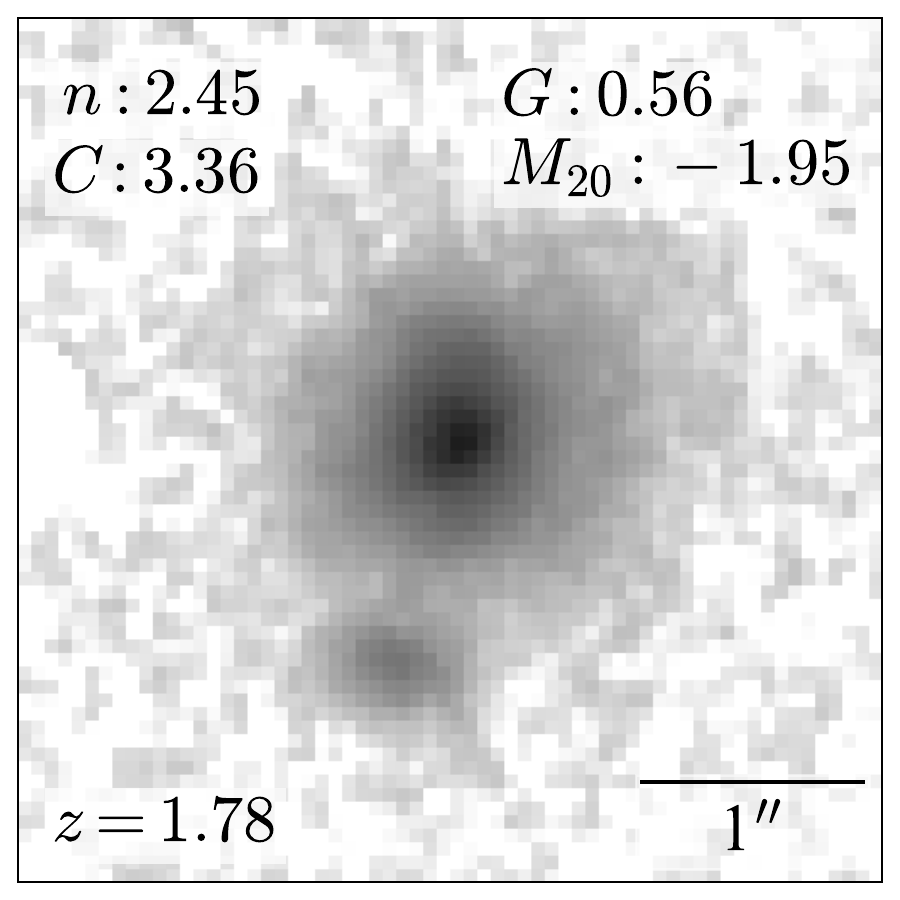}\vspace{0pt}
\includegraphics[width=1\linewidth]{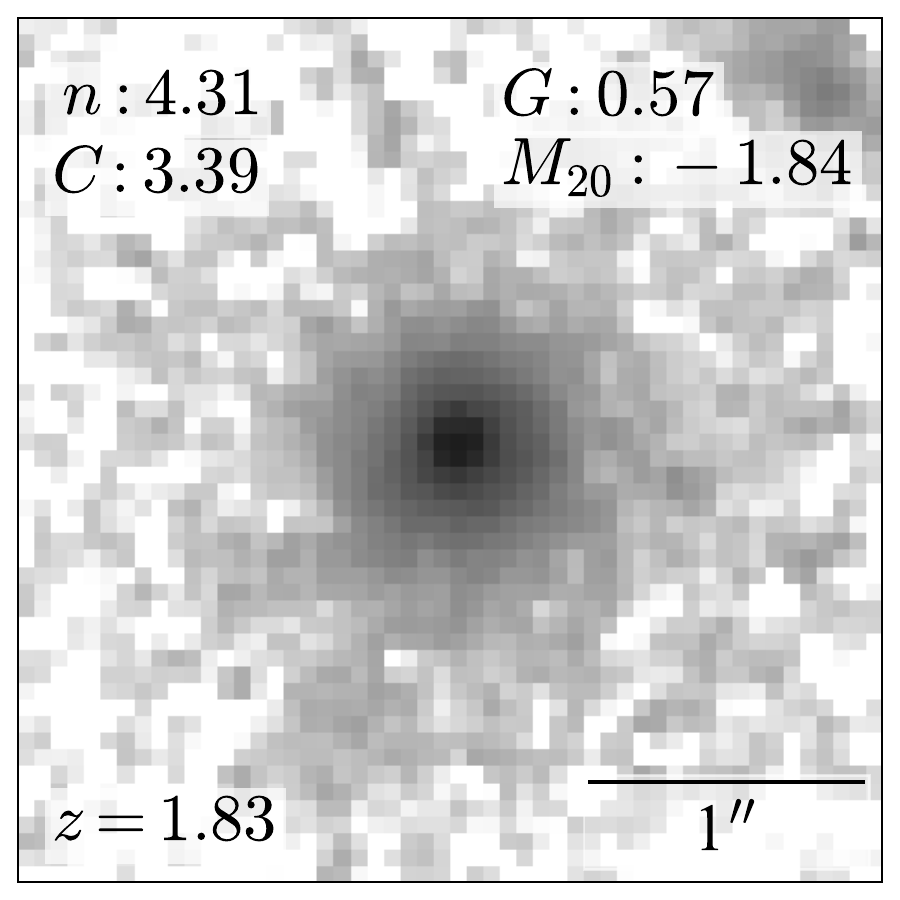}
\end{minipage}}
\caption{Examples of galaxies in the EGS images, with the S\'{e}rsic index increasing from the top to bottom rows (disc to spheroid morphology) and the redshift increasing (from $z \approx$ 0 to 2) from the left to right columns. The filters have been selected to have the similar rest-frame optical wavelength. The measured structural parameters ($n$, $C$, $G$, $M_{20}$) are also listed in each panel. The galaxies on the right two columns are also observed in CEERS, which are shown in the left two columns of Fig. \ref{fig:jwst_images}}
\label{fig:egs_images}
\end{figure*}

\begin{figure*}[htbp]
\centering
\large{\hspace{1em}$z\approx1.25$\hspace{8em}$z\approx1.75$\hspace{8em}$z\approx2.25$\hspace{8em}$z\approx2.75$}\\
\subfigure{
\rotatebox{90}{\scriptsize{\hspace{-28em}\large{$n>4$\hspace{8em}$2<n<4$\hspace{8.5em}$n<2$}}}
% minipage中放置4张图，通过vspace进行换行调整，但是都在subfigure下面，共享一个子标题
\begin{minipage}[t]{0.235\linewidth}
\includegraphics[width=1\linewidth]{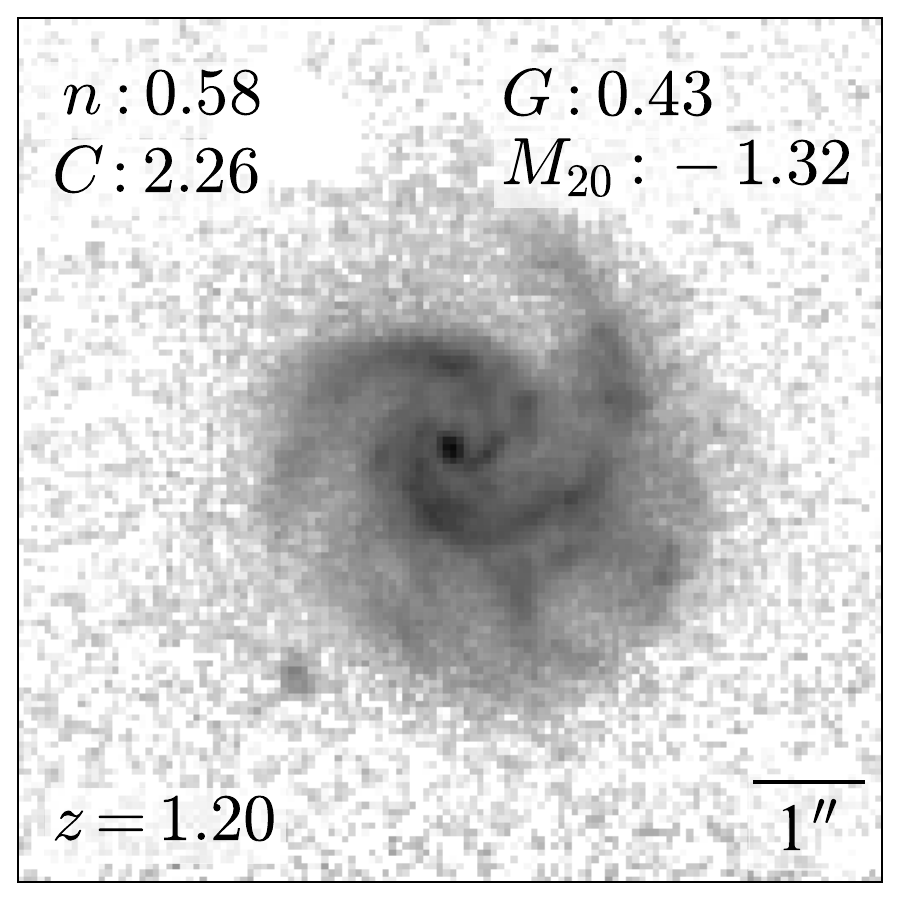}\vspace{0pt}
\includegraphics[width=1\linewidth]{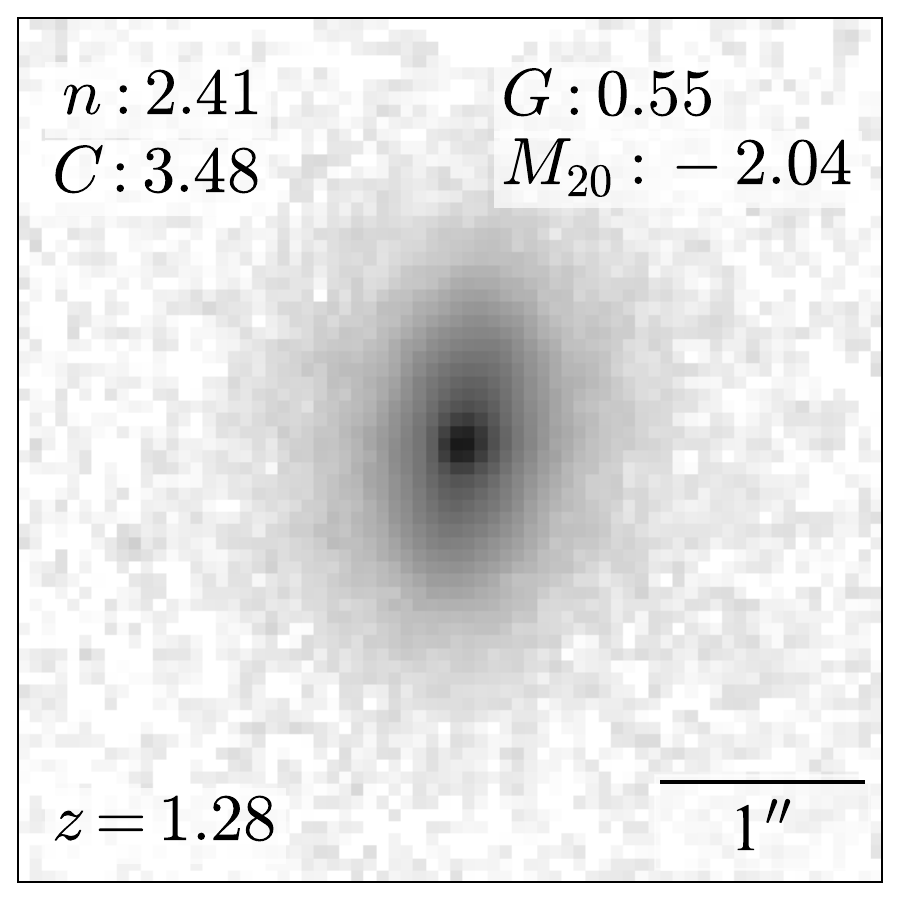}\vspace{0pt}
\includegraphics[width=1\linewidth]{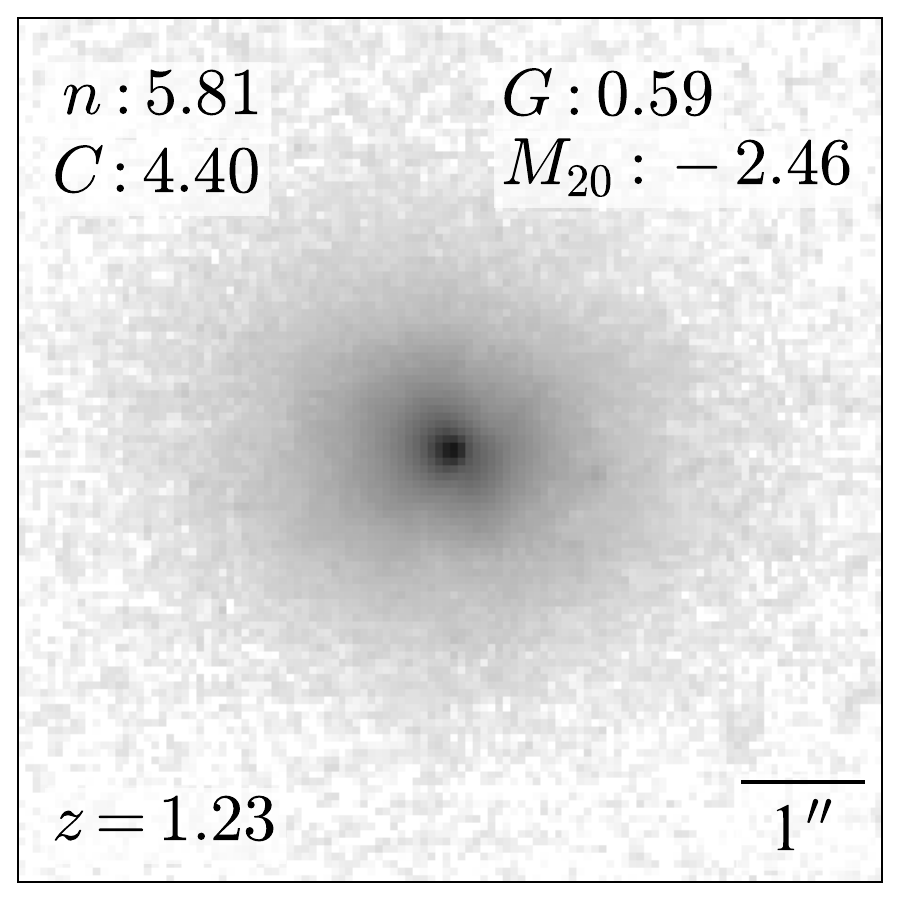}
\end{minipage}}
\hspace{-8pt}
\subfigure{
\begin{minipage}[t]{0.235\linewidth}
\includegraphics[width=1\linewidth]{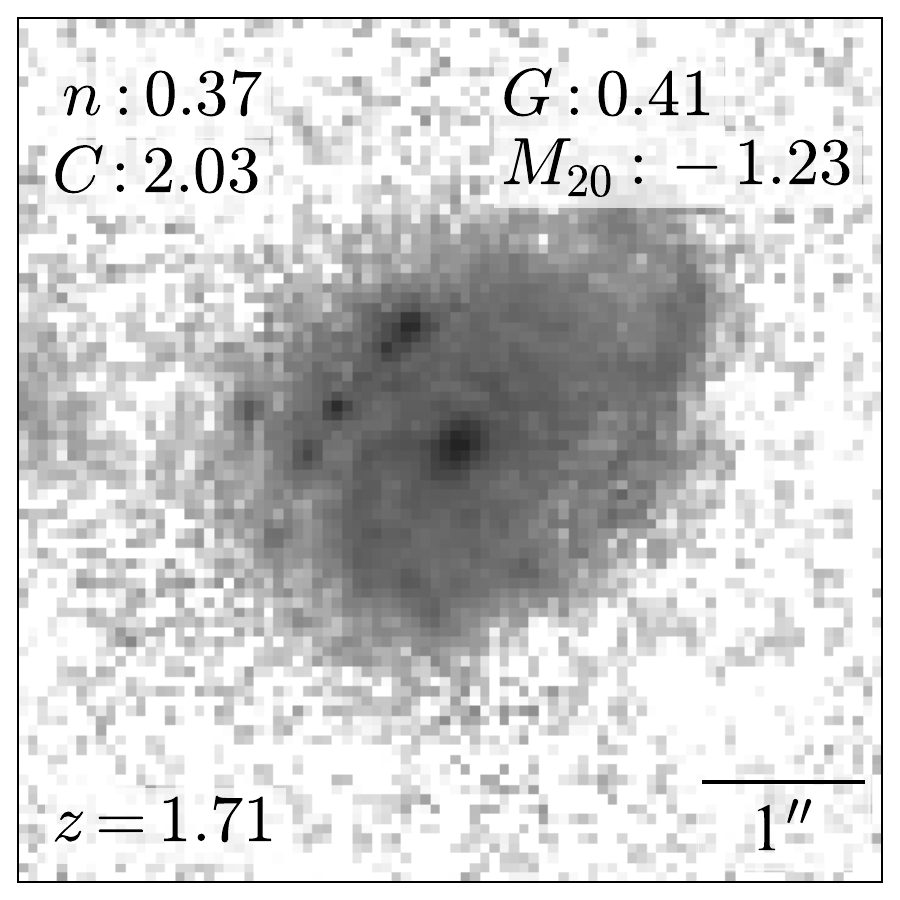}\vspace{0pt}
\includegraphics[width=1\linewidth]{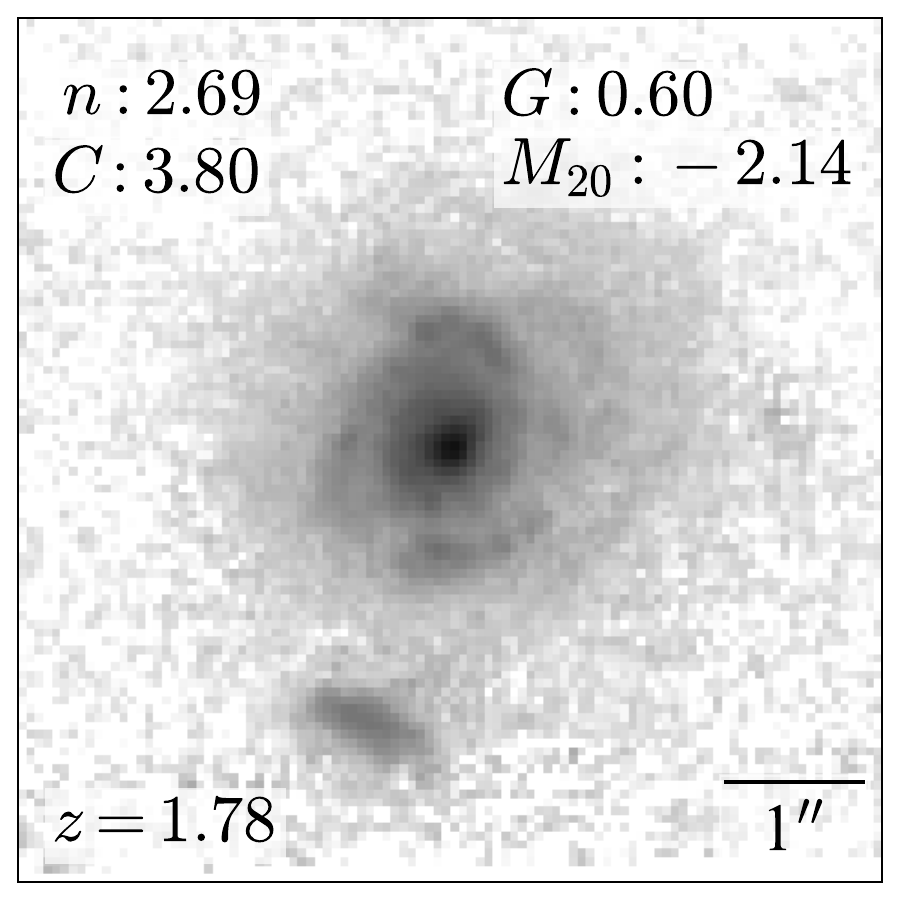}\vspace{0pt}
\includegraphics[width=1\linewidth]{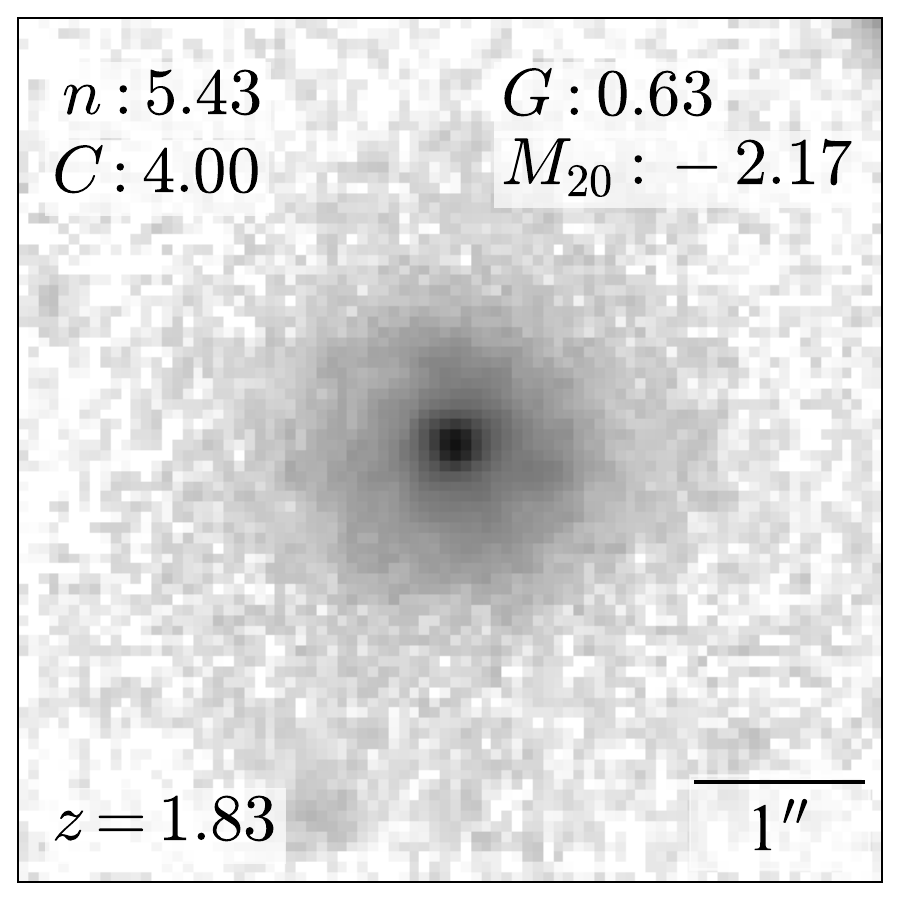}
\end{minipage}}
\hspace{-8pt}
\subfigure{
\begin{minipage}[t]{0.235\linewidth}
\includegraphics[width=1\linewidth]{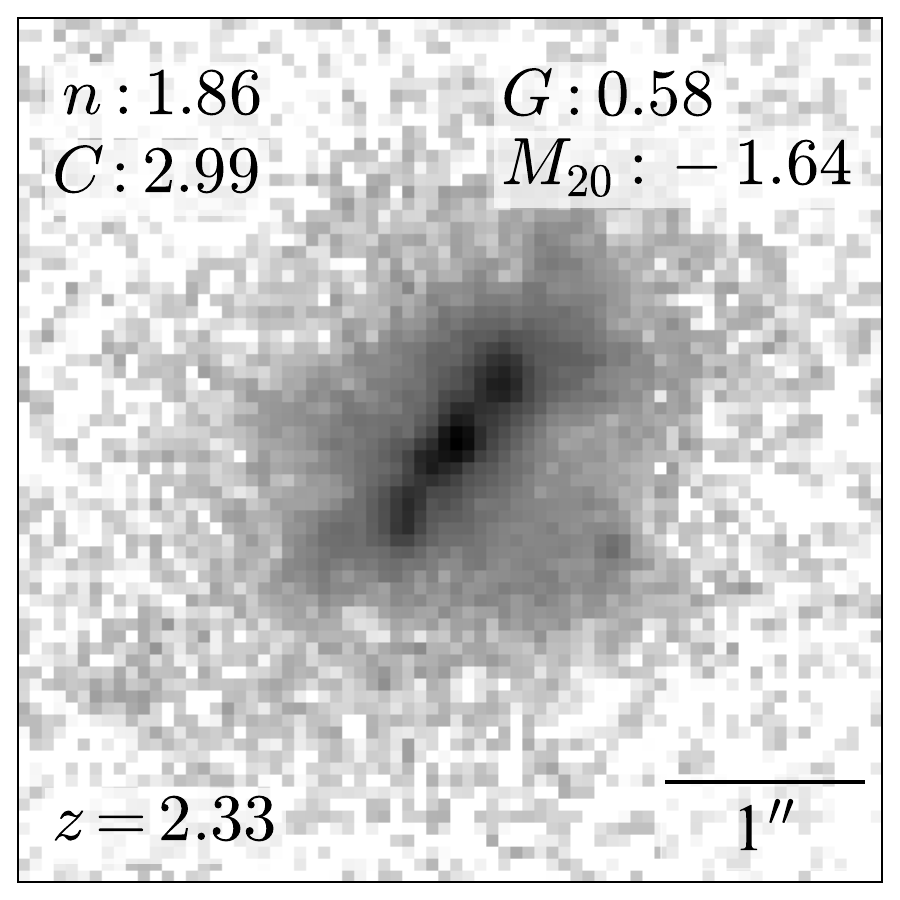}\vspace{0pt}
\includegraphics[width=1\linewidth]{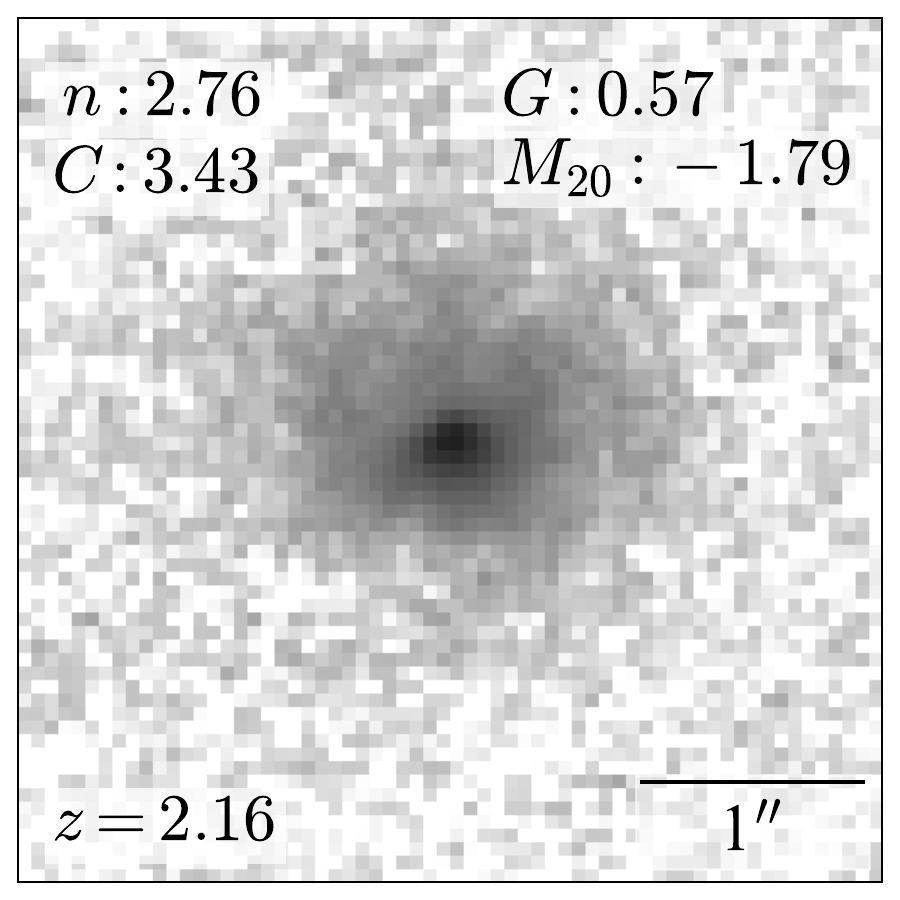}\vspace{0pt}
\includegraphics[width=1\linewidth]{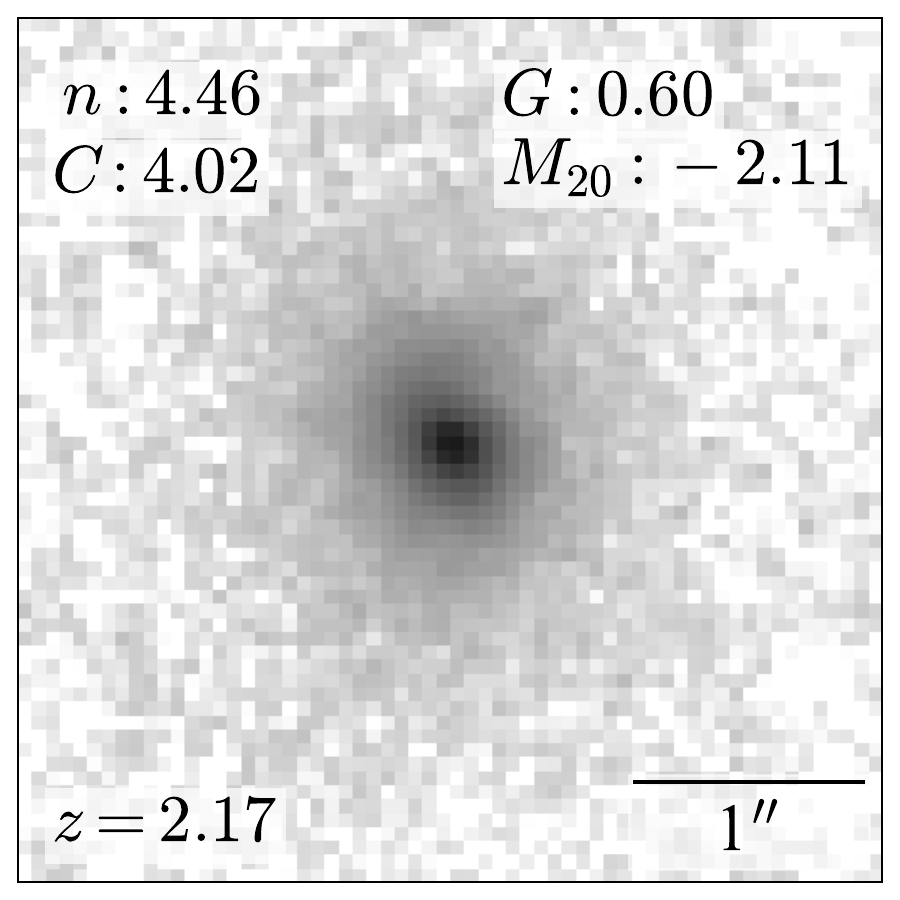}
\end{minipage}}
\hspace{-8pt}
\subfigure{
\begin{minipage}[t]{0.235\linewidth}
\includegraphics[width=1\linewidth]{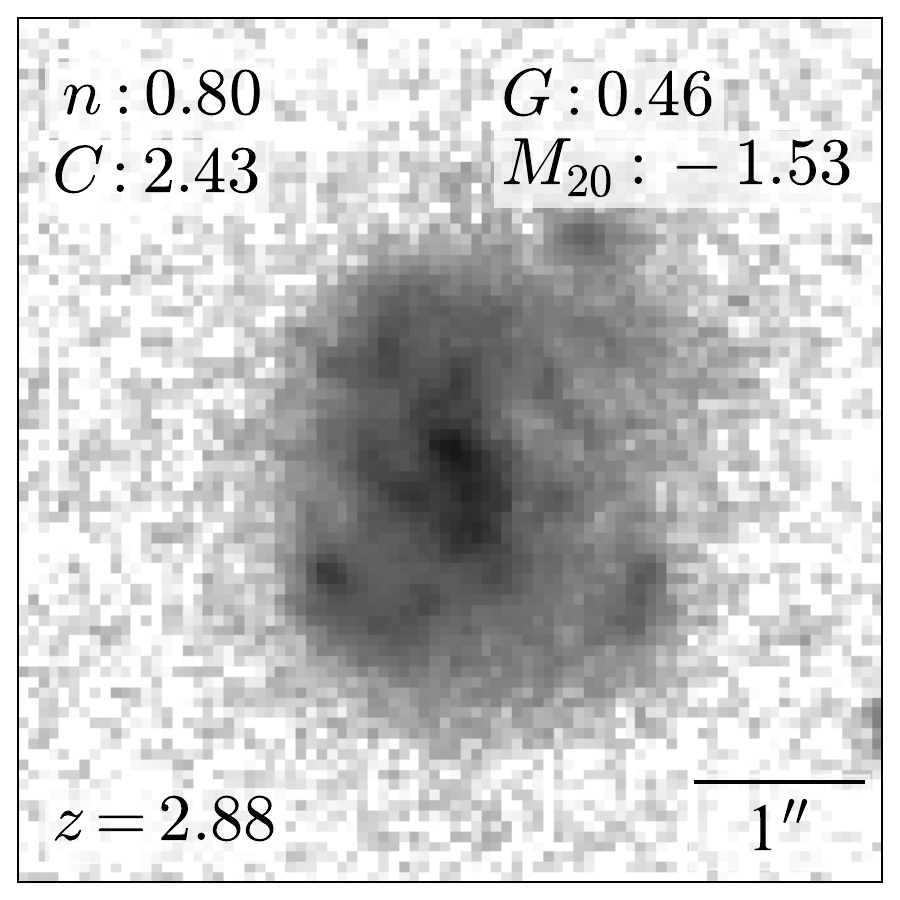}\vspace{0pt}
\includegraphics[width=1\linewidth]{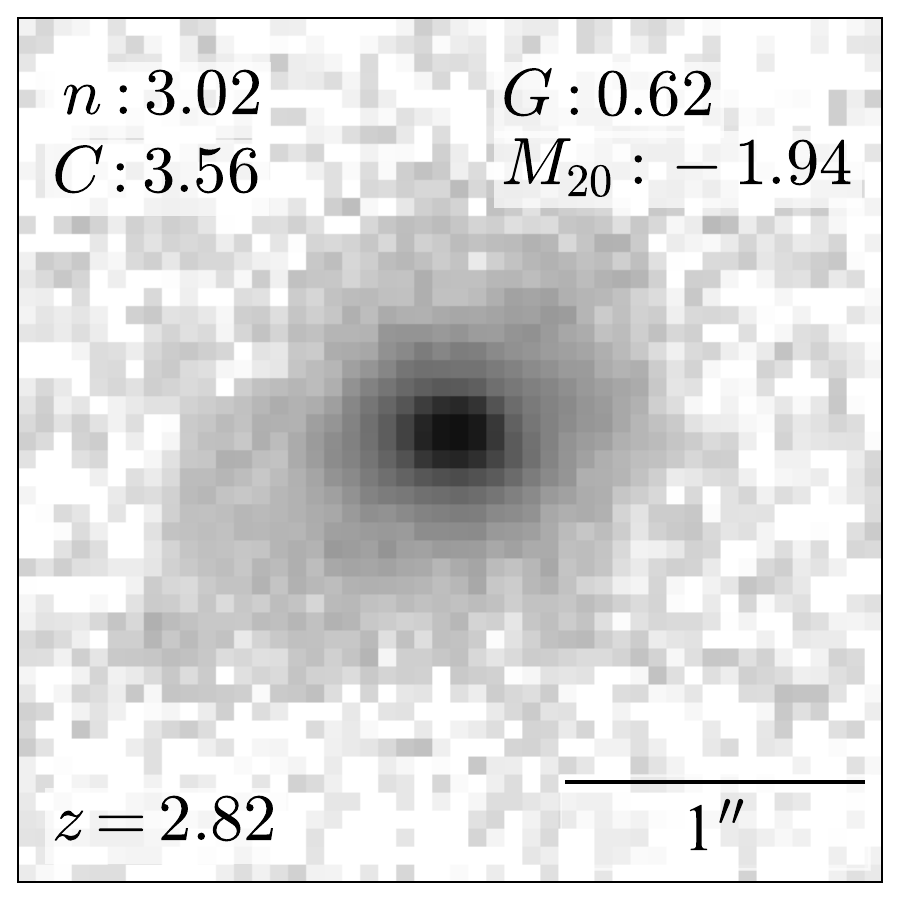}\vspace{0pt}
\includegraphics[width=1\linewidth]{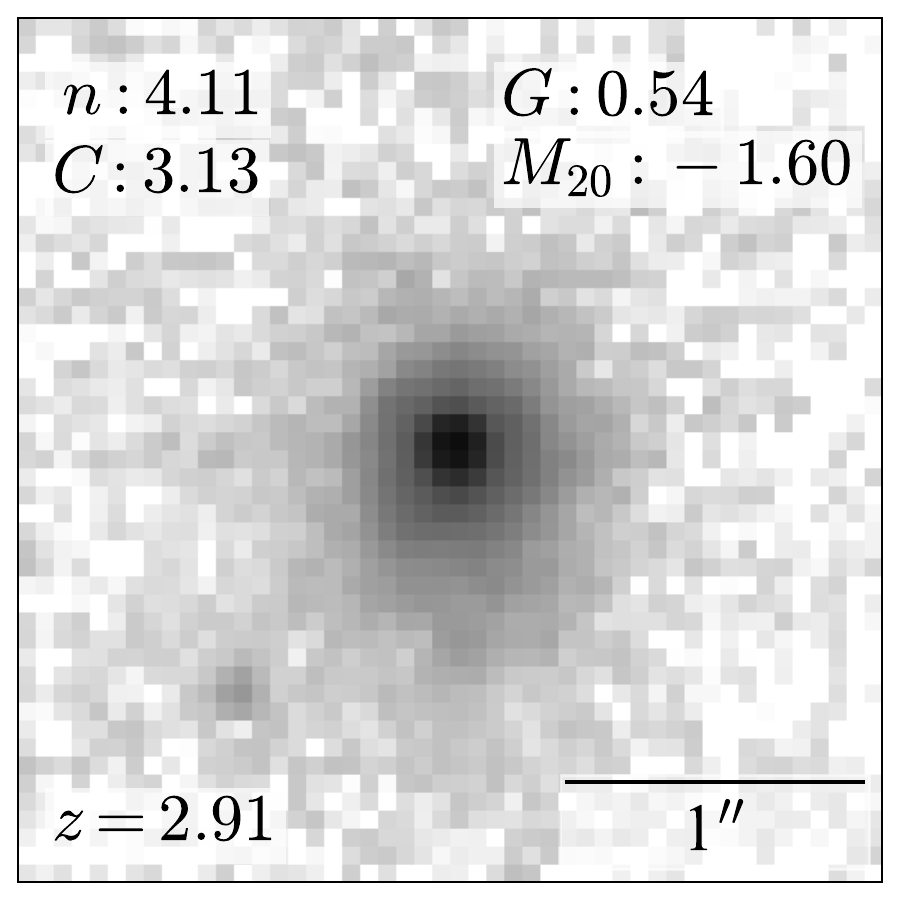}
\end{minipage}}
\caption{Examples of galaxies in the CEERS images, with the S\'{e}rsic index increasing from the top to bottom rows (disc to spheroid morphology) and the redshift increasing (from $z\approx$ 1 to 3) from the left to right columns. The structural parameters are also listed in each panel. The galaxies on the left two columns are also shown in the same as ones in the right two columns of Fig. \ref{fig:egs_images} for the HST observation. Due to the higher resolution and S/N of images in JWST, the galaxies exhibit more details, resulting in higher values of $C$, $G$, and $|M_{20}|$ for galaxies with $n>2$. }
\label{fig:jwst_images}
\end{figure*}

In this section, we study the correlation between the concentration-related parameters and the S\'{e}rsic index in detail in order to understand the PSF smoothing effect. A subset of EGS and CEERS images are shown in Figs. \ref{fig:egs_images} and \ref{fig:jwst_images}, respectively, with the S\'{e}rsic index increasing from top to bottom (disc to spheroid morphology) and the redshift increasing from left to right.

The best-fit S\'{e}rsic index and the measured concentration-related parameters are also labelled in each panel. As shown in Fig. \ref{fig:egs_images}, the concentration indices are similar ($\sim2.4$) for discs with S\'{e}rsic index $n\sim 1$ at $z \approx$ 0 to 2. However, the spheroids ($n>4$) with smaller sizes tend to have lower $C$ values. As the redshift increases, the concentration index drops from $\sim4.3$ to $\sim3.4$ for galaxies with S\'{e}rsic index $n \sim 5$. The PSF smoothing effect is more severe in this case.
For the CEERS images shown in Fig. \ref{fig:jwst_images}, the similar trend can still be observed, although it is not as significant as in the EGS images. This difference is attributed to the much higher spatial resolution provided by JWST.

\subsection{PSF smoothing effect on the concentration index}\label{sec:c_n}

\subsubsection{The correlation between \texorpdfstring{$C$}{C} and \texorpdfstring{$n$}{n}}

\begin{figure*}[htbp]
\centering
\begin{minipage}[t]{0.92\linewidth}
\includegraphics[width=1\linewidth]{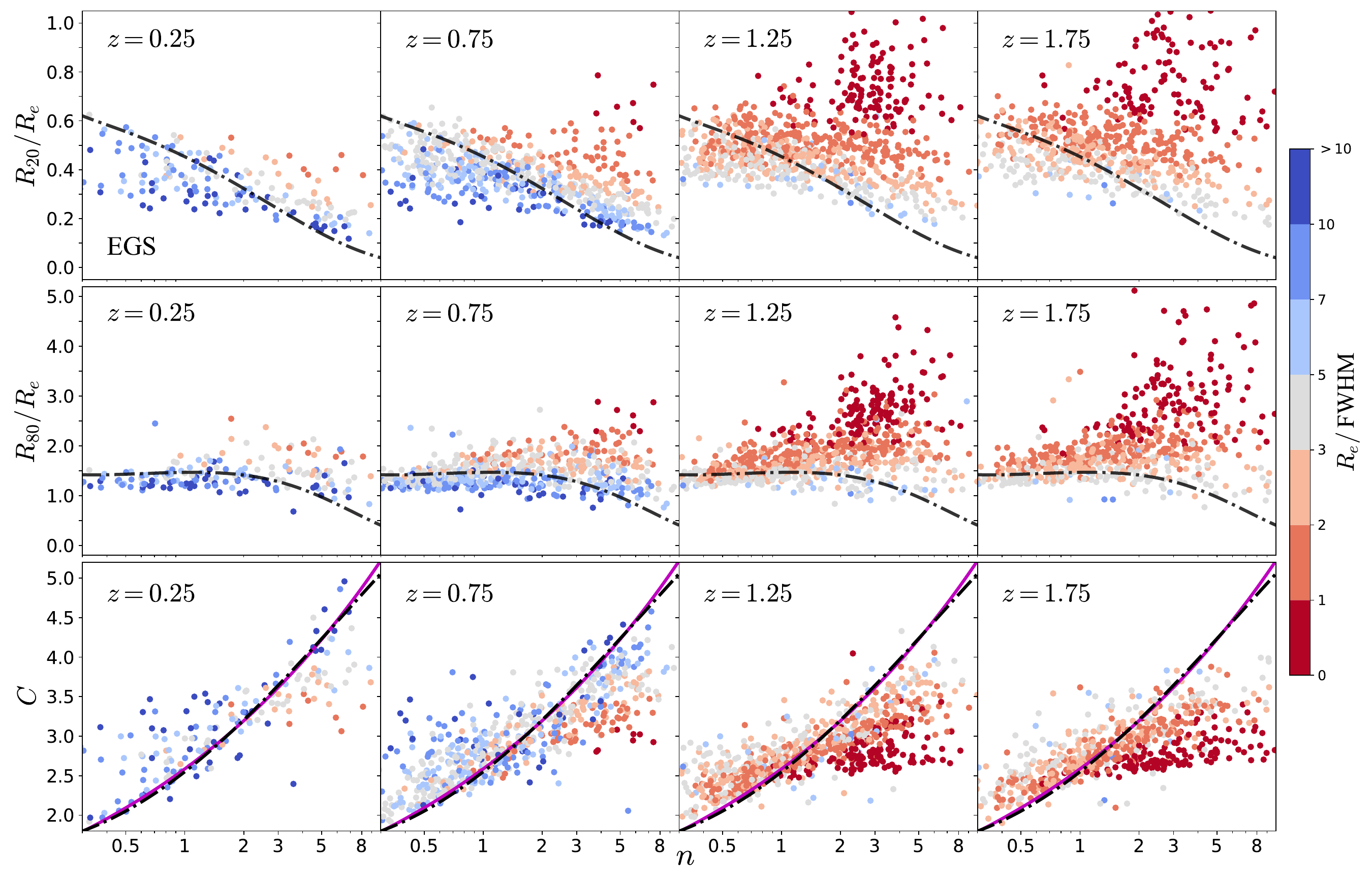}%\vspace{-3pt}
\caption{$R_{20}/R_e$, $R_{80}/R_e$ and $C$ as the S\'{e}rsic index a function of $n$ for the EGS images at $0<z<2$. The black dot-dashed curve in each panel is the numerical correlations derived by integrating the S\'{e}rsic profile. In the bottom row, the purple solid curve is the empirical relation of Eq. (\ref{cn_equ}) given by \cite{Andrae_2011}. The colour of the points marks the relative sizes of the galaxies with $R_e/{\rm FWHM}$.}
\label{fig:egs_cn}
\vspace{10pt}
\end{minipage}
\begin{minipage}[t]{0.92\linewidth}
\includegraphics[width=1\linewidth]{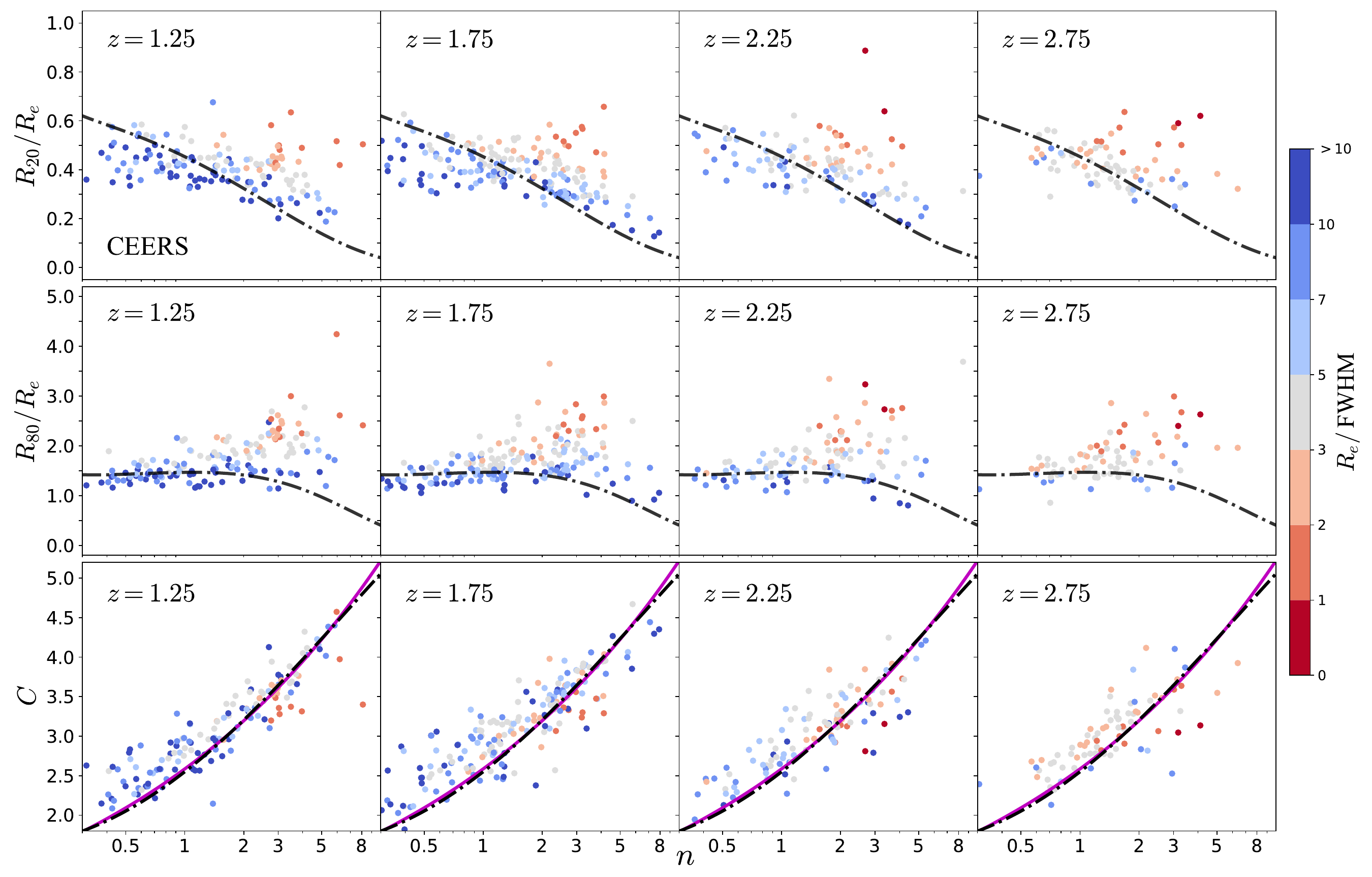}%\vspace{-3pt}
\caption{Same as Fig. \ref{fig:egs_cn}, but for the CEERS images at $1<z<3$. There is very few galaxies with extremely small size ($R_e/{\rm FWHM}\leq 1$). The overall agreement between $C$ and $n$ are quite good consistent with the numerical relationship different redshifts.}
\label{fig:jwst_cn}
\end{minipage}
\end{figure*}

The measured $R_{20}/R_e$, $R_{80}/R_e$ and $C$ as a function of $n$ with the EGS images at different redshifts are shown in Fig. \ref{fig:egs_cn}. Note that $R_{80}/R_e$ can be less than 1 for the galaxies with steep light profiles ($n\geq 5$). This is because $R_{20}$ and $R_{80}$ are measured out to $1.5R_p$ which may not be large enough to enclose the total flux of the galaxy \citep{Graham_2005}. Additionally, due to the default circular aperture used in {\tt Statmorph}, the $R_{20}$ and $R_{80}$ of galaxies with small axis ratio (close to edge-on) tend to be underestimated. This effect is further explored in Sect. \ref{sec:c_e}; when we adopt the elliptical apertures with the same axis ratio as the galaxy outskirt, the underestimation is significantly reduced, especially for galaxies with low S\'{e}rsic index.

The numerical relations between the structural parameters ($R_{20}/R_e$, $R_{80}/R_e$, $C$) and the S\'{e}rsic index are over-plotted with a dot-dashed line\footnote{The numerical relation is derived by integrating the S\'{e}rsic model out to $1.5R_p$.}. The empirical relation of Eq. (\ref{cn_equ}) is also plotted as the purple solid line. The agreement between the two curves is remarkable. The colour of each point indicates the $R_e/{\rm FWHM}$ ratio, representing the relative size of the galaxy compared to the PSF (red for smaller galaxies and blue for bigger ones). 

At lower redshifts ($z \leqslant 1$), galaxies with $R_e/{\rm FWHM}\geq5$ (blue points) generally follow the theoretical relation between $R_{20}/R_e$ ($R_{80}/R_e$) and $n$. As the redshift increases, the galaxies become smaller with much lower $R_e/{\rm FWHM}$, with $R_{20}/R_e$ overestimated up to a factor of 5 for smaller galaxies (i.e., the red points). $R_{80}/R_e$ is also overestimated but not as significantly as $R_{20}/R_e$. Such deviation is less severe for galaxies with smaller S\'{e}rsic indices, even at higher redshifts.

Galaxies at lower redshifts with $R_e/{\rm FWHM} >3$ exhibit a strong correlation between $C$ and $n$, consistent with the theoretical expectation. In addition, $C$ and $n$ of the higher redshift galaxies with $n < 2$ also have a good relationship, consistent with the theoretical curve. Only at higher S\'{e}rsic indices, the measured $C$ values start to deviate from the theoretical expectation. It is noticeable that $R_{20}/R_e$ is more significantly overestimated by the PSF smoothing effect than $R_{80}/R_e$ for the smaller galaxies. According to Eq. (\ref{equ:c}), the concentration index is underestimated. As the galaxy becomes even smaller with $R_e/{\rm FWHM}<1$, the galaxy light profile is overshadowed by the PSF. The $C$ values of the galaxies with different S\'{e}rsic indices all converge to roughly the same value ($\sim 2.7$), corresponding to the PSF.

Our results confirm that the relative size and the S\'{e}rsic index are important parameters in the determination of the concentration index.
Previous literature usually applied a correction factor to the directly measured $C$ values of high redshift galaxies to account for such effect. However, for the CEERS images, the relationship between $C$ and $n$ still holds up to redshift $z\sim3$, as shown in Fig. \ref{fig:jwst_cn}. This is mainly caused by the smaller PSF in JWST, with very few galaxies with $R_e/{\rm FWHM}<2$ in our sample. For JWST, therefore, a correction of the concentration index may not even be needed for high redshift galaxies ($1<z<3$).

\subsubsection{The correlation between \texorpdfstring{$C_{59}$}{C} and \texorpdfstring{$n$}{n}}

\begin{figure*}[htbp]
\centering
\begin{minipage}[t]{0.92\linewidth}
\includegraphics[width=1\linewidth]{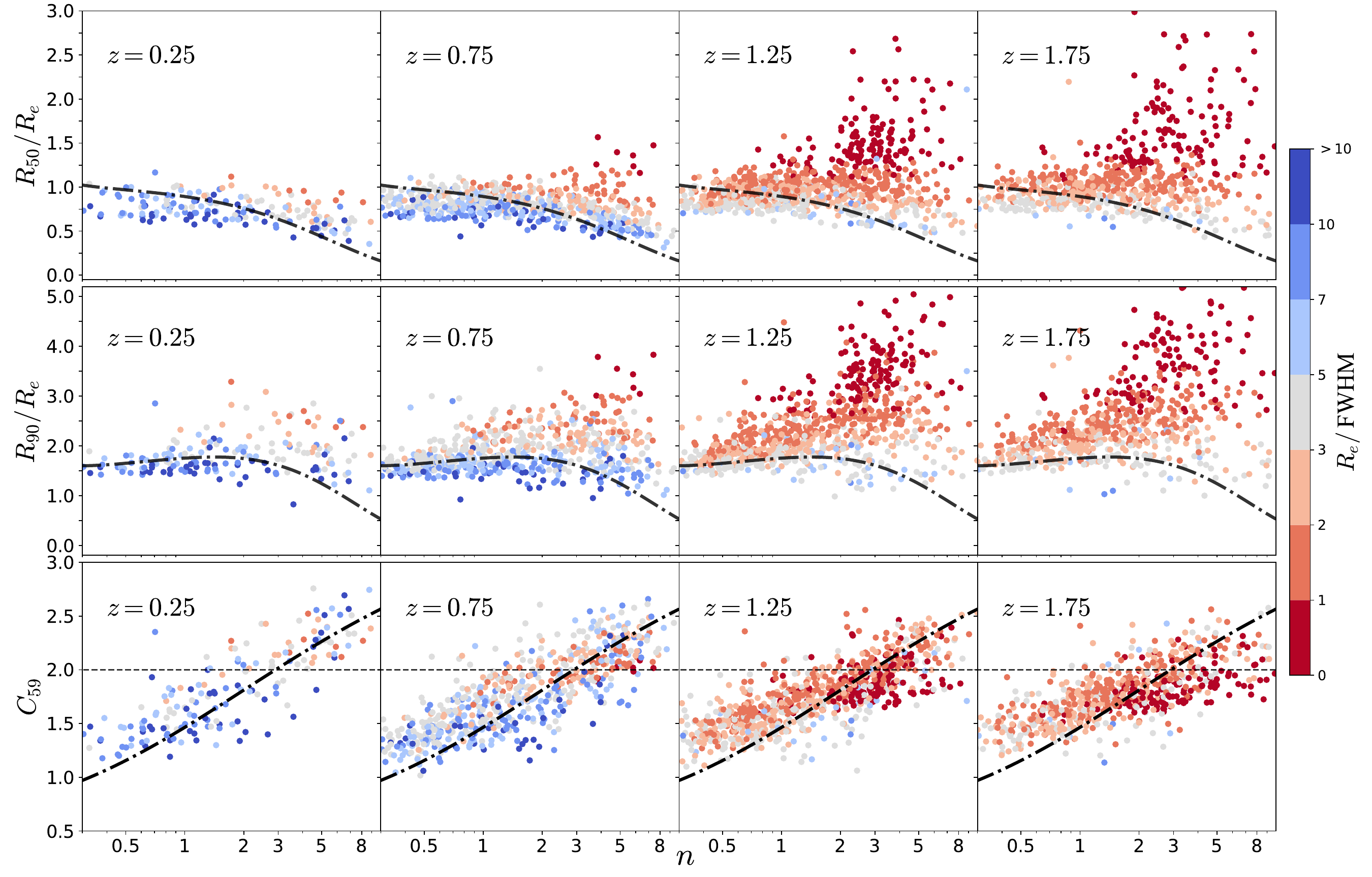}%\vspace{-3pt}
\caption{$R_{50}/R_e$, $R_{90}/R_e$ and $C_{59}$ as a function of the S\'{e}rsic index $n$ for the EGS images at $0<z<2$. The black dot-dashed curve in each panel is the numerical correlations derived by integrating the S\'{e}rsic profile. In the bottom row, the horizontal dashed line $C_{59}= 2$ indicates the empirical boundary between spheroids and discs. \label{fig:egs_c59}}
\vspace{10pt}
\end{minipage}
\begin{minipage}[t]{0.92\linewidth}
\includegraphics[width=1\linewidth]{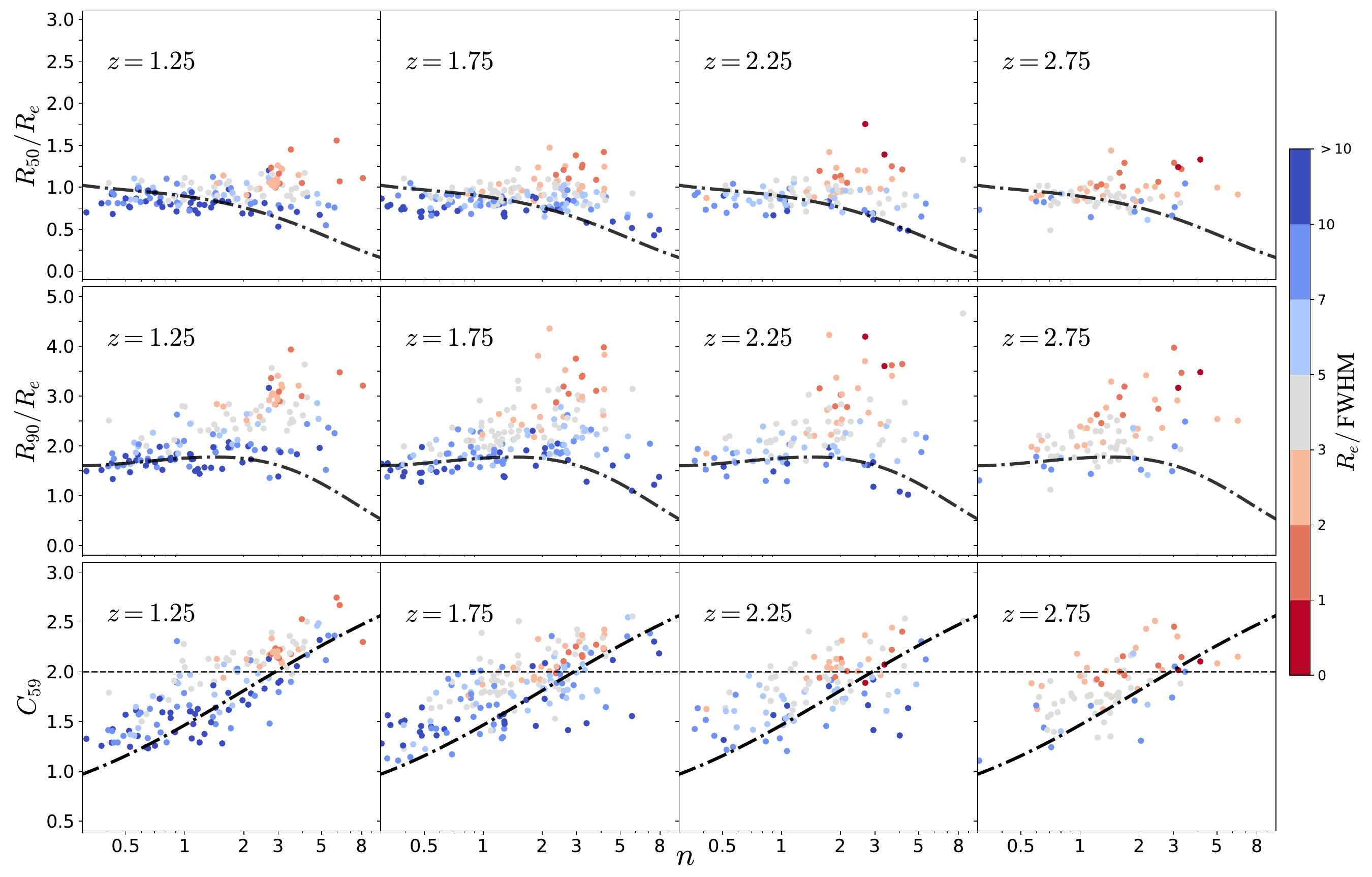}%\vspace{-3pt}
\caption{Same as Fig. \ref{fig:egs_c59}, but for the CEERS images at $1<z<3$. \label{fig:jwst_c59}}
\end{minipage}
\end{figure*}

Another concentration index commonly used in literature is $C_{59} = 5 \,{\rm log}(R_{90}/R_{50})$, with $R_{90}$ representing the radius enclosing 90\% of the Petrosian flux \citep{Blanton_2001}. It can also be used to distinguish discs from spheroids \citep{Strateva_2001,Kauffmann_2003}. 
The correlation between $C_{59}$ and $n$ for the EGS images at different redshifts is shown in Fig. \ref{fig:egs_c59}. Similar to $R_{20}$ and $R_{80}$, $R_{50}$ and $R_{90}$ are overestimated for small galaxies. However, $R_{50}$ and $R_{90}$ exhibit comparable levels of overestimation, resulting in the $C_{59}$ values less affected by the PSF smoothing effect compared to the $C$ values discussed before.
Objects with $R_e/{\rm FWHM}<1$ are generally underestimated. We notice that $C_{59}$ for galaxies with lower $n$ are overestimated, mainly due to single S\'{e}rsic component fitting might not adequately describe these galaxies. Moreover, $C_{59}$ is less affected by the aperture axis ratio.

The empirical boundary between spheroids and discs is indicated by the horizontal dashed line at $C_{59}= 2$ (corresponding to $n\sim3$) in the bottom rows of Fig. \ref{fig:egs_c59}. \cite{Strateva_2001} suggested $r_{90}/r_{50} = 2.6$ ($C_{59}\sim 2.07$ according to Eq. (\ref{equ:c})) as the criterion, with $r_{90}/r_{50} > 2.6$ for spheroids, and $r_{90}/r_{50}< 2.6$ for disks. Here $r_{90}$ and $r_{50}$ are the radii of 90\% and 50\% flux within $2R_p$ (instead of $1.5R_p$ used in this work). It seems that this criteria works for galaxies with $R_e/{\rm FWHM}\geq2$.

As shown in Fig. \ref{fig:jwst_c59}, the $R_{50}$ and $R_{90}$ for the CEERS images are consistent with the numerical relationships, except for the smaller objects that are also slight overestimated. In this case, it seems that $C_{59}$ is robust for galaxies with $R_e/{\rm FWHM}>1$ with no correction needed.

\subsection{PSF smoothing effect on the Gini and \texorpdfstring{$M_{20}$}{M} statistics}\label{sec:gm}

\begin{figure*}[p]
\centering

\begin{minipage}[t]{0.92\linewidth}
\includegraphics[width=1\linewidth]{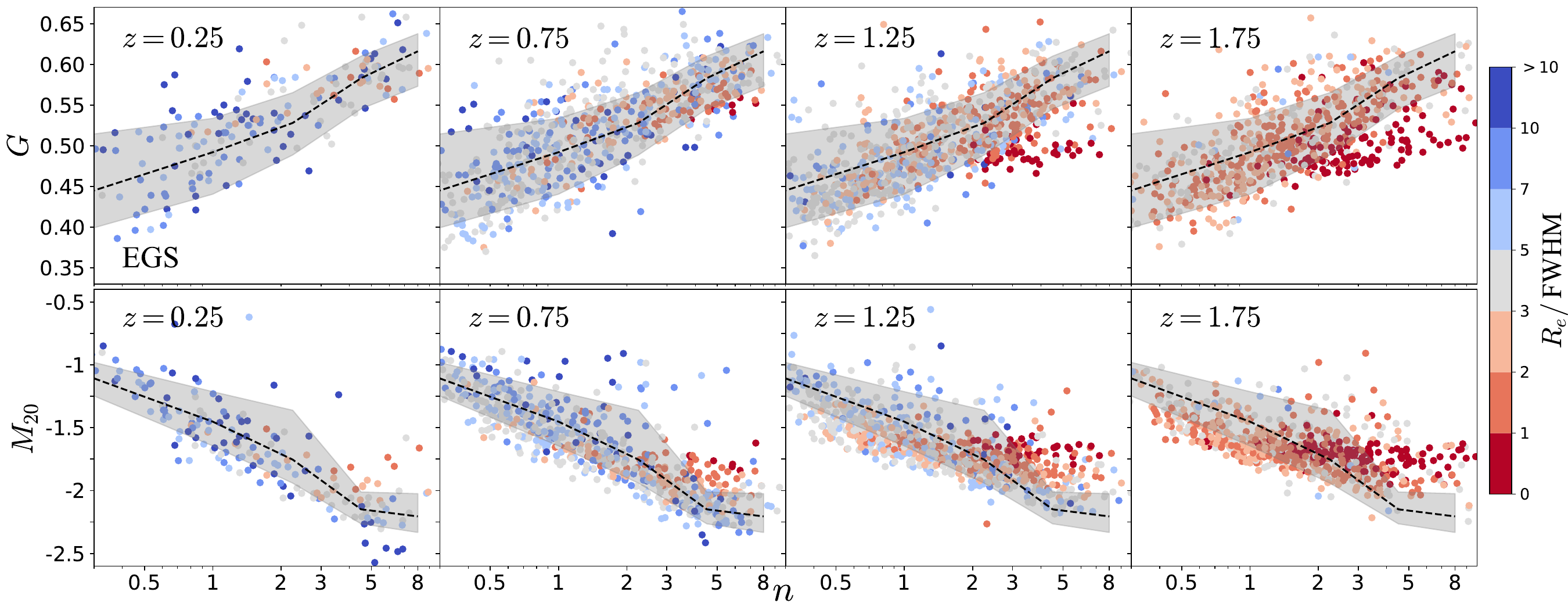}
\includegraphics[width=1\linewidth]{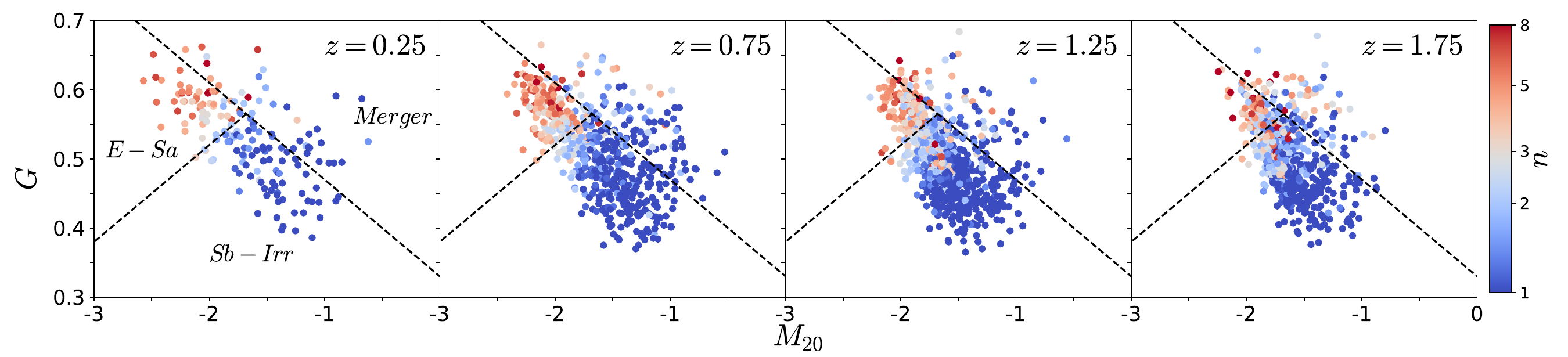}
\caption{The measured $G$ (top row) and $M_{20}$ (middle row) for the EGS images at $0<z<2$ as a function of $n$. In each panel, the black dashed line shows the median trend for the $R_e/{\rm FWHM>5}$ sub-samples at $z<1$ (left two column), with the grey shaded region indicating the 16th to 84th percentile range of the variation. The distributions of galaxies in our sample at different redshift ranges in the $G$-$M_{20}$ space are shown in the bottom row, colour-coded by the S\'{e}rsic index. The dashed lines are the boundaries between different galaxy types (Merger, E-Sa, Sb-Irr) derived from \cite{Lotz_2008}.
\label{fig:egs_gm}}
\vspace{10pt}
\end{minipage}
\begin{minipage}[t]{0.92\linewidth}
\includegraphics[width=1\linewidth]{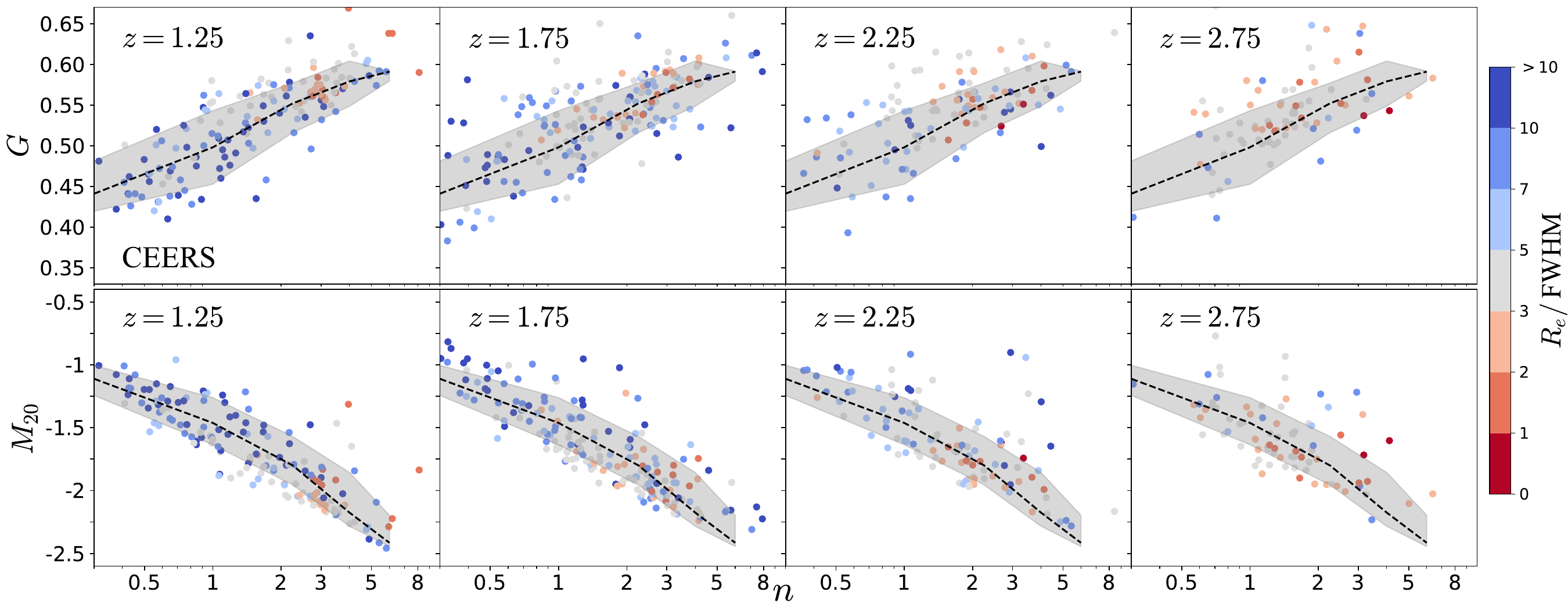}
\includegraphics[width=1\linewidth]{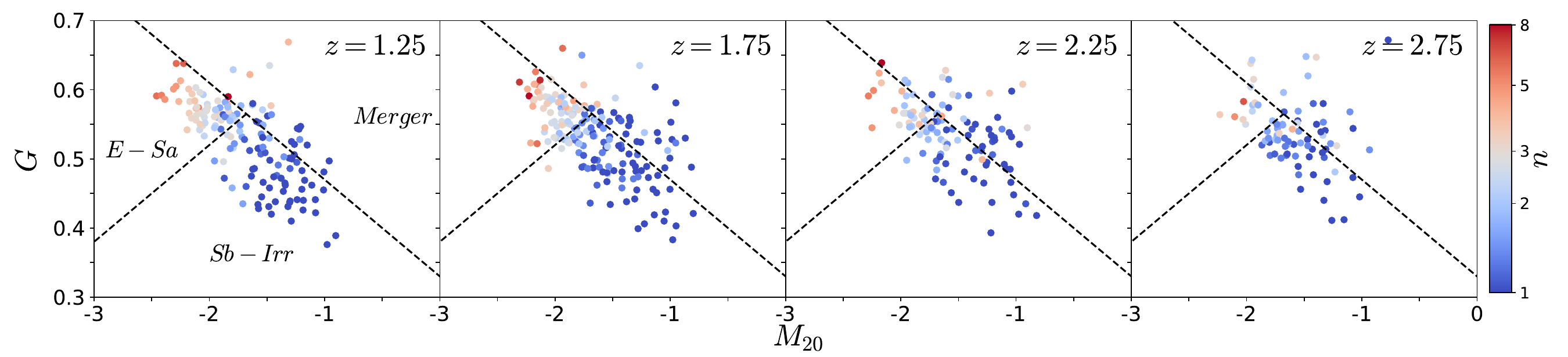}
\caption{Same as Fig. \ref{fig:egs_gm} but for the CEERS images at $1<z<3$. In each panel, the black dashed line shows the median trend for the $R_e/{\rm FWHM>3}$ sub-sample at $z=1.25$ (left column), with the grey shaded region indicating the 16th to 84th percentile range of the variation. The parameters are less affected by the PSF smoothing effect. The E-Sa and Sb-Irr galaxies in the $G$-$M_{20}$ are well separated in CEERS. \label{fig:jwst_gm}}
\end{minipage}
\end{figure*}

The relationship between $G$ and $M_{20}$ with respect to $n$ for the EGS images is shown in the top two rows of Fig. \ref{fig:egs_gm}.
The galaxies at higher redshifts roughly follow the dashed line of the low redshift sample, as indicated by the black dashed line for $z<1$\footnote{In practice, a numerical relationship between $G$ ($M_{20}$) and $n$ cannot be directly obtained. The two indicators are derived from the pixel values of the galaxy image, which depend on the structure of the galaxy. $M_{20}$ is also sensitive to the off-centre bright clumps in the discs. Note that we have selected galaxies with $R_e/{\rm FWHM}>5$ in this redshift range to ensure less PSF smoothing effect in these two parameters.}, but with substantial scatters, especially for the galaxies with lower $R_e/{\rm FWHM}$. 
At $z\geq1$, the $G$ values of the high S\'{e}rsic index galaxies with $R_e/{\rm FWHM}\leq2$ are generally lower than the low redshift relation (i.e., the dashed line). Generally speaking, due to the PSF smoothing effect, the Gini coefficient is underestimated, but not as significant as the concentration index shown in Fig. \ref{fig:egs_cn}. 

At $z=0.25-0.75$, $M_{20}$ exhibits a strong correlation with the S\'{e}rsic index, whereas the correlation becomes flattened at higher redshifts.
Note that for galaxies with $n \leq 2$, $M_{20}$ is slightly lower (with higher absolute values) than the low redshift relationship as shown in the left panel. As the galaxy image loses the fine structures at higher redshifts due to the surface brightness dimming and poor spatial resolution, the bright clumps distributed across the galaxy have been washed out to result in lower $M_{20}$. 
Good examples are shown in right two columns of Fig. \ref{fig:egs_images} and left two columns of Fig. \ref{fig:jwst_images}, as the discs observed by HST look much more smooth than those observed by JWST. At higher S\'{e}rsic indices, as the PSF smoothing effect is more significant, $M_{20}$ is systematically higher (with lower absolute values) than the low redshift one.

The distribution of the galaxies at different redshifts in the $G$-$M_{20}$ space is shown in the bottom panel of Fig. \ref{fig:egs_gm}. The black dashed lines are the boundaries between different galaxy types from \cite{Lotz_2008}, for $0.2 < z < 1.2$ EGS galaxies as following:
\begin{eqnarray}
\label{equ:g_m}
    {\rm Merger} &:& G>-0.14M_{20}+0.33, \nonumber\\
    {\rm E}-{\rm Sa} &:& G\leq -0.14M_{20}+0.33, \,G> 0.14M_{20}+0.80, \nonumber\\
    {\rm Sb}-{\rm Irr} &:& G\leq -0.14M_{20}+0.33, \,G\leq 0.14M_{20}+0.80. \nonumber \\
\end{eqnarray}

Focusing on the E-Sa and Sb-Irr regions, at $z\sim0.5$, the dashed line provides a clear demarcation between the two regions. Most points with $n\leq3$ are located in the Sb-Irr region. With the redshift increasing, the Sb-Irr region is gradually populated by galaxies with high S\'{e}rsic indices. For all $n>4$ galaxies, at $z=0.25$, no object resides in the Sb-Irr region. And the fraction of the $n>4$ galaxies in this region becomes 5\% for $z=0.75$ (5 in 99), 26\% for $z=1.25$ (20 in 77), and 52\% for $z=1.75$ (37 in 71). 
It seems that the empirical classifications of the galaxy morphology works well at lower redshifts. At higher redshifts, due to the observational bias, a significant fraction of the galaxies with high S\'{e}rsic indices would be misclassified as discs in the $G$-$M_{20}$ space. 
\cite{Lotz_2008} has shown that, as the redshift increases, the observed Sb-Irr fraction rises, while the observed fraction of E-Sa declines. Consistent with our result, this trend suggests that the distribution of the galaxies in $G$-$M_{20}$ space is affected by the PSF smoothing effect, leading to inaccurate interpolations of their morphologies. In this case, simply correcting the $G$ or $M_{20}$ values (e.g. a systematic shift for all the $G$ and $M_{20}$ values measured at the high redshift) would not help to improve the separation of the galaxies with different morphologies, since these galaxies are mixed together in the $G$-$M_{20}$ space. 

For the CEERS images, both $G$ and $M_{20}$ are less affected by the PSF smoothing effect at each redshift bin ($1<z<3$), as shown in Fig. \ref{fig:jwst_gm}. This is attributed to the higher spatial resolution in JWST. The empirical separation between the $n>2.5$ and $n<2.5$ galaxies in the $G$-$M_{20}$ space also works well at different redshifts.

\section{Discussion}\label{sec:discussion}
As we have shown, the PSF smoothing effect could significantly affect the measured non-parametric morphology indicators, especially for the galaxies with smaller sizes and higher S\'{e}rsic indices. The traditional correction strategy by adding a single term to all the galaxies in a given high redshift range could not properly account for such a complexity. In this section, the observational bias and problems with the previous correction methods are investigated and discussed in detail.

\subsection{Effect of the galaxy axis ratio} \label{sec:c_e}

\begin{figure*}[htbp]
\centering
\subfigure{
\begin{minipage}[t]{0.235\linewidth}
\includegraphics[width=1\linewidth]{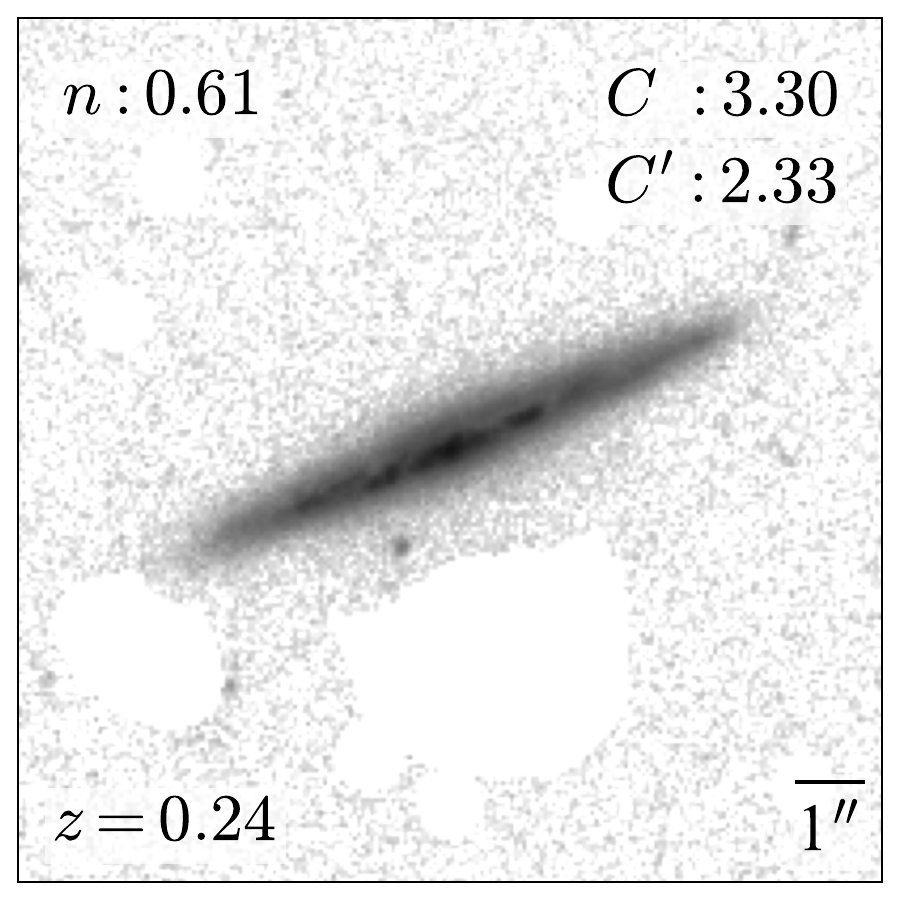}
\end{minipage}}
\hspace{-8pt}
\subfigure{
\begin{minipage}[t]{0.235\linewidth}
\includegraphics[width=1\linewidth]{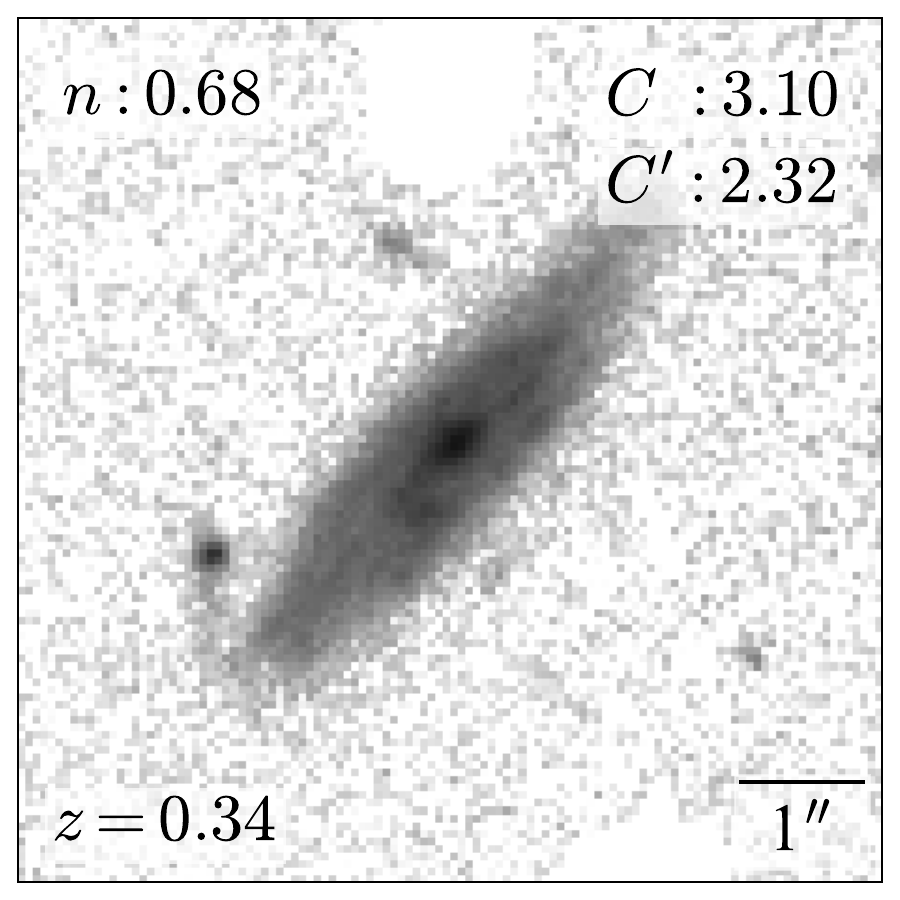}
\end{minipage}}
\hspace{-8pt}
\subfigure{
\begin{minipage}[t]{0.235\linewidth}
\includegraphics[width=1\linewidth]{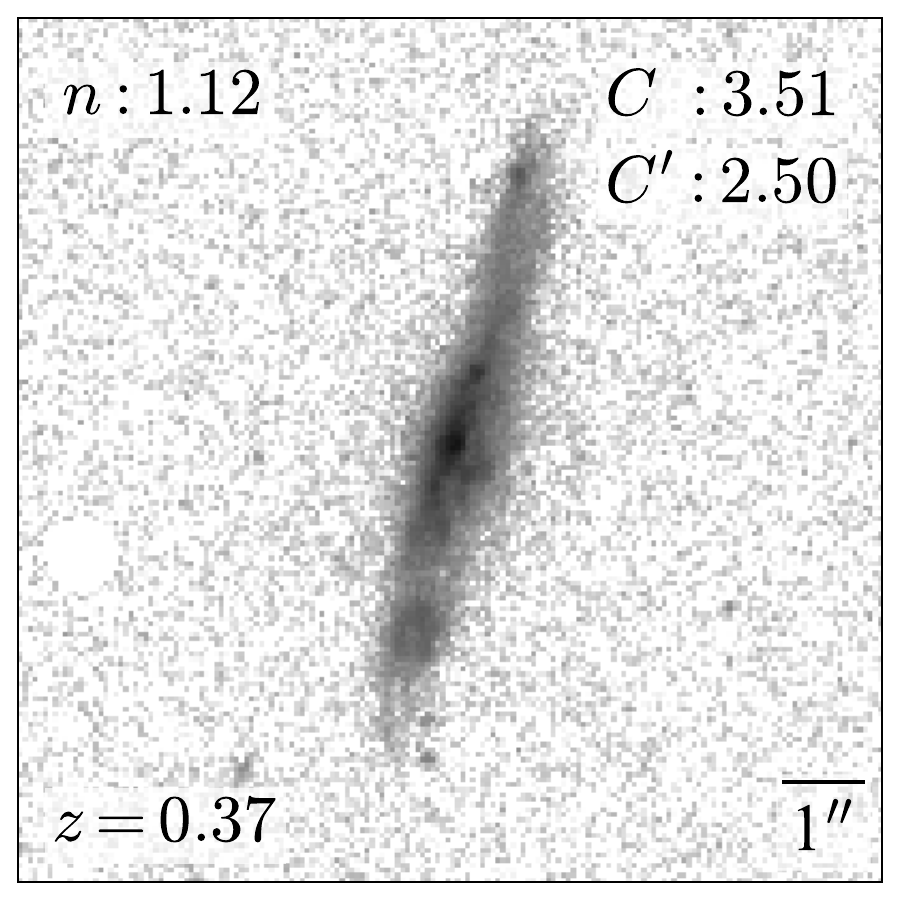}
\end{minipage}}
\hspace{-8pt}
\subfigure{
\begin{minipage}[t]{0.235\linewidth}
\includegraphics[width=1\linewidth]{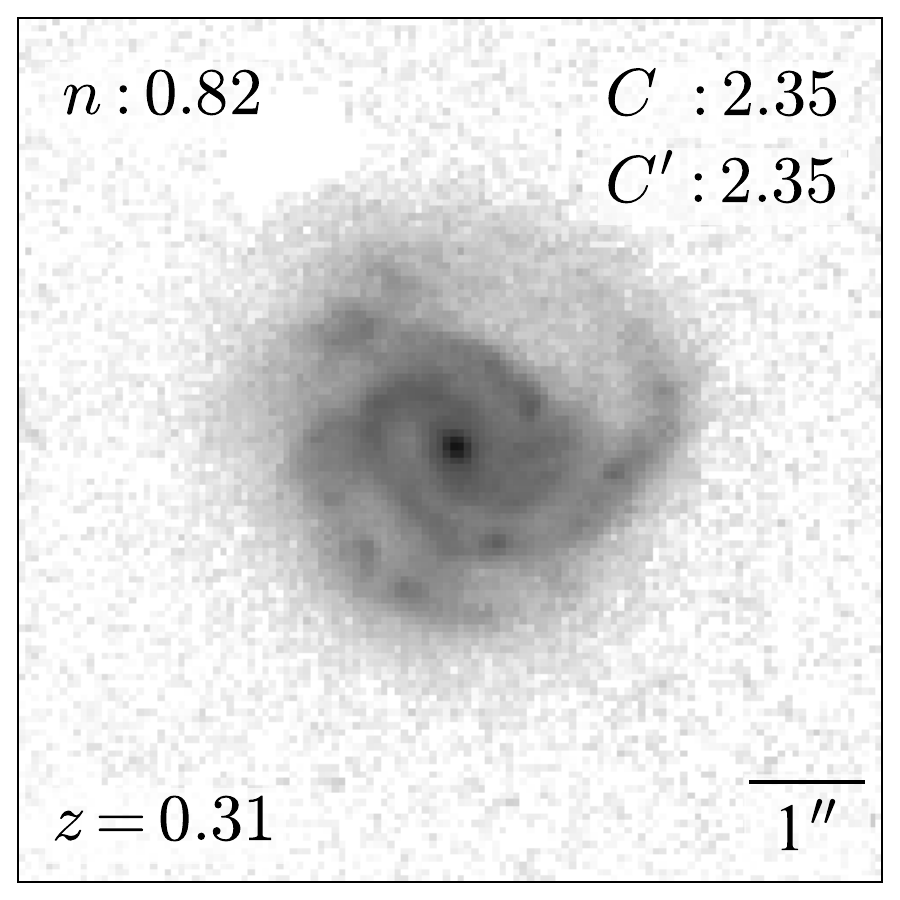}
\end{minipage}}
\caption{Examples of the edge-on (left three panels) and face-on (right panel) galaxies at $z=0.25$ for the EGS images. The left three edge-on galaxies with small axis ratio ($q<0.3$) have $C$ (from the circular aperture) overestimated compared to $C^{\prime}$ (from elliptical apertures with the same axis ratio as the galaxy outskirt). For the face-on galaxy, $C$ and $C^{\prime}$ are close to each other. \label{fig:e_figure}}
\end{figure*}

\begin{figure*}[ht!]
\centering
\begin{minipage}[t]{0.98\linewidth}
\includegraphics[width=1\linewidth]{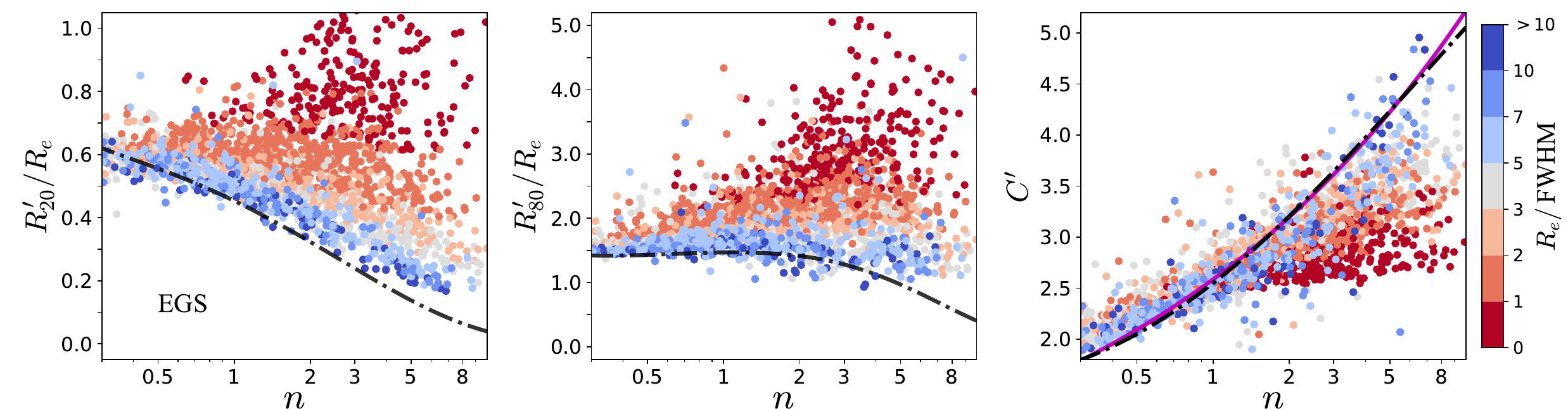}\vspace{-2pt}
\includegraphics[width=1\linewidth]{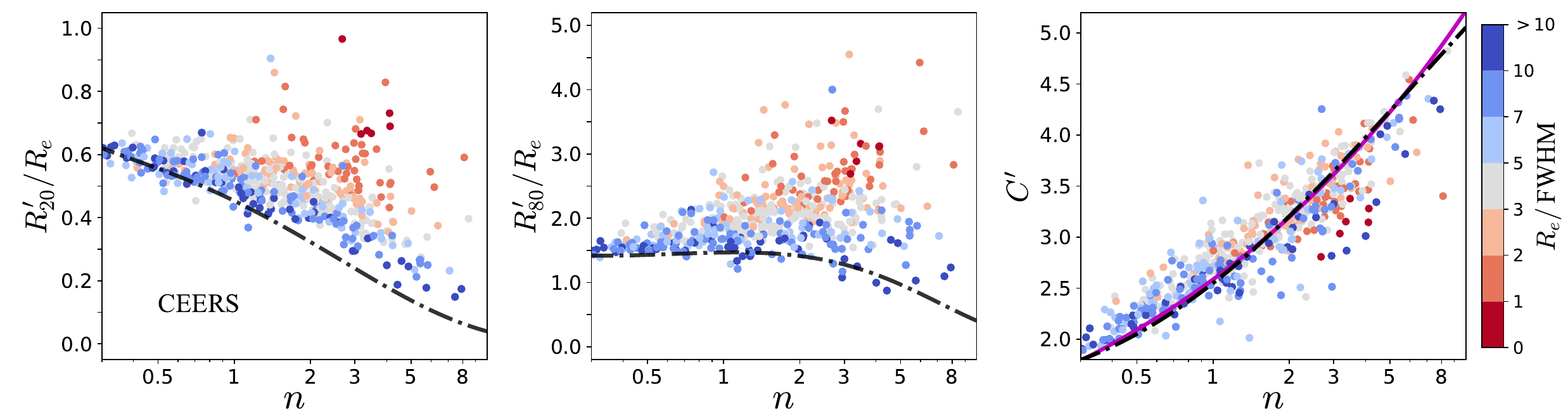}
\end{minipage}
\caption{Similar to Figs. \ref{fig:egs_cn} and \ref{fig:jwst_cn}, the structural parameters of all the EGS (top) and CEERS (bottom) images are measured with the elliptical apertures with the same axis ratio as the galaxy outskirt. The scatter is much more reduced compared to Figs. \ref{fig:egs_cn} and \ref{fig:jwst_cn} for the galaxies with low $n$ and large $R_e/{\rm FWHM}$. \label{fig:ce}}
\end{figure*}

In the galaxy size measurement, the $R_{20}$ and $R_{80}$ values are determined using a circular annulus via the curve of growth method (out to $1.5R_p$) in {\tt Statmorph}, while $R_e$ is derived from the single S\'{e}rsic component fit using G{\small ALFIT}. \cite{Andrae_2011} suggested that the concentration index can be influenced by the aperture axis ratio. For $q < 0.5$, the concentration index measured with the circular aperture could be substantially overestimated (even up to 30\% for galaxies with $n = 1$). 

It is noticeable that in the $C$-$n$ space, the distribution of galaxies with $n\leq2$ scatter around the expected correlation in Fig. \ref{fig:egs_cn}. Most of these objects are close to edge-on, with examples shown in the left three panels of Fig. \ref{fig:e_figure}. Here $C^\prime$ is the concentration index measured in the elliptical apertures with the same axis ratio as the galaxy outskirt. 
Galaxies in the left three panels with small axis ratios ($q<0.3$) have $C > C^\prime$, while the face-on galaxy in the right panel has similar $C^\prime$ and $C$. Note that the four galaxies have the similar S\'{e}rsic index ($n\sim1$), implying that their intrinsic $C$ values should also be similar. Indeed, after adopting the ellipse aperture, the $C^\prime$ values of these galaxies are all $\sim2.5$. It is necessary to use elliptical isophotes to get an accurate concentration index measurement to verify our previous result in Sect. \ref{sec:c_n}. 

The relationship between the measured $C^{\prime}$ and $n$ is shown in Fig. \ref{fig:ce}. Compared to $C$, it is observed that most blue points ($R_e/{\rm FWHM} > 5$) for $C^{\prime}$ align more closely with the expected correlation. The measured $R^{\prime}_{20}/R_e$ for the galaxies at small S\'{e}rsic indices with the elliptical apertures are slightly higher above the theoretical curve. 
This is mainly caused by the fact that the galaxy inner region is usually rounder than the outskirts due to the existence of the bulge (or AGN), with the PSF further enhancing this effect. Our test indicates that the axis ratio of the aperture is important to get the accurate concentration measurement, consistent with \cite{Andrae_2011}.
Using the elliptical apertures can improve the accuracy of the concentration index measurement, as it prevents some edge-on galaxies from exhibiting higher $C$ values in the case of the circular aperture. Nonetheless, our previous conclusions about the PSF smoothing effect on the concentration index are not affected, as shown in the right column of Fig. \ref{fig:ce}.

\subsection{Understanding the PSF smoothing effect with mock images}

\subsubsection{Test with the idealised mock images}\label{sec:c_mock}
\begin{figure*}[htbp]
\centering
\includegraphics[width=0.95\textwidth]{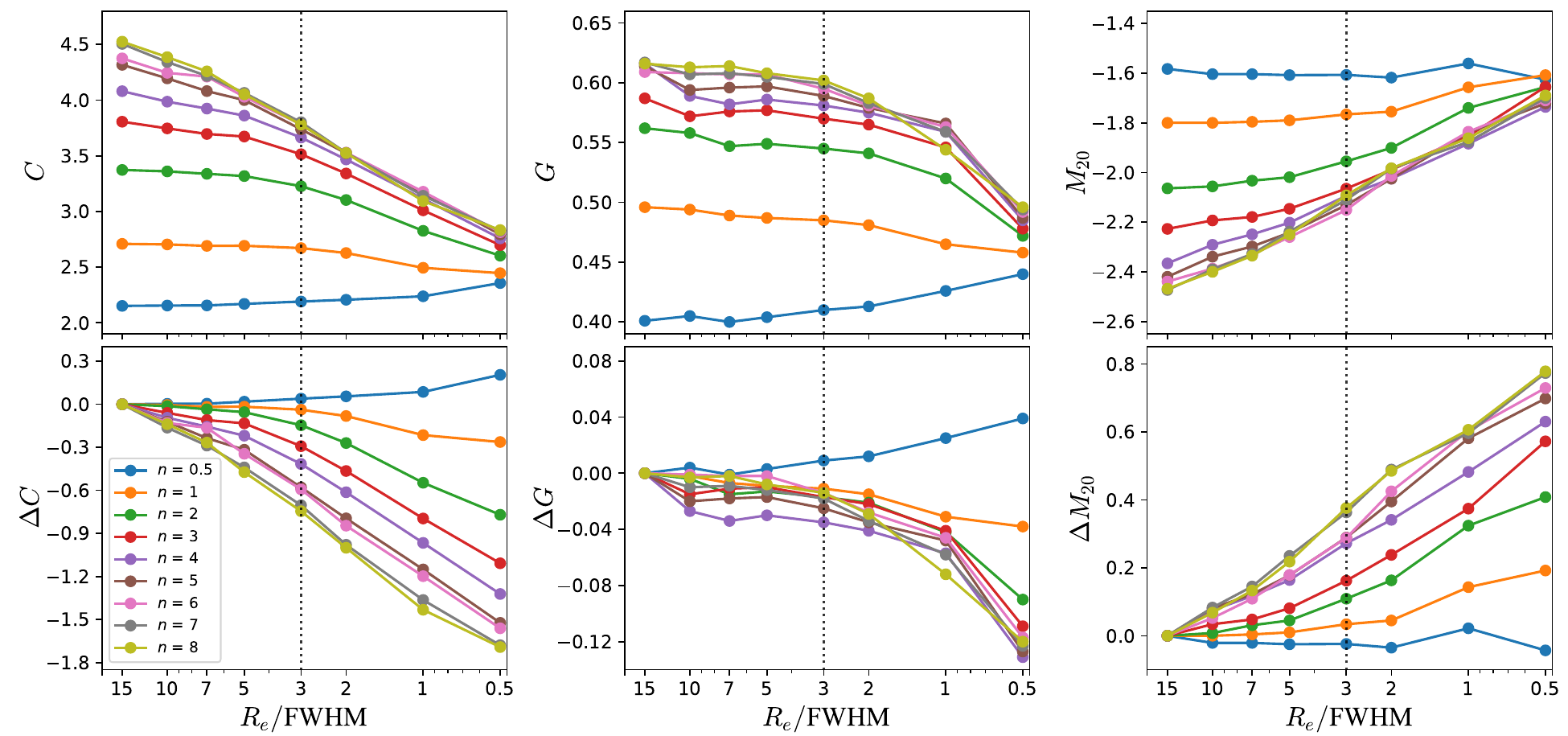}
\caption{The measured $C$ (left), $G$ (middle) and $M_{20}$ (right) as a function of $R_e/{\rm FWHM}$, with different colour representing different S\'{e}rsic indices. The top and bottom rows show the directly measured structural parameter ($C$, $G$, $M_{20}$) and the deviation from the intrinsic values ($\Delta C$, $\Delta G$, $\Delta M_{20}$), respectively. As the galaxy becomes smaller, $C$, $G$ and $M_{20}$ gradually converge to the corresponding values of the PSF. \label{fig:hst_gal}}
\end{figure*}

To mimic the PSF smoothing effect on the galaxies in the F160W band with different sizes and S\'{e}rsic indices, we create a series of mock images on a $500\times 500$ pixel grid with $0.5\leq n\leq 8$ and $0.5 \leq R_e/{\rm FWHM}\leq 15$ ($0.1\arcsec\leq R_e\leq3.0\arcsec$) with the pixel scale of 0.06 \arcsec/pixel, and the total magnitude $m=18.5$ (with ${\rm S/N}>20$). These idealised images are measured with {\tt Statmorph} for the non-parametric morphology indicators in the first place, then convolved with the PSF in F160W and added with the sky background in the same filter. The mock images are measured again {\tt Statmorph} for the non-parametric morphology indicators, and compared with the intrinsic structural parameters.
The variation of the measured $C$ values with the galaxy sizes and S\'{e}rsic indices are shown in the top left panel of Fig. \ref{fig:hst_gal}. The bottom left panel shows the relative deviation, $\Delta C = C-C_{\rm int} $, where $C_{\rm int}$ is the intrinsic concentration index. It is clear that the $C$ value remains roughly the same for low S\'{e}rsic indices ($n\leq1$) with little or no impact from the PSF. For the galaxy with higher S\'{e}rsic index, as $R_e/{\rm FWHM}$ decreases, the concentration index estimated from the mock image decreases, with larger decrement for the galaxies of higher S\'{e}rsic indices.

The vertical dashed line marks the position at $R_e/{\rm FWHM} = 3$, where the measured concentration indices of the mock images with $n\geq4$ are roughly the same ($\sim 3.8$). At even smaller $R_e/{\rm FWHM}$, the $C$ values of these mock images with high S\'{e}rsic indices (such as $n \geq 4$) are quite similar. This agrees with \cite{Andrae_2011}, that the $C$ values of the galaxies with higher S\'{e}rsic indices are more affected than those with smaller S\'{e}rsic indices.

For the extremely small mock galaxies with different S\'{e}rsic indices and $R_e/{\rm FWHM} \sim0.5 $, the $C$ values converge to $\sim 2.7$, corresponding to the value of the PSF. The results are in agreement with the $C$ values of the smaller galaxies (red dots) in the bottom rows of Fig. \ref{fig:egs_cn}.
The concentration index directly measured from these images would be significantly underestimated, up to 30\% for these small galaxies with high S\'{e}rsic indices ($n\geq3$). This suggests that the galaxy light profile has been overshadowed by PSF. In other words, the measured $C$ value does not reflect the shape of the galaxy light profile, but the shape of the PSF profile.

At different S\'{e}rsic indices, the $G$ values roughly remain the same for the mock images with $R_e/{\rm FWHM}\geq3$, and then decrease at smaller $R_e/{\rm FWHM}$, which eventually converge at $\sim 0.5$. Note that at smaller $R_e/{\rm FWHM}$, the measured G values of mock images with $n\sim 0.5$ actually increase to reach 0.5.
Compared to the concentration index in the left column, the Gini coefficient measurement is less affected by the PSF smoothing effect.
Moreover, for galaxies with $n>3$, their $G$ values are similar, with slightly larger $G$ values for higher S\'{e}rsic indices.
The $M_{20}$ statistic also shows the increasing pattern (with its absolute value decreasing) similar to the concentration index, which converges to $\sim -1.70$.

The results of mock images for the CEERS F200W observation are shown in Appendix \ref{app:a}, which are similar to the results of the EGS F160W mock images. Note that for extreme cases ($R_e/{\rm FWHM}=0.5$), $R_e$ is only one pixel large. This galaxy can be considered as point source. The radial light profile actually reflects the PSF. However, in CEERS SW observations, it is rare to find $M_{\ast}>10^{9.5}{\rm M_{\odot}}$ galaxies with $R_e/{\rm FWHM}<1$ ($R_e\sim 0.06^{\prime \prime}$). In this case, these galaxies with higher $n$ and smaller $R_e$ do not show large deviation from the theoretical expectation between $C$ and $n$.

\subsubsection{Test with the artificially redshifted galaxies}

To correct for the observational bias in the concentration-related parameters of the high redshift galaxies, the common approach is to create the mock images for the high
redshift galaxies from the images of the low redshift ones, which are then measured to derive the corresponding correction factors \citep{Giavalisco_1996,Conselice_2003,Tohill_2021, Yu_2023}. The basic procedure of the mock image generation includes resolution degradation, surface brightness dimming, size and luminosity evolution, PSF matching and noise addition. This strategy has been widely used to verify and correct the possible observational bias in the detection and measurement of bars and spiral arms in discs \citep{van-den-Bergh_2002,Sheth_2008,Yu_2018}, the non-parametric indicators \citep{Abraham_1996,Petty_2014,Whitney_2021,Yu_2023}, the S\'{e}rsic index and size measurement of high redshift galaxies \citep{Barden_2008,Paulino-Afonso_2017}.

\begin{figure*}[htbp]
\centering
\subfigure{
\begin{minipage}[t]{0.235\linewidth}
\includegraphics[width=1\linewidth]{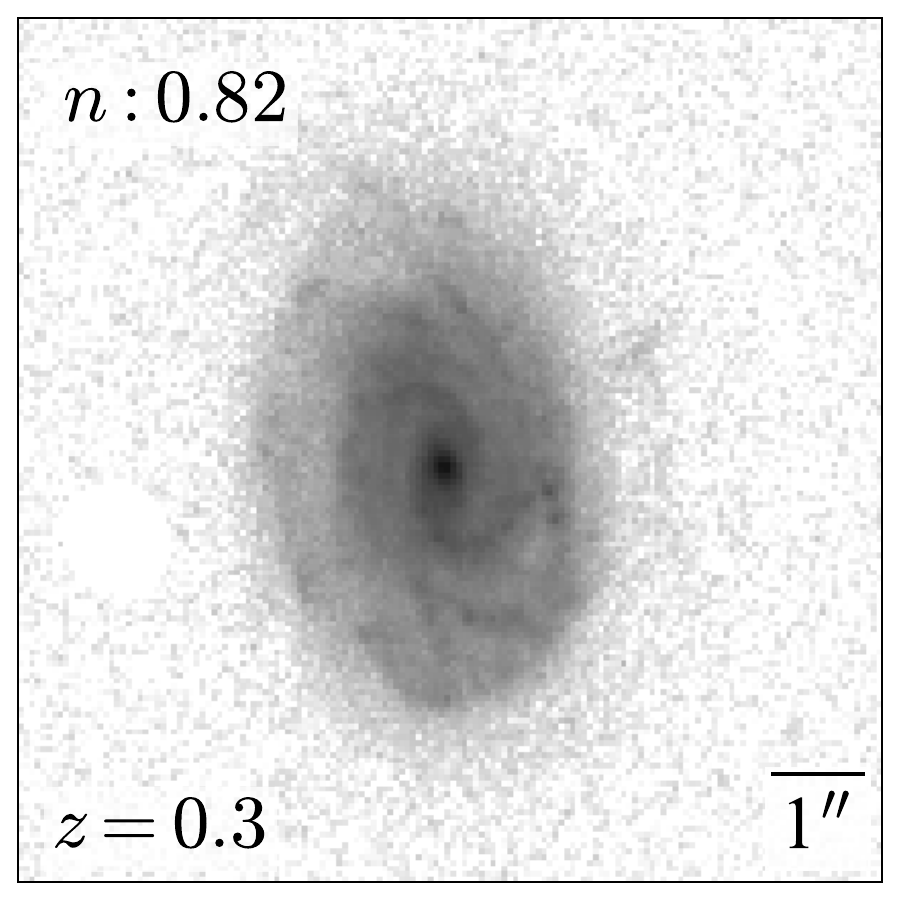}\vspace{0pt}
\includegraphics[width=1\linewidth]{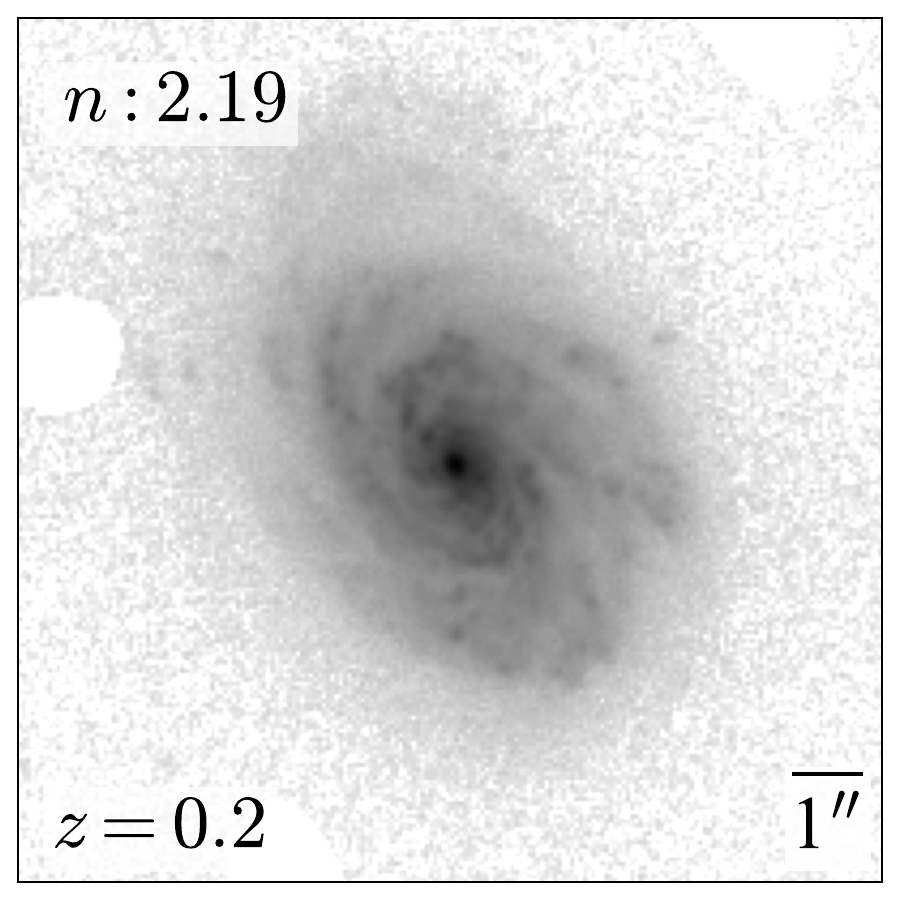}\vspace{0pt}
\includegraphics[width=1\linewidth]{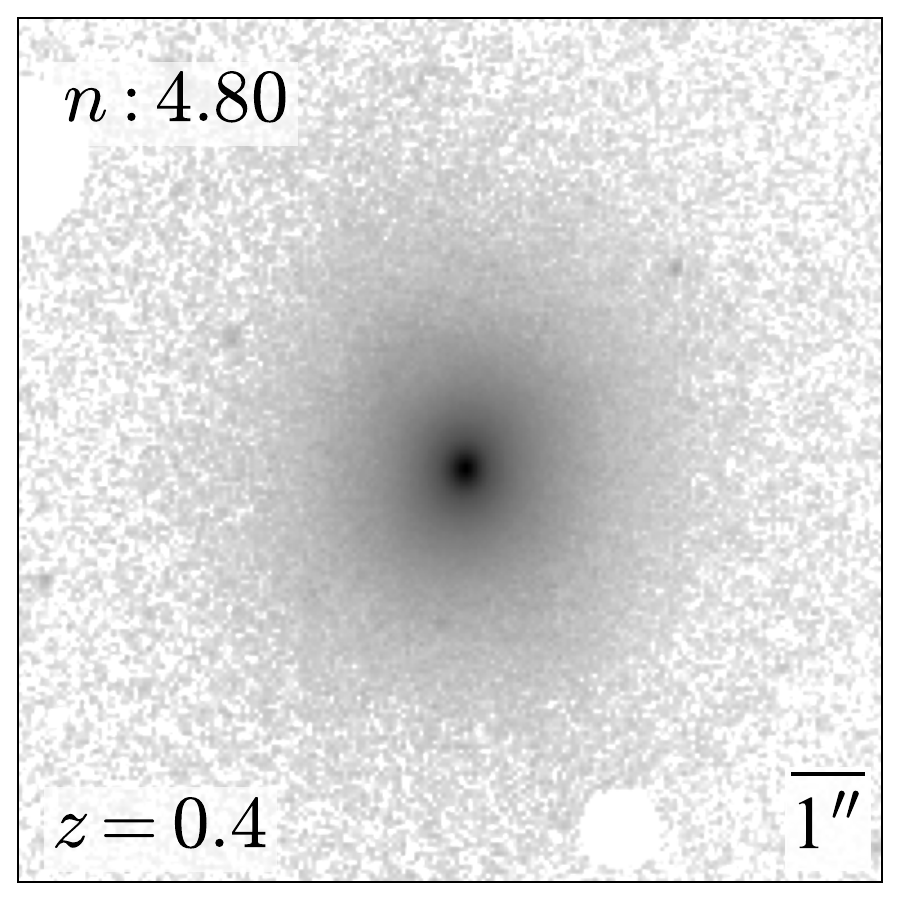}
\end{minipage}}
\hspace{-8pt}
\subfigure{
\begin{minipage}[t]{0.235\linewidth}
\includegraphics[width=1\linewidth]{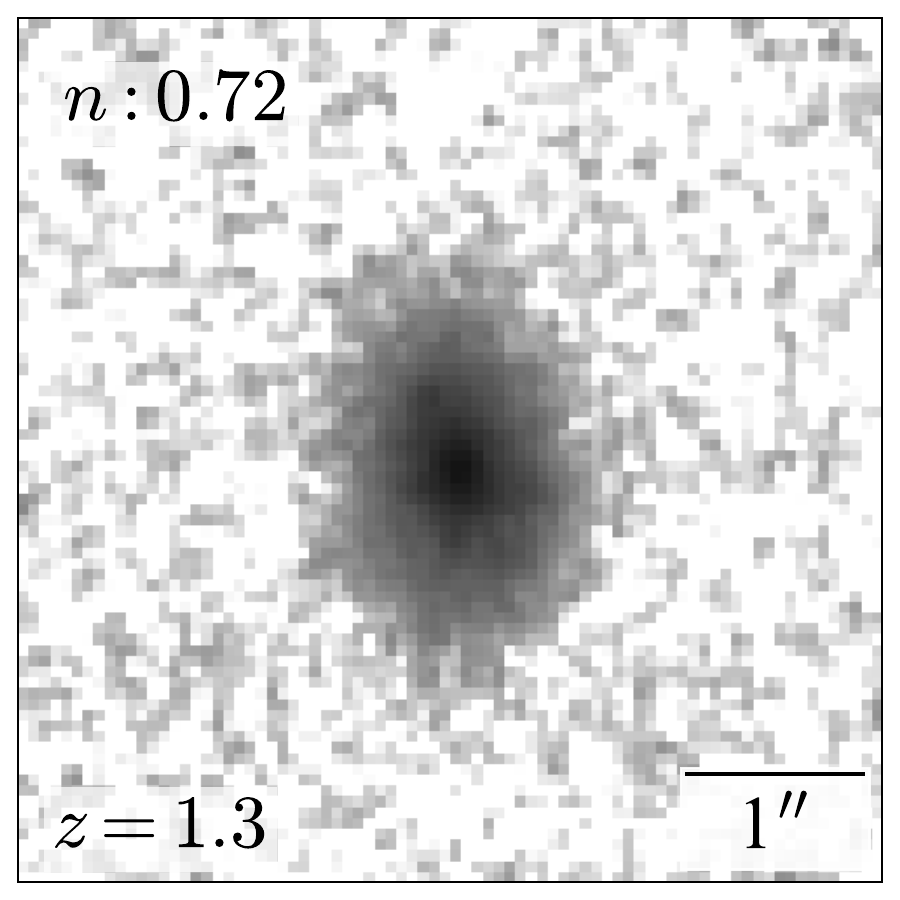}\vspace{0pt}
\includegraphics[width=1\linewidth]{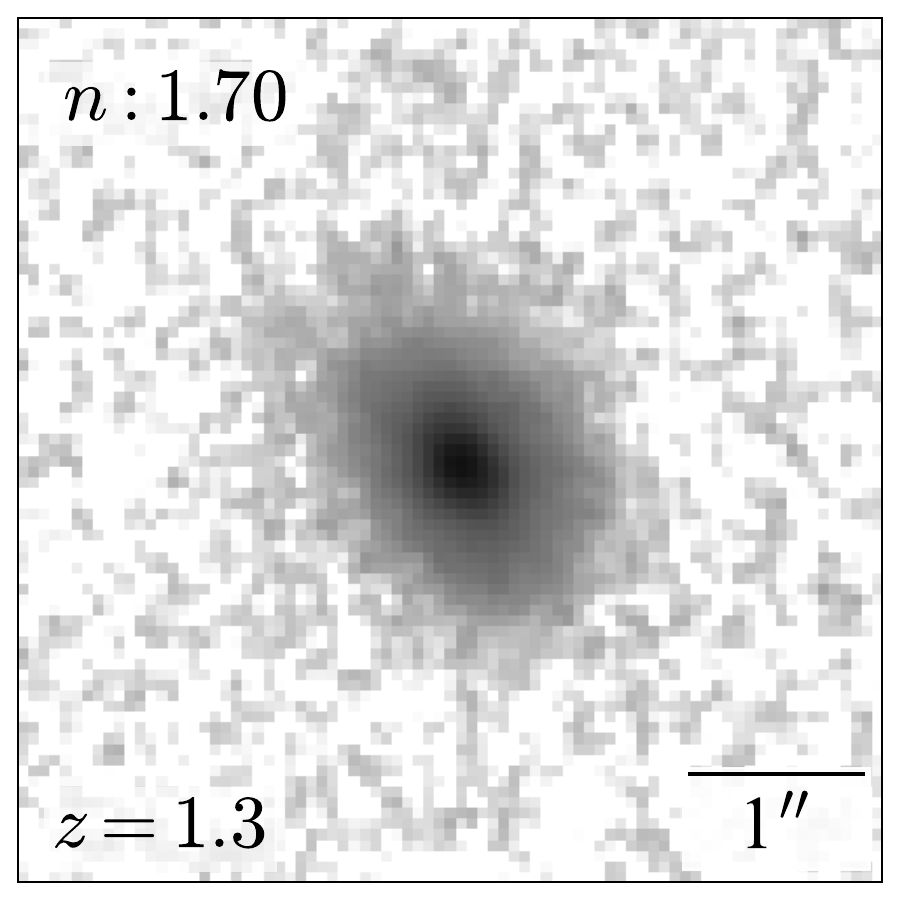}\vspace{0pt}
\includegraphics[width=1\linewidth]{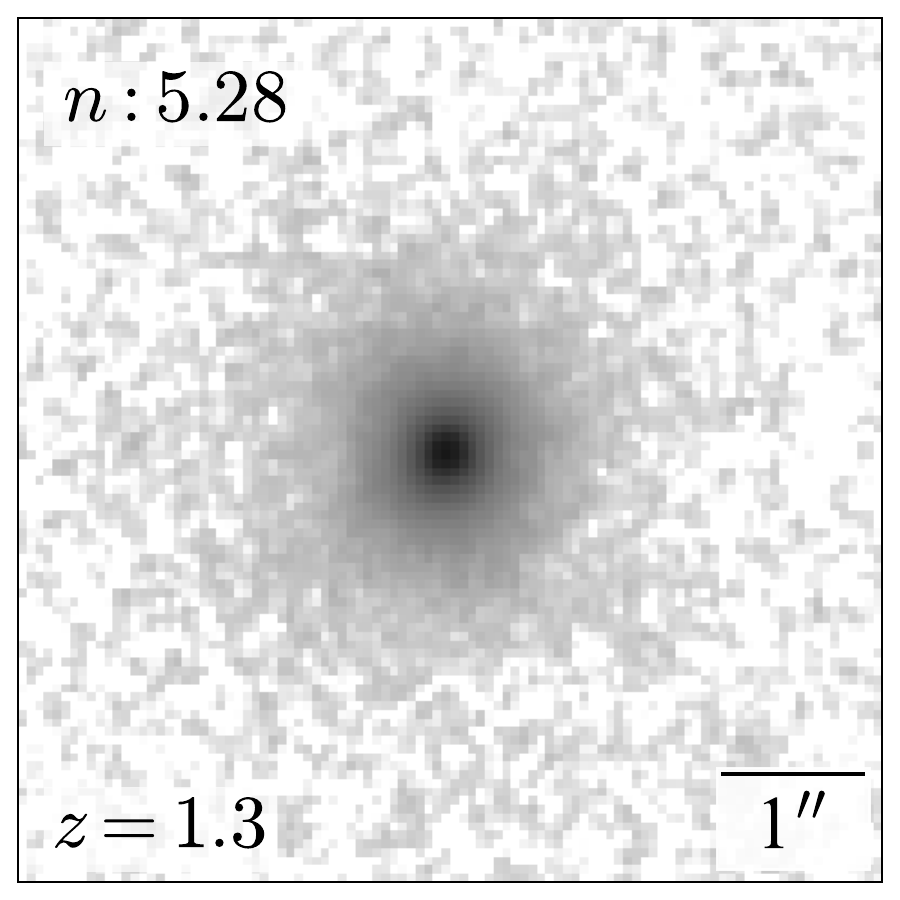}
\end{minipage}}
\hspace{-8pt}
\subfigure{
\begin{minipage}[t]{0.235\linewidth}
\includegraphics[width=1\linewidth]{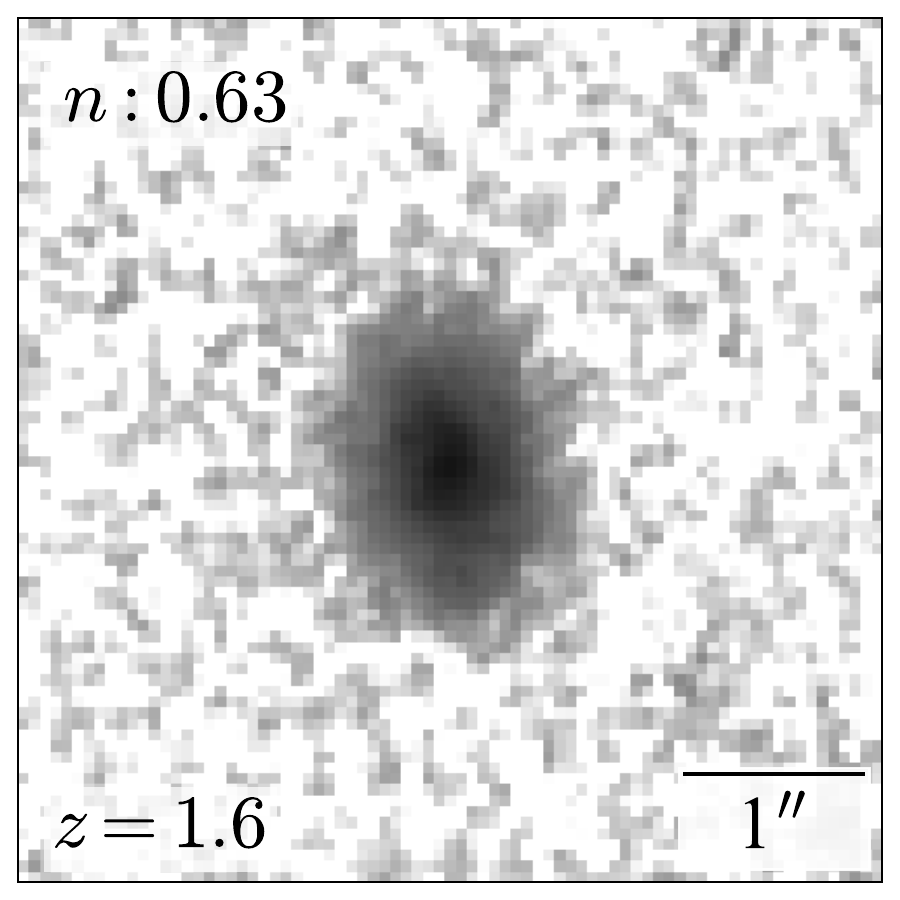}\vspace{0pt}
\includegraphics[width=1\linewidth]{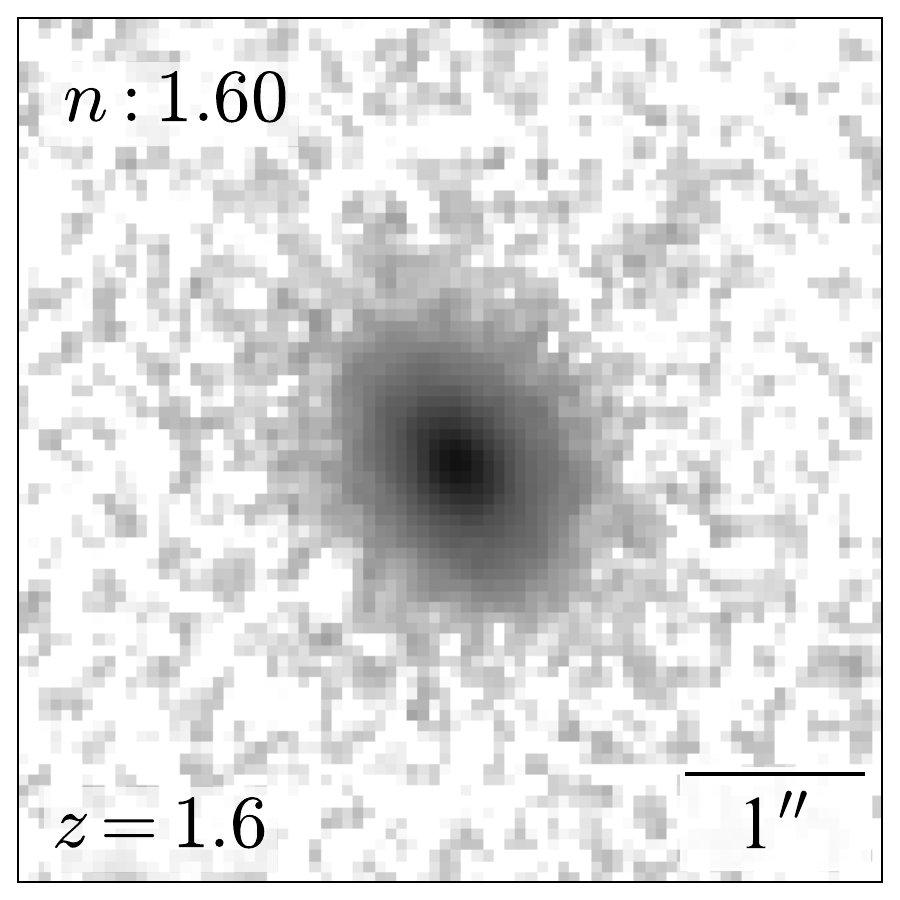}\vspace{0pt}
\includegraphics[width=1\linewidth]{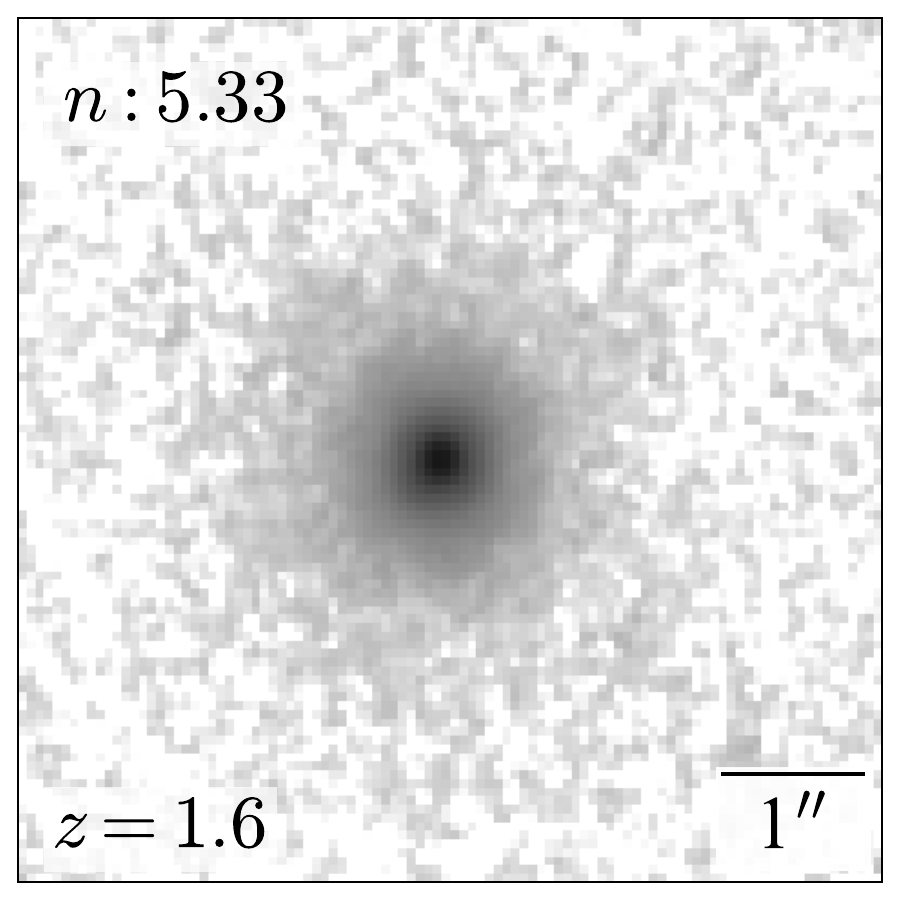}
\end{minipage}}
\hspace{-8pt}
\subfigure{
\begin{minipage}[t]{0.235\linewidth}
\includegraphics[width=1\linewidth]{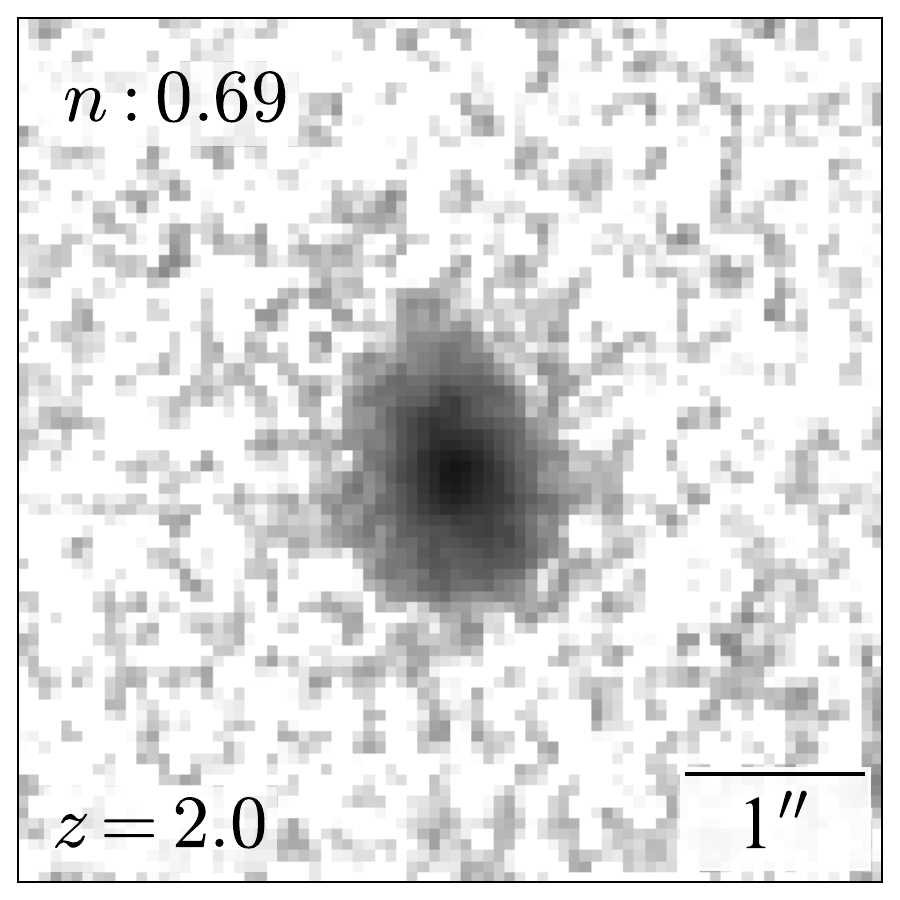}\vspace{0pt}
\includegraphics[width=1\linewidth]{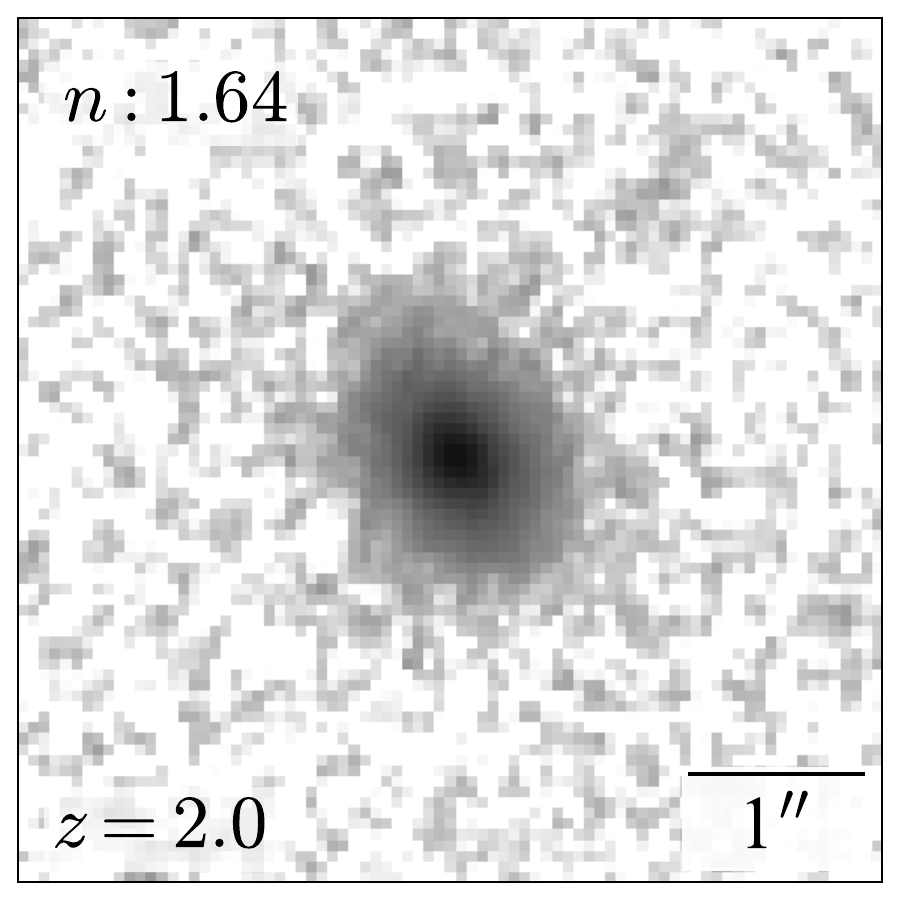}\vspace{0pt}
\includegraphics[width=1\linewidth]{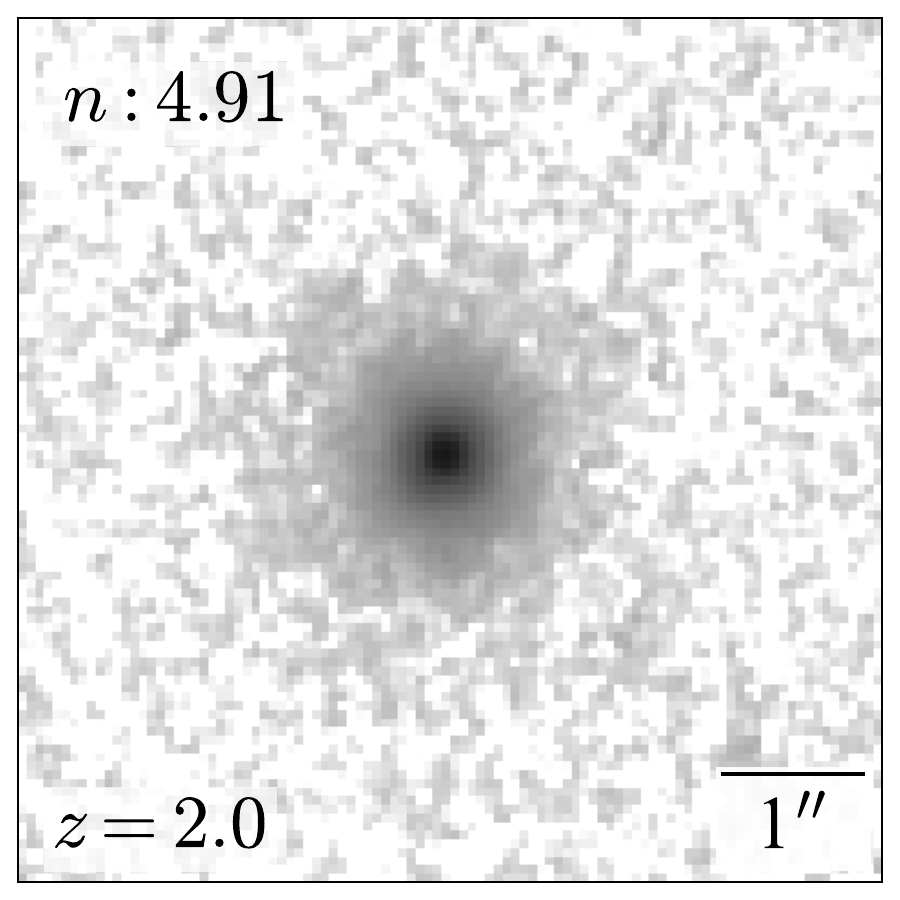}
\end{minipage}}
\caption{Examples of the artificially redshifted galaxies in the HST observation. The original low redshift galaxy is shown in the left column, with the right three columns showing the high redshift mock images. The best-fit S\'{e}rsic index is also labelled in each panel. \label{fig:mock}}
\end{figure*}

In order to evaluate the redshift effect, we also use the low redshift galaxies in the EGS images ($z\sim0.3$) to generate the mock images of the HST observed higher redshift galaxies ($1.1\leq z \leq2.0$). Similar to \cite{Yu_2023}, first, we consider both the angular size variation of the galaxies at high redshifts and size evolution through the cosmic time to re-scale the galaxy images with a proper rebinning factor. The evolution of the average effective radius is considered as $R_e \varpropto(1+z)^{\beta} $ with $\beta=-0.75$ for the discs and ${\beta=-1.48}$ for the spheroids \citep{van-der-Wel_2014}. The discs and spheroids are separated at $n=3$. The binned images are then scaled with $(1+z)^{-4}$ to account for the surface brightness dimming on the redshift \citep{Tolman_1930} and $(1+z)$ for the cosmological compression of the frequency.
We also adopt the luminosity evolution given by $(1+z)^{\alpha}$, where $\alpha$ is estimated for different rest-frame wavelengths, galaxy types, and stellar masses for the CANDELS galaxies (see \cite{Yu_2023} for more details). The surface brightness ($\mu_f$) for the mock galaxy at the final redshift ($z_f$) is 
\begin{equation}
    \mu_f = \mu_i + 2.5{\rm log}\left(\frac{1+z_f}{1+z_i}\right)^3- 2.5{\rm log}\left(\frac{1+z_f}{1+z_i}\right)^{\alpha}- 2.5{\rm log}\left(\frac{1+z_f}{1+z_i}\right)^{-2\beta}
    %\mu_f = \mu_i &+& 2.5{\rm log}\left(\frac{1+z_f}{1+z_i}\right)^3\nonumber\\ &-& 2.5{\rm log}\left(\frac{1+z_f}{1+z_i}\right)^{\alpha}\nonumber \\ &-& 2.5{\rm log}\left(\frac{1+z_f}{1+z_i}\right)^{-2\beta},
\label{mu_equ}
\end{equation}
where $z_i$ and $\mu_i$ are the redshift and the surface brightness of the initial galaxy, respectively. We do not consider the morphological or magnitude k-correction, since our samples are at the similar rest-frame waveband, where the structural parameters do not vary too much \citep{Conselice_2003,Baes_2020,Nersesian_2023}. More details of these procedures are described in Appendix \ref{app:b}.

\begin{figure*}[htbp]
\centering
\includegraphics[width=1\textwidth]{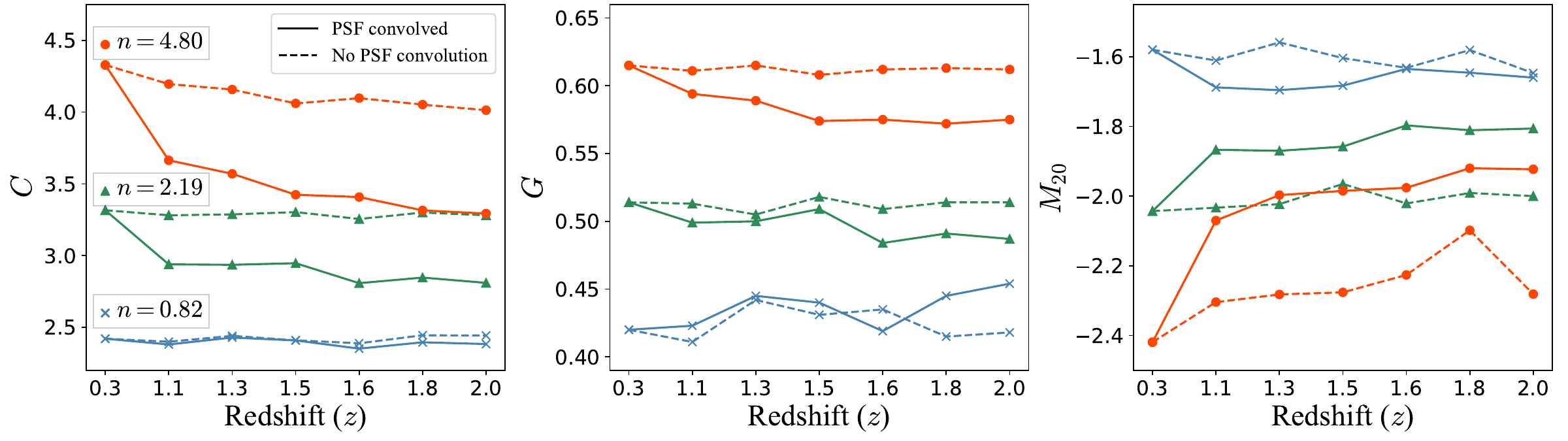}
\caption{The structural parameters of the three galaxies and the corresponding high redshift mock images in Fig. \ref{fig:mock} are measured. The colour indicates the galaxies with different S\'{e}rsic indices ($n=0.82$, $2.19$, and $4.80$) for the top, middle, bottom rows in Fig. \ref{fig:mock}, respectively. The solid lines are the results of the PSF convolved mock galaxies, and the dashed lines are the results of the mock galaxies without the PSF convolution. \label{fig:egs_mock}}
\end{figure*}

In order to mimic the PSF effect, we rebin the original PSF in the filter at a lower redshift, similar to the galaxies binning procedure. Then we utilise the {\tt create\_matching\_kernel} function in {\tt Photutils} \citep{larry_bradley_2022_6825092} to create the PSF matching kernel. We choose the {\tt TukeyWindow} to remove high frequency noise from the PSF matching kernel with $$\sigma_{\rm kernel} = \sqrt{\sigma_{\rm target}^2-\sigma_{\rm rebin}^2},$$ where $\sigma_{\rm kernel}$, $\sigma_{\rm target}$ and $\sigma_{\rm rebin}$ are the standard deviations of the PSF matching kernel, the target PSF and the rebinned PSF, respectively.
Examples of mock images are shown in Fig. \ref{fig:mock}. Note that for the galaxies at $z=1.1$, $1.3$, and $1.5$, the target PSF and the background are adopted from WFC3/F125W, while for the galaxies at $z=1.6$, $1.8$, and $2.0$, the target PSF and the background are adopted from WFC3/F160W. We choose three typical galaxies with $n=0.82$, $2.19$, and $4.80$ to represent the morphological variation from the disc and the spheroid. At higher redshifts, the best-fit S\'{e}rsic index slightly reduces for the galaxies in the top and middle rows, but increases slightly for the galaxy in the bottom row. %Nevertheless, the S\'{e}rsic index is still accurate enough to describe the light profile of galaxies.

Both mock images before and after the PSF convolution (with the sky background added) are both generated; the difference of the structural parameters between these two mock images can be used to quantify the PSF smoothing effect.
We calculated the structural parameters for these mock galaxies (with and without PSF convolution). Results of the $C$, $G$, and $M_{20}$ are shown in the left, middle and right columns of Fig. \ref{fig:egs_mock}, respectively. The solid and dashed lines are the results of the mock images with and without the PSF convolution, respectively. 
The non-parametric morphology indicators of the $n=0.82$ galaxy (blue line) do not change much for the mock images with or without PSF convolution. While the Gini coefficient is lower at $z=1.1$ and increases at the higher redshifts, primarily influenced by the noise.
For the $n=2.19$ galaxy (green line), the structural parameters of the images without the PSF convolution remain the same with increasing redshift, while the $C$, $G$ and absolute value of $M_{20}$ in the PSF convolved images decrease at higher redshift. 
For the $n=4.80$ galaxy (red line) without the PSF convolution, the $G$ value remains consistent across all redshifts, while the $C$ and $M_{20}$ values exhibit slight variations. At $z=1.1$, the $C$ value of the PSF convolved images decreases by 18\% and then slightly reduces at higher redshifts. The $G$ value and the absolute $M_{20}$ decrease at $z=1.1$ and remain unchanged at higher redshifts. 
Clearly, these structural parameters of the discs seem not sensitive to the PSF effect. While the structural parameters of the spheroids are significantly affected by the PSF, consistent with our previous results.

\subsection{Comparison with previous works on the structural parameter measurement of high redshift galaxies}

\cite{Whitney_2021} investigated how the concentration and asymmetry indicators evolve using the CANDELS data. They artificially redshifted a sample of the low redshift galaxies ($0.5 < z < 1$) to different high redshift ranges (up to $z \sim 2.75$) to derive the corresponding correction factor for the structural parameters. They found that the higher redshift galaxies are more asymmetric and more concentrated. 
However, our tests indicate that the non-parametric indicators are affected differently by the PSF smoothing effect, depending on the S\'{e}rsic index and relative size of the galaxy. The $C$ values of the galaxies with the higher S\'{e}rsic indices (spheroids) are usually more affected (underestimation) than the galaxies with the lower S\'{e}rsic indices (discs). For the galaxies with $n\leq2$, the concentration index does not even need any correction. Without distinguishing different types of the galaxies (with different S\'{e}rsic indices), simply applying the same correction factor to all the measured $C$ values of the high redshift galaxies would result in systematic overestimation of the intrinsic concentration index for discs and systematic underestimation for the spheroids at higher redshifts.

\begin{figure}[htbp]
\centering
\subfigure{
\begin{minipage}[t]{0.48\linewidth}
\includegraphics[width=1\linewidth]{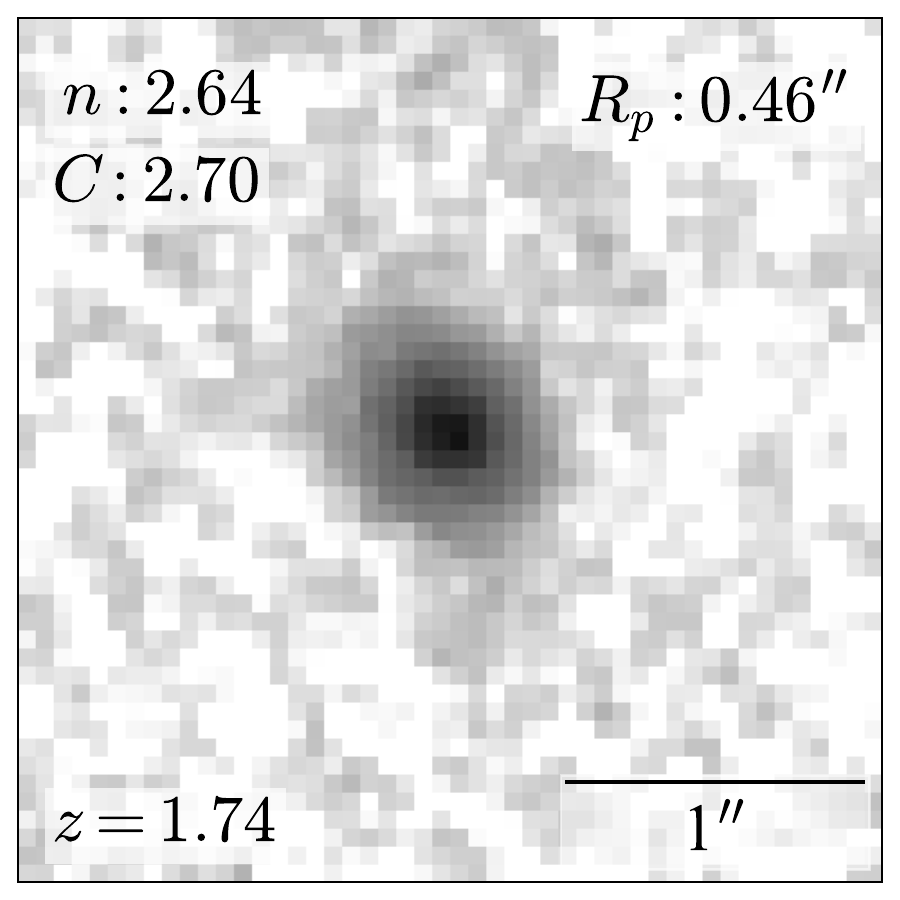}
\end{minipage}}
\hspace{-8pt}
\subfigure{
\begin{minipage}[t]{0.48\linewidth}
\includegraphics[width=1\linewidth]{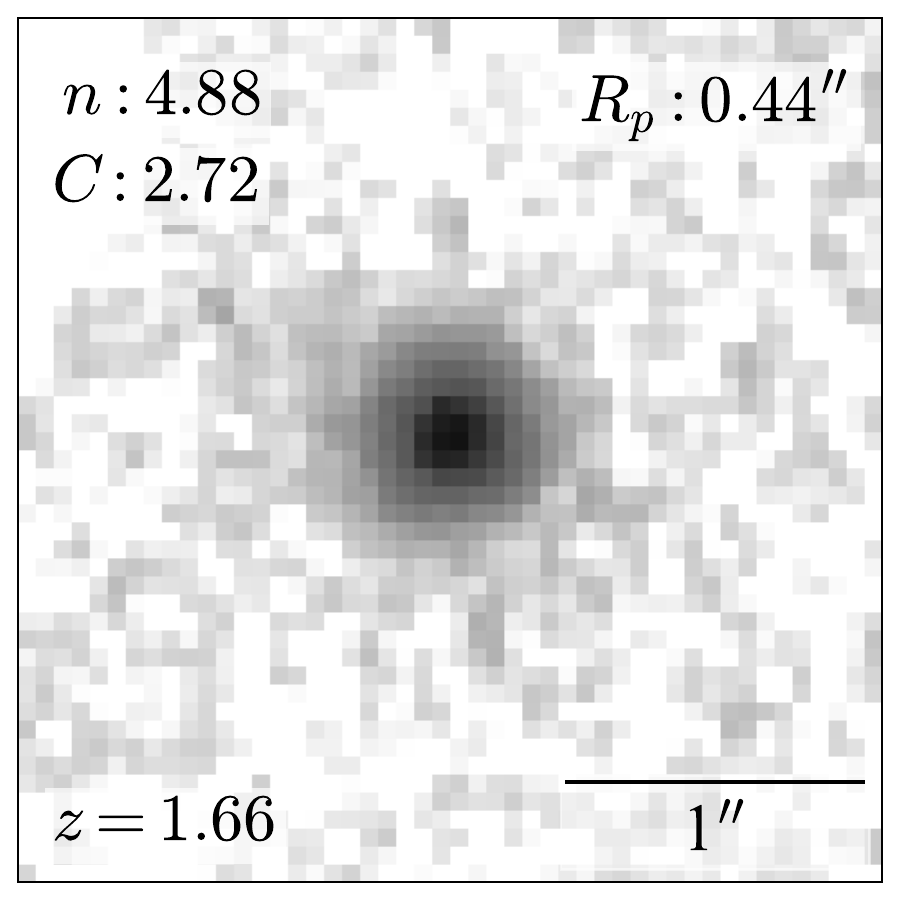}
\end{minipage}}
\hspace{-8pt}
\caption{Example of two galaxies in the EGS images at $z\sim1.7$ with different $n$ values, but similar $C$ and $R_p$. Traditional correction method would not properly reflect the intrinsic difference in the light profile between the two galaxies.  \label{fig:rp_figure}}
\end{figure}

Recently, \cite{Yu_2023} utilised the high quality DESI images of the low redshift galaxies to create the mock images for the CEERS high redshift galaxies. They considered $R_p/{\rm FWHM}$ as an additional parameter for the correction, which is more accurate than the traditional methods. This correction works well at large $R_{p}/{\rm FWHM}$, since the PSF smoothing effect is less severe. However, as shown in the left panel of Fig. 9 in their paper, the correction for small galaxies with $R_{p,True}/{\rm FWHM} =1.98$ still exhibits large scatter. This could only get worse for even smaller galaxies. For example, the two galaxies in EGS shown in Fig. \ref{fig:rp_figure} have similar $R_{p}/{\rm FWHM}$ and measured concentration values, but quite different intrinsic light profiles (i.e., the S\'{e}sic index). The two galaxies would receive the same correction factor for $C$ to produce incorrect values.
Since JWST provides images with higher resolution than HST, most galaxies with $M_{\ast}>10^{9.5}{\rm M_{\odot}}$ in our sample have $R_{p}/{\rm FWHM}>5$ and the PSF smoothing effect is much weaker. The previous correction method is less useful for JWST SW images that do not even need any correction for the concentration index.

According to our results, the non-parametric method should be mainly used in conjunction with the model-dependent parameters, to ensure more robust measurement on the galaxy structural parameters. Note that the numerical relationship between $C$ and $n$ could not be simply adopted as the baseline of the correction for the concentration-related parameters, since the light profile of the real galaxy is usually too complicated to be accurately described by a single S\'{e}rsic model. 
For $R_e/{\rm FWHM} \leq3$ galaxies, the non-parametric method could not accurately reflect the light profile concentration. The traditional correction method without considering the light profile shape is unable to recover the true values. In fact, it is recommended to directly use the model-dependent parameters ($R_e$ and $n$) from the 2D images modelling.

\cite{Yao_2023} investigated the non-parametric morphology indicators of 1376 galaxies across the redshift ranges of $z \simeq 0.8-3.0$ in the CEERS survey. Notice that no correction on the concentration index for high redshift galaxies was performed. Within this redshift range ($0.8 < z < 3$), the angular resolutions of the galaxy images are similar ($7.5\mbox{-}8.5 \ {\rm kpc\ arcsec^{-1}}$). Moreover, they matched the PSF sizes across different wavelengths with the F200W PSF (FWHM $\sim 0.70\arcsec$) at the rest-frame optical wavelength, and with the F444W PSF (FWHM $\sim 0.14\arcsec$) at longer wavelength. They found that in the rest-frame optical wavelength bands, the measured $C$ values of the massive galaxies ($M_{\ast}\geq 10^{10}{\rm M_{\odot}}$) decrease with increasing redshift, and vice versa for low-mass galaxies. This result is different from \cite{Whitney_2021} who adopted the correction for the concentration index. As \cite{Yao_2023} has mentioned, such a difference may arise from the difference in the observational wavelength and the data processing procedures. In addition, according to our tests, matching the PSF of different filters to the lowest resolution would reduce the relative sizes of the galaxies ($R_e/{\rm FWHM}$), resulting in inaccurate concentration-related parameters.

%\subsection{The structural parameters of high redshift galaxies }
Recently, \cite{Kartaltepe_2023} investigated the morphologies of 850 galaxies with $M_{\ast} > 10^9 {\rm M_{\odot}}$ at $z=3-9$ detected in both HST/WFC3 and CEERS JWST/NIRCam images. The empirical PSF FWHM of LW images are usually $\gtrsim 0.12\arcsec$ \citep{Finkelstein_2023,Zhuang_2023}, similar to the PSF FWHM of HST. In this case, the resolution of JWST detected $z>3$ galaxies is comparable to that of HST detected $1<z<2$ galaxies at the rest-frame optical wavelength, as shown in Fig. 2 of \cite{Kartaltepe_2023}. In the left panel of Fig. 8 in their work, a significant fraction of the galaxies with higher S\'{e}rsic indices show lower $C$ values ($C<3$). 
According to our results, this phenomenon indicates severe PSF smoothing effect.
The same issue may also exist in \cite{Ferreira_2023}, where the spheroids and discs with quite different S\'{e}rsic indices show very similar $C$ values at different redshifts. About the galaxy morphological classification in the $G$-$M_{20}$ space, as shown in the right panel of Fig. 8 in \cite{Kartaltepe_2023}, the S\'{e}rsic indices of some galaxies in the E-Sa and merger regions are large, consistent with our results. This alignment confirms that the $G$ and $M_{20}$ parameters are also affected by the PSF smoothing effect, but to a less extent that $C$.

The higher redshift galaxies are substantially more compact with smaller $R_e$ than the lower redshift counterparts \citep{Daddi_2005,Cassata_2013,van-der-Wel_2014}, which should naturally be more affected by PSF to exhibit lower $C$ values, especially for those with higher S\'{e}rsic indices. For the $z>3$ galaxies observed by JWST in LW, the PSF smoothing effect on the non-parametric indicators is also significant, and the model dependent method should be considered.

\subsection{Comparison between JWST and HST structural parameters}\label{sec:dif}
\begin{figure*}
\includegraphics[width=1\textwidth]{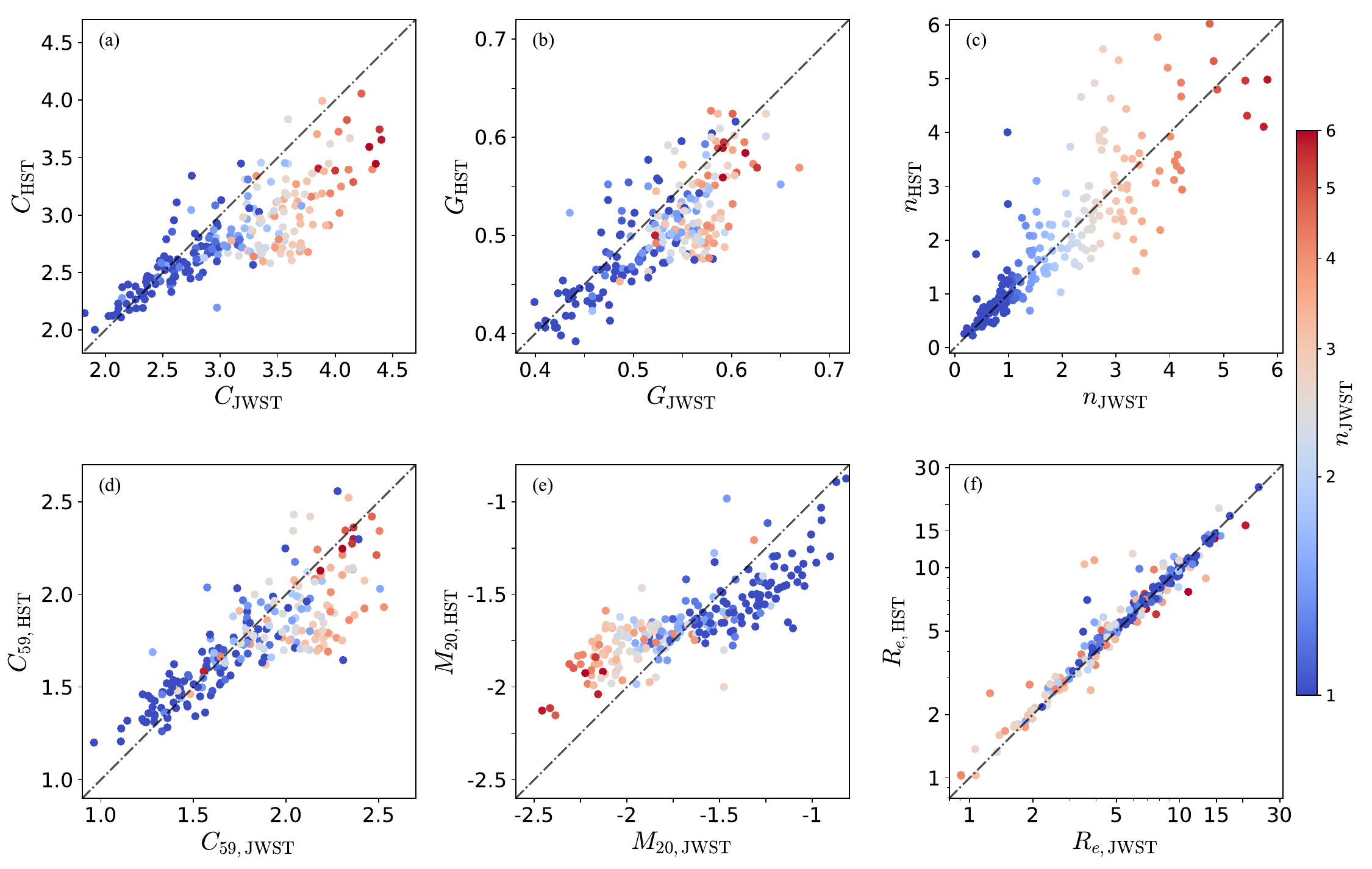}
\caption{Comparison of the structural parameters of the galaxies in our sample both observed by the JWST ($X$-axis) and HST ($Y$-axis) at the rest-frame optical wavelength, colour-coded by $n_{\rm JWST}$ measured from the CEERS images. \label{fig:dif}}
\end{figure*}

\cite{Kartaltepe_2023} and \cite{Ferreira_2023} have compared the morphologies of the same galaxies in both HST and JWST observations. Some galaxies exhibit the distinct spiral structures in JWST images that are not as pronounced as the HST images, potentially leading to misinterpretation as mergers in the HST observations. In addition, many high redshift galaxies previously identified as spheroids in HST are later confirmed as discs in JWST observations. So the structural parameters between JWST and HST observations are expected to be different.

In the comparison of the structural parameters for galaxies observed by both HST and JWST at the rest-frame optical wavelength (Fig. \ref{fig:dif}), we find that at lower concentrations (and smaller S\'{e}rsic indices), $C_{\rm HST}$ and $C_{\rm JWST}$ (top left panel) are in agreement. However, for galaxies with higher S\'{e}rsic indices, $C_{\rm JWST}$ tends to be larger than $C_{\rm HST}$.
$C_{59}$ (bottom left panel), $G$ (top middle panel) show similar pattern as the concentration index but with less scatter.
Compared to the JWST results, $M_{\rm 20, HST}$ (bottom middle panel) is overestimated at lower values while underestimated at higher values. This maybe due to the fact that $M_{20}$ is more sensitive to the resolution of images, not only for spheroids, but also for discs (though $C$, $G$ and $n$ for discs are less affected by the systematic effects).
The S\'{e}rsic indices for both HST and JWST observations are consistent but exhibiting a large scatter at higher $n$ values. The effective radii ($R_e$) are well aligned with each other. This confirms that the 2D image modelling of the HST galaxy images can produce robust S\'{e}rsic index and size measurement consistent with the JWST observations.

\subsection{The evolution of galaxy structural parameters}

\begin{figure}[ht!]
\centering
\includegraphics[width=0.45\textwidth]{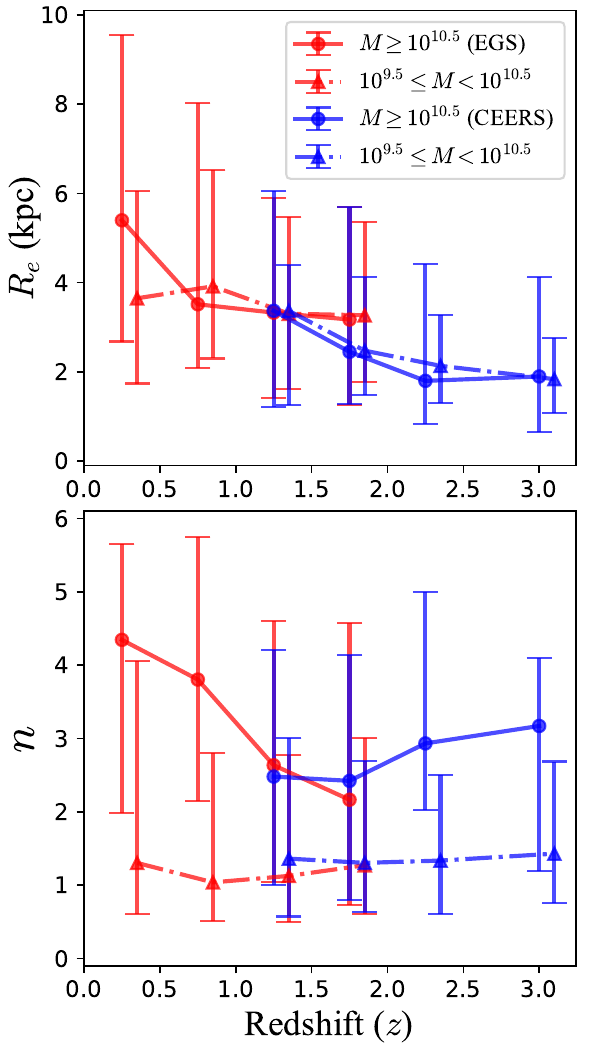}
\caption{Evolution of the effective radius (top) and S\'{e}rsic index (bottom) for galaxies in our sample. The triangle with dashed solid lines are the median values for low-mass galaxies ($10^{9.5}{\rm M_{\odot}}\leq M_{\ast} < 10^{10.5}{\rm M_{\odot}}$), while the circles with solid lines are the means for high-mass galaxies ($M_{\ast} \geq 10^{10.5}{\rm M_{\odot}}$). Error bars define the 16th and 84th percentiles of the distributions. The EGS and CEERS images are colour-coded by the red and blue lines, respectively. \label{fig:re_n_z}}
\end{figure}

Here we investigate the evolution of the galaxy light profiles for low-mass ($10^{9.5}{\rm M_{\odot}}\leq  M_{\ast} < 10^{10.5}{\rm M_{\odot}}$) and high-mass ($10^{10.5}{\rm M_{\odot}}\leq M_{\ast} < 10^{11.5}{\rm M_{\odot}}$) galaxies in our sample. We mainly focus on the S\'{e}rsic index and the effective radius derived by fitting the single S\'{e}rsic model to the galaxy mosaic with G{\small ALFIT}. As shown in Fig. \ref{fig:re_n_z}, both mass sub-samples follow similar decreasing trend of galaxy size ($R_e$) with increasing redshift. In certain redshift ranges, the typical sizes of the high-mass and low-mass galaxies are similar.
The S\'{e}rsic index of low-mass galaxies are $\sim1$ at different redshift ranges. For the high-mass galaxies, the S\'{e}rsic index decreases from $n\sim4$ at lower redshifts to $\sim3$ at higher redshifts.
Our result confirms that galaxies at higher redshifts are more compact with smaller $R_e$  \citep[e.g.][]{Daddi_2005,Trujillo_2007,Buitrago_2008,Dokkum_2010,Weinzirl_2011,Barro_2013,Williams_2014,van-der-Wel_2014,Davari_2017}. In addition, higher mass galaxies at $z<1$ have larger S\'{e}rsic indices, due to a high fraction of spheroids present at lower redshifts in our sample. While for lower mass galaxies, the S\'{e}rsic indices at each redshift bin is roughly the same ($n\sim1$), suggesting the prevalence of discs in these lower mass systems.

When describing the galaxy light distribution, it is important to distinguish between `concentrated' and `compact'. The former implies that most of the galaxy's flux is centrally distributed, while the latter implies a smaller size and higher density. Our results, along with the findings from \cite{Yao_2023}, suggest that high-mass galaxies at higher redshifts are less concentrated compared to those at lower redshifts, but they appear more compact. Various mechanisms, such as an inside-out formation process could contribute to this phenomenon. In the future, more space based large scale surveys (CSST, Euclid, Roman) will help to answer this question of galaxy growth and structure evolution.

\section{Conclusion}\label{sec:conclusion}
In this work, we investigate the PSF smoothing effect on the concentration-related structural parameters ($C$, $G$, $M_{20}$) in observations and mock images. We select a mass-limited sample ($10^{9.5}{\rm M_{\odot}}\leq M_{\ast} \leq 10^{11.5}{\rm M_{\odot}}$) of  2305 galaxies at redshift $0 < z < 2$ from the EGS field in the CANDELS survey and 524 galaxies at $1 < z < 3$ from the CEERS survey. 
The images in different filters are used for different redshift ranges to ensure similar rest-frame optical wavelength. 
It has been shown that the S\'{e}rsic index can be robustly derived from 2D image fitting with PSF convolution, therefore it can be used as a reliable indicator for the intrinsic shape of the light profile.
To understand the PSF smoothing effect on the non-parametric morphology indicators for high redshift galaxies, we investigate the dependence of the non-parametric morphology indicators on the S\'{e}rsic index and the relative size of the galaxy ($R_e/{\rm FWHM}$) with HST and JWST observations, and mock images.

Our conclusions are listed as follows.

\begin{enumerate}
  \item The concentration index is generally underestimated due to the PSF smoothing effect, especially for the smaller galaxies with higher S\'{e}rsic index. For the galaxies with lower S\'{e}rsic index ($n\leq2$) or larger relative size ($R_e/{\rm FWHM}>3$), the concentration index is almost unaffected. With the idealised mock images, we confirm that the underestimation of the $C$ value is mainly due to the more significant overestimation of $R_{20}/R_e$ (compared to $R_{80}/R_e$). Another commonly used concentration index $C_{59}$ (derived from $R_{90}$ and $R_{50}$) is less affected by the PSF smoothing effect, except for the smaller galaxies with $R_e/{\rm FWHM}<1$. The correlation between $C_{59}$ and $n$ is relatively weaker (with larger scatter) than that between $C$ and $n$.
  
  \item The Gini coefficient is less affected by the PSF smoothing effect, while the absolute value of $M_{20}$ is underestimated, in a similar fashion to the $C$-value. For the galaxies observed by HST, the empirical morphological classification in the $G$-$M_{20}$ space works well at lower redshifts, but fails at higher redshifts, with the discs and spheroids mixed together. However, for the galaxies observed by JWST at $1<z<3$, the empirical separation between the E-Sa and Sb-Irr galaxies in the $G$-$M_{20}$ space works well.
  
  \item We discuss the influence of the axis ratio of the aperture used in the concentration index measurement. With the traditional circular aperture, the $C$ values of the edge-on discs are overestimated to deviate from the theoretical expectation between the $C$ and $n$ parameters. When adopting the elliptical apertures with the same ellipticity as the galaxy outskirt, the correlation between $C$ and $n$ is significantly improved to show much tighter correlation. Our main conclusions are not affected by the different aperture shapes.
  
  \item The traditional correction method for the non-parametric morphology indicators may not be accurate enough. Both the relative size of the galaxy ($R_e/{\rm FWHM}$) and the S\'{e}rsic index play important roles here. In fact, the S\'{e}rsic index is a better choice to represent the light concentration for these high redshift galaxies.
  
  \item According to the structural parameters measured with single S\'{e}sic component fitting, we confirm that the galaxies at higher redshifts are more compact with smaller $R_e$. At different redshifts, the lower mass galaxies have S\'{e}rsic index $n\sim1$ (disc dominated). For higher mass galaxies, the S\'{e}sic index decrease from $\sim 4$ to $\sim3$ at higher redshifts.
  
\end{enumerate}

\begin{acknowledgements}
We thank the referee for the helpful comments to improve the paper and Iulia Simion for polishing the language. This work is supported by the National Natural Science Foundation of China under grant No. 12122301, 12233001, by a Shanghai Natural Science Research Grant (21ZR1430600), by the ``111'' project of the Ministry of Education under grant No. B20019, and by the China Manned Space Project with No. CMS-CSST-2021-A04. We thank the sponsorship from Yangyang Development Fund. This work made use of the Gravity Supercomputer at the Department of Astronomy, Shanghai Jiao Tong University. LCH was supported by the National Science Foundation of China (11721303, 11991052, 12011540375, 12233001), the National Key R\&D Program of China (2022YFF0503401), and the China Manned Space Project (CMS-CSST-2021-A04, CMS-CSST-2021-A06). This work has made use of the Rainbow Cosmological Surveys Database, which is operated by the Centro de Astrobiología (CAB/INTA), partnered with the University of California Observatories at Santa Cruz (UCO/Lick,UCSC).
\end{acknowledgements}

% to follow the A&A style
% for the bibliography, at the end
\bibliographystyle{aa} % style aa.bst
\bibliography{ms} % your references Yourfile.bib

\begin{appendix}
%\onecolumn
\section{Idealised mock images for JWST}\label{app:a}

\begin{figure*}[t]
\centering
\includegraphics[width=1\textwidth]{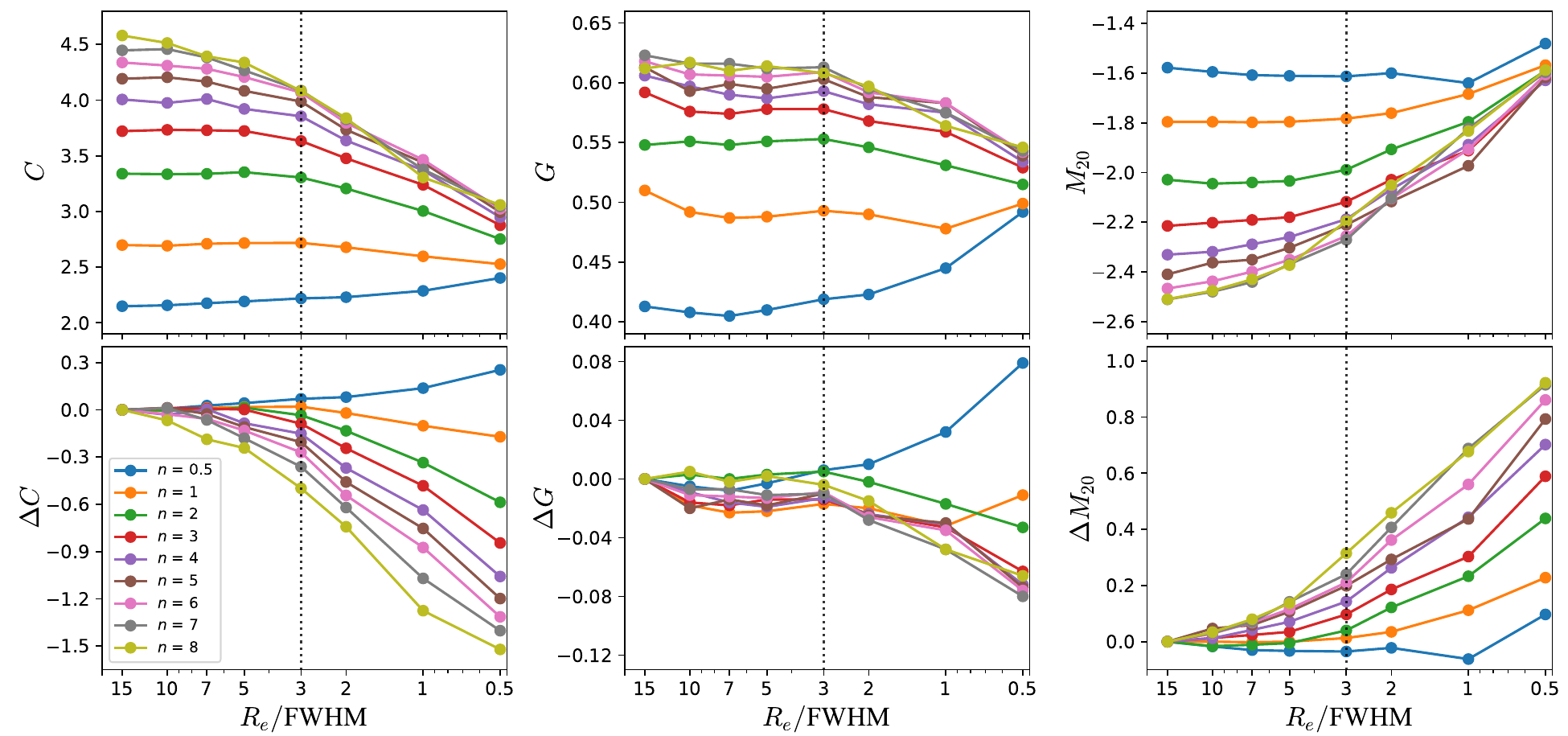}
\caption{The measured concentration index (left), Gini coefficient (middle) and the $M_{20}$ index (right) for the idealised mock images in CEERS survey as a function of the relative size of the galaxy ($R_e/{\rm FWHM}$), with different colours representing different S\'{e}rsic indices. The top and bottom rows show the structural parameter ($C$, $G$, $M_{20}$) and the deviation from the intrinsic values ($\Delta C$, $\Delta G$, $\Delta M_{20}$), respectively. This trend in each panel is similar to Fig. \ref{fig:hst_gal}. \label{fig:jwst_gal}}
\end{figure*}

Similar to the generation of idealised mock galaxies for the EGS F160W images in Sect. \ref{sec:c_mock}, we create a series of mock images on a $500\times 500$ pixel grid with $0.5\leq n\leq 8$ and $0.5 \leq R_e/{\rm FWHM}\leq 15$ ($0.035\arcsec\leq R_e\leq1.05\arcsec$) at a total magnitude of $m=21$ (with ${\rm S/N}>20$). These images are then convolved with the NIRCam/F200W band PSF and added to the sky background in the same filter. The mock images are measured using {\tt Statmorph} to extract the non-parametric morphology indicators.
The impact of the PSF on $C$, $G$ and $M_{20}$ is shown in Fig. \ref{fig:jwst_gal}. The top rows show the variation of indicators with different $R_e/{\rm FWHM}$ for different S\'{e}rsic profiles, and the bottom rows show the relative deviation. 
The trend of decreasing $R_e/{\rm FWHM}$ is consistent with the result of the EGS S\'{e}rsic mock galaxies.
When the PSF FWHM is larger than the $R_e$ ($R_e/{\rm FWHM} =0.5 $), the $C$ values converge to $\sim 3.5$ for all galaxies.
However, for $n=0.5$ cases, the $G$ values increase rapidly as $R_e/{\rm FWHM}$ decreased and converge to $\sim0.5$, close to the results of higher $n$ objects. This behaviour may be attributed to the small FWHM of the PSF from the NIRCam detector, providing higher quality images.

\section{The procedures of artificially redshifted galaxies}\label{app:b}

The steps for artificially redshifted galaxies have been outlined in previous studies \citep{Giavalisco_1996,Conselice_2003,Barden_2008, Yu_2023}. Following their methods, these steps are listed below.
e
\begin{enumerate}
  \item Compute rebinning factor and rebin images
  
  The physical size of a galaxy will remain at higher redshift, so its angular size $a_f$ can be calculated from initial redshift $z_i$ to final redshift $z_f$
  \begin{eqnarray}
    D_{A,i} a_i & = & D_{A,f} a_f , \\
    \frac{D_{L,i} a_i}{{(1+z_i)^2}} & = & \frac{D_{L,f} a_f}{{(1+z_f)^2}} ,
  \end{eqnarray} 
  
  where $D_A$ and $D_L$ are angular distance and luminosity distance at objective redshift respectively. When simulating initial imaging surveys with a pixel scale of $p_i$ transitioning to a target pixel scale of $p_f$, the rebinning factor is given by
  
  \begin{equation}
      B = \frac{a_i/p_i}{a_f/p_f} = \frac{p_f D_{L,f}}{p_i D_{L,i}}\frac{(1+z_i)^2}{(1+z_f)^2}.
  \end{equation}
  Here, the small angle approximation $tan(a)\sim a$ is employed. 

  We also consider the size evolution of galaxies given by $R_e \varpropto(1+z)^{\beta}$ \citep{van-der-Wel_2014}, then the rebinning factor becomes

  \begin{equation}
      B = \frac{p_f D_{L,f}}{p_i D_{L,i}} \cdot \frac{(1+z_i)^2}{(1+z_f)^2} \cdot \frac{(1+z_i)^{-\beta}}{(1+z_f)^{-\beta}},
  \end{equation}
  
  \item Rescale the flux
  
    If the bolometric luminosity of the galaxy remains constant when shifting to different redshifts, we have
    
    \begin{equation}
    4\pi D_{L,i}^2 f_i = 4\pi D_{L,f}^2 f_f.
    \end{equation}
    
    Here, $f_i$ and $f_f$ are the observed fluxes at $z_i$ and $z_f$. The ratio of the flux density per unit solid angle between $z_i$ and $z_f$ is then given by:
    
    \begin{equation}
    \frac{f_f/a_f^2}{f_i/a_i^2} = \frac{D_{L,i}^2}{D_{L,f}^2} \cdot \frac{a_i^2}{a_f^2} = \frac{(1+z_i)^4}{(1+z_f)^4},
    \end{equation}
    
    resulting in the standard surface brightness dependence of $(1+z)^{-4}$ \citep{Tolman_1930}.
  A filter has central wavelength $\lambda$ and width $\Delta \lambda$. Both the observed central wavelength and the observed bandwidth increase by a factor of $(1+z)$. In units of erg ${\rm s}^{-1} {\rm cm}^{-2}$ ${\rm Hz}^{-1}$, the surface brightness scales as $\mu \varpropto (1+z)^{-3}$. 
  Consider the luminosity evolution given by $(1 + z)^{\alpha}$ \citep{Yu_2023}, the final apparent magnitude ($M_{f}$) is
  \begin{equation}
      M_f = M_i + 2.5{\rm log}\left(\frac{1+z_f}{1+z_i}\right)^3 - 2.5{\rm log}\left(\frac{1+z_f}{1+z_i}\right)^{\alpha}.
  \end{equation}
  So the final surface brightness($\mu_f$) becomes
    \begin{eqnarray}
        \mu_f &=& \mu_i + 2.5{\rm log}\left(\frac{1+z_f}{1+z_i}\right)^3 \nonumber \\&-& 2.5{\rm log}\left(\frac{1+z_f}{1+z_i}\right)^{\alpha}- 2.5{\rm log}\left(\frac{1+z_f}{1+z_i}\right)^{-2\beta}.
    \end{eqnarray}
    The morphological or magnitude k-correction is ignored, since our samples are at the rest-frame optical waveband.

  \item PSF convolution
  
    In order to mimic the PSF effect, we rebin the original PSF in the filter at lower redshift, similar to the galaxies rebinning procedure. Then we utilise the {\tt create\_matching\_kernel} function in {\tt Photutils} \citep{larry_bradley_2022_6825092} to create the PSF matching kernel $\kappa$, following 

    \begin{eqnarray}
        {\rm PSF}_i(x,y)\otimes \kappa(x,y) = {\rm PSF}_f(x,y), \\ 
        \kappa(x,y) = \digamma^{-1}\left(\frac{\digamma({\rm PSF}_f(x,y)}{\digamma({\rm PSF}_i(x,y)}\right),
    \end{eqnarray}
    where $\otimes$ is the convolution operator, and $\digamma$ ($\digamma^{-1}$) is (inverse) Fourier transformation operator.
    We choose the {\tt TukeyWindow} to remove high frequency noise from the PSF matching kernel with 
    \begin{equation}
    \sigma_\kappa = \sqrt{\sigma_f^2-\sigma_i^2},  
    \end{equation}
    where $\sigma_\kappa$, $\sigma_f$ and $\sigma_i$ are the stand deviation of the PSF matching kernel, target PSF and initial rebinned PSF, respectively. We subsequently convolve the rebinned image with the PSF matching kernel $\kappa$.
    
  \item Background and noise addition
  
    For the background noise, we utilise the mosaic cutout without sources from actual observations and incorporate it into the images. 

\end{enumerate}

\end{appendix}

\end{document}